\documentclass[12pt]{article}
\usepackage{amsmath}
\usepackage{graphicx}
\usepackage{natbib}
\usepackage{url} % not crucial - just used below for the URL 
\usepackage{xr-hyper} % To reference external documents

\newcommand{\blind}{0}

\usepackage{amsfonts}
\usepackage{amssymb}
\usepackage{lscape}
\usepackage{booktabs}
\usepackage{color}
\usepackage[noend]{algpseudocode}
\usepackage{graphicx}
\usepackage{pgfplots}
\usepackage{pgfplotstable}
\usepackage{epstopdf}
\usepackage{subcaption}
\usepackage{caption}
\usepackage{tikz}
\usepackage[normalem]{ulem}%
\usepackage[ruled]{algorithm2e}
\usepackage{media9}
\usepackage{hyperref} % For hyperlinking references
\usepackage{xr-hyper} % To reference external documents
\usepackage{comment}
\usepackage{color,soul}
\usepackage{soul}
\usepackage{titlesec}
\graphicspath{ {./images/} }

\providecommand{\U}[1]{\protect\rule{.1in}{.1in}}
\graphicspath{ {./images/} }
\pgfplotsset{compat=1.8}
\usepgfplotslibrary{groupplots}
\pgfplotsset{compat=newest}
\pgfplotsset{plot coordinates/math parser=false}
\newlength\figureheight
\newlength\figurewidth
\pgfplotsset{yticklabel style={text width=3em,align=right}}
\pgfplotsset{xticklabel style={text width=2em,align=right}}

\newcommand{\y}{\mathbf{y}}
\newcommand{\z}{\mathbf{z}}

\makeatletter

\def\1{1\!{\rm l}}
\def\BState{\State\hskip-\ALG@thistlm}

\makeatother
\hypersetup{
	pdfborder={0 0 0}, % Remove the red box
	colorlinks=false    % Disable link coloring
}

\begin{document}

\def\spacingset#1{\renewcommand{\baselinestretch}%
{#1}\small\normalsize} \spacingset{1}

%%%%%%%%%%%%%%%%%%%%%%%%%%%%%%%%%%%%%%%%%%%%%%%%%%%%%%%%%%%%%%%%%%%%%%%%%%%%%%

\if0\blind
{
	\title{\bf Robustifying Approximate Bayesian Computation}
	\author{Chaya Weerasinghe\thanks{Weerasinghe has been supported by a Monash Graduate Scholarship. Frazier, Loaiza-Maya, and Drovandi have been supported by Australian Research Council funding schemes DE200101070, DE230100029 and FT210100260, respectively.}, David T. Frazier, Rub\'en Loaiza-Maya\hspace{.2cm}\\
		Department of Econometrics and Business Statistics, Monash University\\
		and \\
		Christopher Drovandi \\
		School of Mathematical Sciences, Queensland University of Technology}
	\maketitle
} \fi
\if1\blind
{
  \bigskip
  \bigskip
  \bigskip
  \begin{center}
    {\LARGE\bf Robustifying Approximate Bayesian Computation}
\end{center}
  \medskip
} \fi

\bigskip
\begin{abstract}
Approximate Bayesian computation (ABC) is one of the most popular `likelihood-free’ methods. These methods have been applied in a wide range of fields by providing solutions to intractable likelihood problems in which exact Bayesian approaches are either infeasible or computationally costly. However, the performance of ABC can be unreliable when dealing with model misspecification. To circumvent the poor behavior of ABC in these settings, we propose a novel ABC approach that is robust to model misspecification. This new method can deliver more accurate statistical inference under model misspecification than alternatives and also enables the detection of summary statistics that are incompatible with the assumed data-generating process. We demonstrate the effectiveness of our approach through several simulated examples, where it delivers more accurate point estimates and uncertainty quantification over standard ABC approaches when the model is misspecified. Additionally, we apply our approach to an empirical example, further showcasing its advantages over alternative methods.
\end{abstract}

\noindent%
{\it Keywords:} Likelihood-free inference; Model misspecification; Robust Bayesian inference; Spike-and-slab priors
\vfil

%\spacingset{1.75} % DON'T change the spacing! -- jcgs site
\spacingset{1.45} %overleaf template

\section{Introduction}
\label{sec:intro}

%\dtf{You need to have a go at cutting this down before I even look at it with a closer eye. There is lots that can be cut here. ABC is very-well developed an understood at this stage, you can just give the highlights. Your introduction should be no more than 2.5 pages in reality. }

Approximate Bayesian methods have gained popularity for conducting reliable inference in complex models where exact Bayesian inference procedures are either infeasible or computationally impracticable. These methods make estimation feasible for ``intractable" statistical models, most notably those for which the likelihood function is not available in closed form. The two most common likelihood-free Bayesian methods in the existing statistical literature are approximate Bayesian computation (ABC) (\citealp{marin2012approximate}; \citealp{sisson2018handbook}) and Bayesian synthetic likelihood (BSL) (\citealp{price2018bayesian}; \citealp{frazier2023bayesian}).

ABC simply requires simulating pseudo-data from an assumed model based on draws from the prior. The parameter draws that produce a ``good" match, in the sense that a chosen distance between the observed and simulated data is small, are ultimately used to conduct posterior inference on the model unknowns. However, due to the curse of dimensionality, the majority of ABC implementations reduce the datasets down to vectors of lower-dimensional summary statistics and conduct the inference based on these summary statistics to reduce the computational burden (\citealp{martin2024approximating}). Whilst ABC implicitly constructs a non-parametric estimate of the likelihood for the summaries, BSL, in contrast, approximates the intractable likelihood by assuming the summaries follow a Gaussian distribution with an unknown mean and variance.

A broad spectrum of inferential studies have been conducted on both ABC and BSL under correct model specification, with fewer considering the case of model misspecification (see \citealp{frazier2020model} and \citealp{frazier2024synthetic}, for examples). By the very nature of the complex scenarios to which approximate Bayesian methods are applied, the class of models used to simulate pseudo-data is highly likely to be misspecified. In such situations where the assumed model is misspecified, which is often the case in practice, both theoretical and empirical evidence from the literature indicates that ABC can produce misleading model inferences. Specifically, \cite{frazier2020model} demonstrate that under model misspecification, while the ABC posterior maintains concentration onto some well-defined pseudo-true value in the parameter space, it does not yield credible sets with valid frequentist coverage and exhibits non-standard asymptotic behaviour. 

Another particular consequence of model misspecification noted in ABC is that different versions of ABC can deliver significantly different point estimators under even mild levels of model misspecification, with more complicated ABC approaches, e.g. post-processing ABC approaches that entail regression adjustments (\citealp{blum2010non}; \citealp{blum2018regression})\textbf{,} performing poorly in comparison with the simpler accept-reject ABC (\citealp{frazier2020model}). In particular, under model misspecification, regression-adjusted ABC posteriors can concentrate posterior mass on parameter values that have no meaningful interpretation, and this behaviour is particularly concerning because these methods are often used to mitigate the curse of dimensionality in ABC inference, and are popular in empirical applications. More importantly, unreliable ABC inference can also have a significant impact on predictions in misspecified models, as discussed in \cite{WEERASINGHE2025270}.

Thus, there are clear reasons to reconsider how to apply ABC when the model specified is in doubt. With this goal in mind, we propose a novel ABC method that caters for possible model misspecification and delivers inferences that are robust to such settings. This new approach takes inspiration from the robust BSL (R-BSL) method proposed by \cite{frazier2021robust}, which augments the BSL posterior with a vector of adjustment parameters that ``soak-up" the model misspecification. Our approach relies on the fairly innocuous assumption that the summary statistics used in ABC can be partitioned into two sets: first, are summaries for which the assumed model can match the observed data, and which are informative for at least one unknown parameter; second, are summaries that the researcher believes the model may not be able to match. Building on this partition, we propose a two step ABC method: we localize the posterior approximation using the initial set of summaries that can be matched under the model; second, we implement a robust ABC procedure that can adjust the location of the second set of summaries when model misspecification is present. Since this approach is designed to deliver robust inferences in misspecified models, we refer to it as ``robust ABC" (R-ABC).

As evidenced by a series of simulation and empirical examples, this R-ABC approach has at least two main advantages over standard ABC methods. Firstly, this new two-step procedure is less sensitive to model misspecification than standard ABC and local post-processing ABC approaches. Second, R-ABC has an in-built mechanism for diagnosing model misspecification, allowing the researchers to detect which summary statistics cannot be matched by the assumed data generating process (DGP). By taking a two-step approach, R-ABC circumvents parameter proliferation, which is a severe issue in ABC. This R-ABC approach can be used in similar situations as the R-BSL approach in \cite{frazier2021robust}, however, it is well-known that reliable inference in BSL ultimately requires the Gaussian approximation to the likelihood of the summaries to be reasonably accurate (\citealp{frazier2023bayesian}). In contrast to BSL, ABC is largely unaffected by the specific distribution of the summaries. Therefore, R-ABC can be applied to deliver reliable inferences in cases where R-BSL may be unreliable or difficult to implement due to the violation of the Gaussian approximation for the summaries.

The remainder of the paper is organized as follows. In Section \ref{abc_and_model_misspec}, we give a brief overview of ABC and discuss the consequences of model misspecification in ABC. Section \ref{rabc} presents our robust approach to ABC. Section \ref{simulation_ex} then illustrates the performance of our proposed robust ABC approach in an extensive set of simulation experiments, including comparative analysis with the R-BSL approach. An empirical illustration that demonstrates the performance of this robust ABC approach is included in Section \ref{empirical_exercise}. The paper concludes in Section \ref{discussion}.

\section{ABC and model misspecification}
\label{abc_and_model_misspec}
\subsection{An outline of ABC}

We observe data $\mathbf{y}:=(y_{1},...,y_{n})^{\prime }$ from some true distribution $P_{0}$. Since $P_{0}$ is unknown, we model $\mathbf{y}$ using class of parametric models $\mathcal{P}:=\{P_{\theta}: \theta\in\Theta \subseteq \mathbb{R}^{d_{\theta}}\}$ that depends on unknown parameters $\theta$, where $P_{\theta}$ denotes the probability measure of the model, $p(\cdot\mid{\theta})$ is the corresponding density function, and our prior beliefs over $\Theta$ are given by the density $\pi(\theta)$. %Using the observed data $\mathbf{y}$, the model $P_{\theta}$, and the prior density $\pi(\theta)$, Bayes rule delivers the posterior density $\pi(\mathbf{\theta |y})\propto p(\mathbf{y}|\theta) \pi(\mathbf{\theta}).$

Drawing samples from the Bayesian posterior $\pi(\mathbf{\theta |y})\propto\pi(\theta)p(\y|\theta)$ requires that  $p(\mathbf{y}|\theta)$ is analytically tractable, or can be unbiasedly estimated. In settings where $p(\mathbf{y}|\theta)$ is intractable,  likelihood-free methods like ABC or BSL replace likelihood evaluation with model simulations, to obtain an approximation to the exact posterior; for a review of such methods see \citealp{martin2024approximating}.

The goal of methods like ABC and BSL is to produce draws from an approximation to $\pi(\mathbf{\theta |y})$ in cases where $p(\mathbf{y}|\theta)$ is intractable. ABC conducts inference on the unknown $\theta$ by first drawing $\theta$ from $\pi(\theta)$, then simulating pseudo-data $\mathbf{z}:=(z_{1},...,z_{n})^{'}$ (with support $\mathcal{Z}$) from $P_{\theta}$,  and ``comparing" $\mathbf{z}$ with the observed data $\mathbf{y}$. In most cases, this comparison is carried out using a vector of summary statistics $\eta(\cdot): \mathbb{R}^{n} \to \mathbb{R}^{d_{\eta}}, d_{\eta} \geq d_{\theta}$ (where $d_{\eta} = \mathrm{dim}(\eta)$ and $d_{\theta} = \mathrm{dim}(\theta)$) and a metric $d\{\cdot,\cdot\}$. The draws of $\theta$ that yield a small distance $d\{\eta(\z),\eta(\y)\}$ relative to a pre-defined tolerance level, $\epsilon$, can be shown to be draws from the ABC posterior:
\begin{flalign}
	\pi_{\epsilon }[{\theta |\eta (\y)}]&=\int_{\mathcal{Z}}\pi_{\epsilon}[\theta,\z|\eta(\y)]d\z,\quad 	\pi_{\epsilon}[\theta,\z|\eta(\y)] =  \frac{\pi(\theta)p(\mathbf{z}|\theta)\1_{\epsilon}(\mathbf{z})}{\int_{\Theta}\int_{\mathcal{Z}} \pi(\theta)p(\mathbf{z}|\theta)\1_{\epsilon}(\mathbf{z}) d\z d\theta},  \label{abc_post}
\end{flalign}
where $\1_{\epsilon}(\mathbf{z}) = \1\{  d\{\eta(\mathbf{z}),\eta(\mathbf{y})\}\leq \epsilon  \}$ is one if $d\{\eta(\mathbf{z}),\eta(\mathbf{y})\}\leq \epsilon$ and zero elsewhere. 

The simplest (accept-reject) ABC algorithm (\citealp{tavare1997inferring}; \citealp{pritchard1999population}) proceeds as per Algorithm \ref{ABC} (see Appendix \ref{appendix_A}). Among the extensions to the basic accept-reject ABC algorithm, the most common post-processing approach is the so-called local linear regression adjustment (\citealp{beaumont2002approximate}; see \citealp{blum2018regression} for a review of this common regression post-processing approach), and it is represented in Algorithm \ref{ABC-Reg} in Appendix \ref{appendix_A}.

Whilst, there are many algorithms for sampling from the ABC posterior, in this article we use the sequential Monte Carlo (SMC)-ABC algorithm described by \cite{drovandi2011estimation} which is also known as the ``ABC-SMC replenishment algorithm". This method generates samples (referred to as particles in the SMC context) from an adaptive sequence of ABC posteriors with decreasing ABC thresholds. It eliminates particles with the highest discrepancies at each iteration and rejuvenates the particle population through a resampling and move step. The move step uses an Markov chain Monte Carlo (MCMC) ABC kernel to maintain the particle distribution. The number of MCMC iterations for each particle adapts based on the overall MCMC acceptance rate. In this article, the SMC-ABC algorithm is stopped once the MCMC acceptance rate falls below 1\%.

\subsection{Model misspecification in simulation-based inference}

%\dtf{This can all be drastically shortened. We don't need any of this except really the Last paragraph. I've drastically shortened this.}

In ABC, the researcher implicitly assumes that there are values of $\theta$ such that $\eta(\z)$ is ``close to''  $\eta(\y)$. However, this is unlikely to be the case in practice: due to the inherent complexity of the DGP to which ABC is applied, the class of models $\mathcal{P}$ used to simulate pseudo-data $\z$ is highly likely to be misspecified.  Following \cite{frazier2020model}, the meaningful concept of model misspecification in ABC is whether or not $\eta(\z)$ can match $\eta(\y)$ for some $\theta\in\Theta$. In particular, for $b_0:=\text{plim}_n\eta(\y)$ and $b(\theta):=\text{plim}_n \eta(\mathbf{z})$, where $\text{plim}_n$ denotes probability limit as $n\rightarrow\infty$, we say that model is misspecified, in the ABC {\it{sense}} if
	$$\inf_{\theta\in\Theta}\|b(\theta)-b_0\|>0.$$

Model misspecification in ABC means that asymptotically for any value of $\theta \in \Theta$, $\eta(\y)$ cannot be replicated by $\eta(\z)$ where $\z$ is simulated under $P_\theta$. As shown in \cite{frazier2020model}, under model misspecification, ABC-based inference can be unreliable: the ABC posterior displays non-standard asymptotic behavior and does not produce reliable credible sets. Further, it is now well-known that common regression post-processing approaches, like linear regression adjustment may lead to very poor inferences in misspecified models. 

\subsection{R-BSL}
%\dtf{You can't apply something in a paper without first giving the background on it. If you ware going to }
Given the poor behavior of ABC and ABC-Reg in misspecified models, \cite{frazier2021robust} proposed a method to fit misspecified models using BSL. The approach, called robust BSL (R-BSL) is based on the idea of model expansion, whereby additional parameters are artificially added to `soak-up' the underlying model misspecification. The R-BSL mean (R-BSL-M) adjustment approach adds to $\eta(\z)$ a vector of adjustment components $\Gamma\in\mathbb{R}^{d_\eta}$ to create a new vector of simulated summaries:
$$\eta_m(\theta,\Gamma):= \frac{1}{m}\sum_{i=1}^{m}\eta(\z^i) +\text{diag}\big[\Sigma_m(\theta)^{1/2}\big]\Gamma,\quad \z^i\stackrel{iid}{\sim} P_\theta .$$
The R-BSL-M posterior is constructed by assuming a Gaussian likelihood for $\eta(\y)$ with mean $\eta_m(\theta,\Gamma)$ and variance $\Sigma_m(\theta)=\frac{1}{m}\sum_{i=1}^{m}[\eta(\z^i)-\frac{1}{m}\sum_{i=1}^{m}\eta(\z^i)][\eta(\z^i)-\frac{1}{m}\sum_{i=1}^{m}\eta(\z^i)]^{'}$. Given that $\Gamma$ is a random parameter, the joint vector of unknowns becomes $\zeta:=(\theta^{'}, \Gamma^{'})^{'}$. An alternative approach, also proposed in \cite{frazier2021robust}, is the R-BSL variance approach (R-BSL-V), which is also based on a Gaussian likelihood for $\eta(\y)$, but with mean $\frac{1}{m}\sum_{i=1}^{m}\eta(\z^i) $ and variance 
$$
\Sigma_m(\theta,\Gamma):=\Sigma_m(\theta)+\Sigma_m(\theta)^{-1/2}\text{diag}(\Gamma)\Sigma_m(\theta)^{-1/2}, 
$$ where now $\Gamma\in\mathbb{R}^{d_\eta}_+$. Across many different numerical experiments and empirical examples, the R-BSL-V posterior has demonstrated remarkable accuracy even in highly misspecified models (see, e.g., \citealp{frazier2024synthetic} for examples).

\subsection{Brief motivating example}
\label{motivating_ex}
While R-BSL is a useful approach for fitting intractable models in a robust manner, it is not a panacea. In particular, it is well-known that ABC methods are more robust than BSL methods when the underlying summaries are non-Gaussian, owing to the Gaussian likelihood assumption used in BSL. In the following brief example, we compare the behavior of ABC and R-BSL and show that R-BSL cannot accurately capture model misspecification in cases where the summaries display non-Gaussian behavior.

\begin{comment}
\dtf{This is not a ``brief motivating example''. This needs to be at most a page or two. The following is a list of all you want in this section. The rest should go in the more general analysis section. 
	\begin{itemize}
	\item We consider a series of financial returns, which are well-known to exhibit heavy tails, and thus will likely deliver non-Gaussian summaries. 
	\item We model the series using an $\alpha$-stable SV model (Citation goes here), but for brevity we defer the model details until section xxx. Following \cite{martin2019auxiliary}, we use as summary statistics those from an auxiliary GARCH model with student-t errors. 
	\item We compare the ABC, ABC-Reg and R-BSL posteriors for the tail index parameter, which governs the thickness of the tails captured in the data. 
	\item Then, give a very brief synopsis of the results along with Figure 2. 
	\item Move the rest of this to Section 5 where we explain the example in more detail. Even then, you will need to cut this material down quite a bit. This section needs to be bare bones. 
	\end{itemize}
}
\end{comment}

\begin{figure}[h!]
	\centering 
	\includegraphics[width=0.5\linewidth]{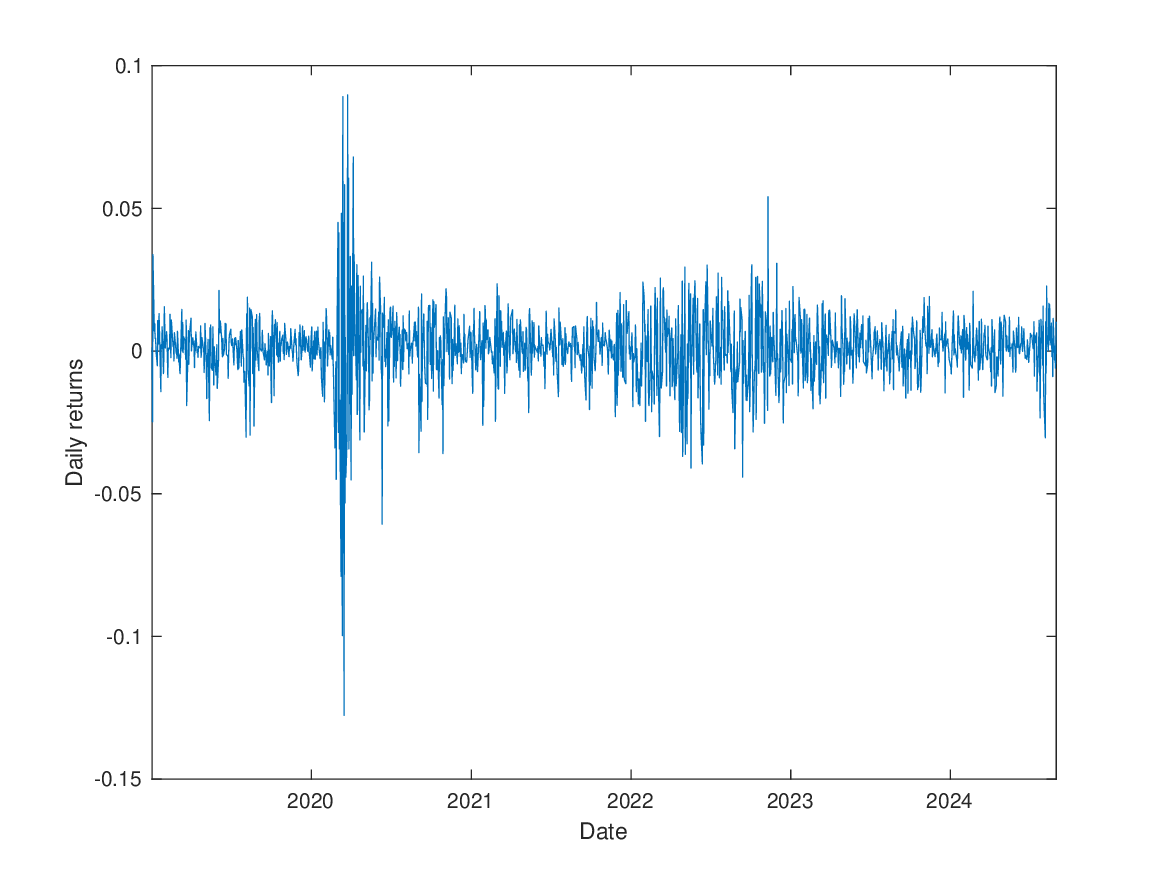} 
	\caption{Daily returns data of S\&P500 index from 2 January 2019 to 30 August 2024.} 
	\label{emp_returns} 
\end{figure}

We use daily close-to-close S\&P500 returns data (sourced from Global Financial Data) from 2 January 2019 to 30 August 2024, comprising 1425 observations, and the data is shown in Figure \ref{emp_returns}. The series exhibits significant volatility clustering, with a relatively calm period interrupted by sudden jumps in volatility, associated with the impact of the COVID-19 pandemic on global financial markets. The heavy-tailed nature of the series is suggesting the presence of non-Gaussian summaries, and it is evidenced by Figure \ref{emp_est_sum_goodparam} in Appendix \ref{appendix_D}, where further discussion on this is also provided.

We model the series using an $\alpha$-stable stochastic volatility (SV) model (\citealp{lombardi2009indirect}; \citealp{martin2019auxiliary}), but for brevity we defer the model details until Section \ref{empirical_exercise}. Following \cite{martin2019auxiliary}, we choose summary statistics derived from an auxiliary GARCH model with Student-t errors, where the summary statistics are defined as in (\ref{summary_emp}). Here, we conduct inference on $\theta$ using two different versions of ABC: the SMC-ABC approach (hereafter, ABC-SMC), where we choose $d\{x,y\}=\|x-y\|$ to be the Euclidean norm, and a local linear regression adjustment approach to SMC-ABC (hereafter, ABC-SMC-Reg).\footnote{Following \cite{beaumont2002approximate}, we take as the kernel function, $K_\epsilon(t)$, the Epanechnikov kernel: $K_\epsilon(t)=  c \epsilon ^ { - 1 } \left( 1 - ( t / \epsilon ) ^ { 2 } \right) $, if $t\le\epsilon$, and zero else, where $c$ is a normalizing constant.}  

\begin{figure}
	\centering
	\begin{subfigure}{.5\textwidth}
		\centering
		\includegraphics[width=1\linewidth]{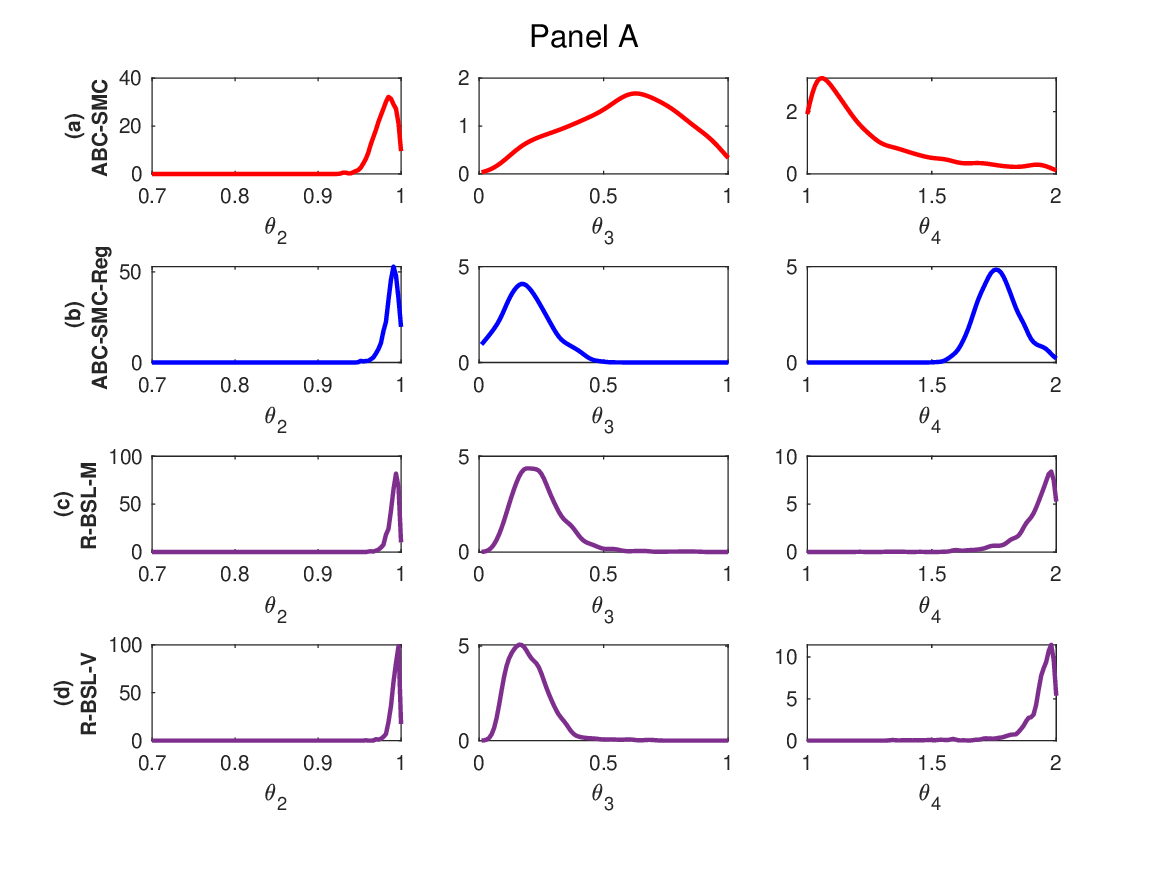} 
		\label{emp_stdabc_rbsl} 
	\end{subfigure}%
	\begin{subfigure}{.5\textwidth}
		\centering
		\includegraphics[width=1\linewidth]{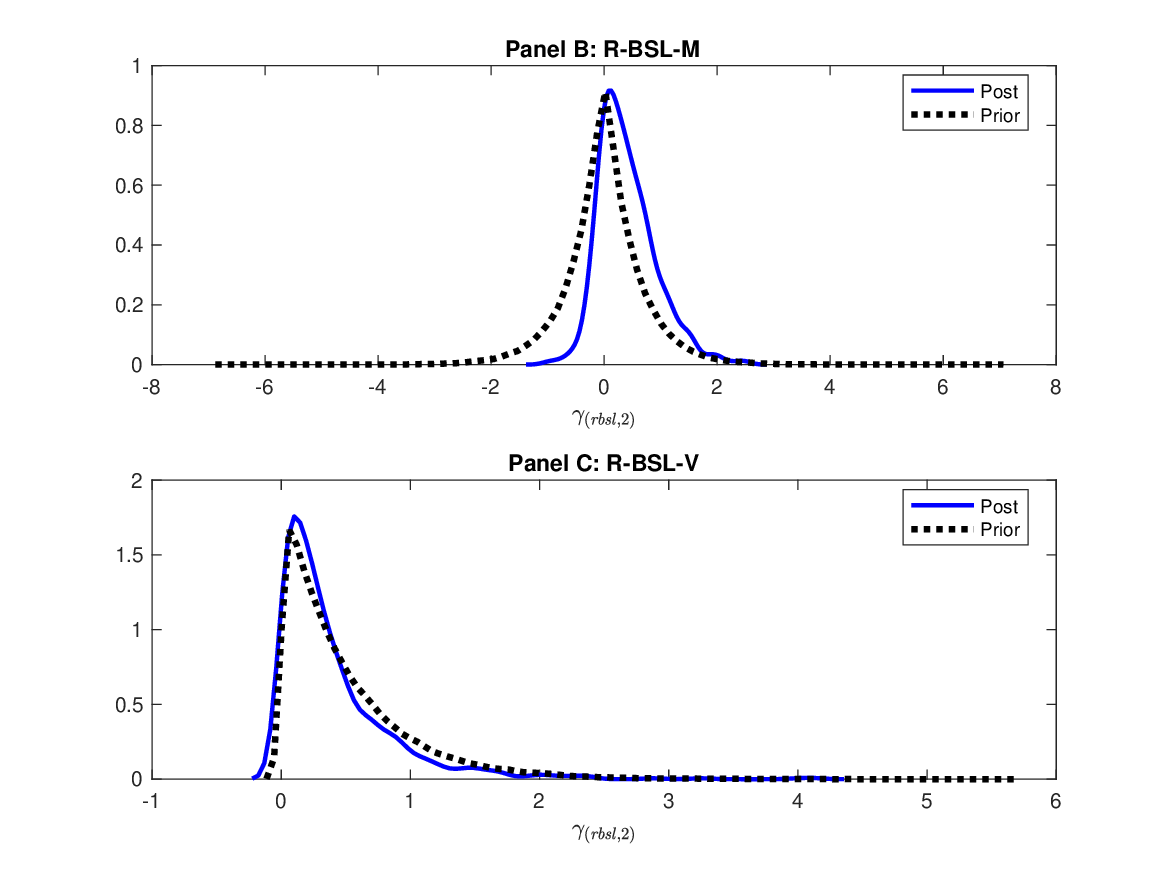} 
		\label{emp_gamma_S2} 
	\end{subfigure}
\captionsetup{skip=-20pt}
		\caption{In Panel A, rows (a) and (b) respectively provide the posteriors for $\theta_{2}, \theta_{3}$ and $\theta_{4}$ using ABC-SMC and ABC-SMC-Reg, while the rows (c) and (d) respectively provide the same information for R-BSL-M and R-BSL-V. Panels B and C display the posteriors for the adjustment component associated with the summary $S_{2}$ (as defined in (\ref{summary_emp})) of R-BSL-M and R-BSL-V, respectively.} 
	\label{fig_emp_theta_gam}
\end{figure}

We compare the ABC-SMC, ABC-SMC-Reg and R-BSL posteriors for the tail index parameter $\theta_{4}$, which governs the tail thickness in the data. Detailed implementation procedures for these methods are provided in Section \ref{empirical_exercise}. In Panel of A of Figure \ref{fig_emp_theta_gam}, we provide the posterior distributions for $\theta_{4}$ along with other unknowns $\theta_{2}, \theta_{3}$ for each estimation method. The posterior distributions of ABC-SMC-Reg and R-BSL are largely similar across all parameters, but differ significantly compared to those obtained using ABC-SMC, particularly for $\theta_{4}$. While ABC-SMC yields posteriors for $\theta_{4}$ that are centered near unity, aligning with the extremely heavy-tailed nature of the data, R-BSL and ABC-SMC-Reg yield posteriors for $\theta_{4}$ that are heavily skewed towards to 2, indicating a tail structure resembling that of a Gaussian distribution. 

When inferences from ABC and regression-adjusted ABC differ, it suggests evidence of model misspecification (\citealp{frazier2020model}). Further, as shown in Figure \ref{emp_post_pred_sum} in Appendix \ref{appendix_D}, the observed summaries are not within the support of the posterior predictive distribution (of the summaries), particularly for the summaries  $S_{2}$ and $S_{3}$ (see Section \ref{empirical_exercise} for the definition of these summaries), which provides additional evidence for potential model misspecification. Panel B of Figure \ref{fig_emp_theta_gam} presents the posteriors for the adjustment component corresponding to $S_{2}$ in R-BSL-M, where it can be observed that they deviate from their prior (a slight deviation also seen in R-BSL-V for the corresponding adjustment component in Panel C), further giving the evidence for model misspecification.

Due to the heavy-tailed nature of the returns data, the assumption that the summary statistics follow a Gaussian distribution is violated and highlights the limitations of R-BSL in the context of heavy-tailed data. However, as detailed in Section \ref{empirical_exercise}, the proposed R-ABC approach produces reliable point estimates for all parameters in this empirical example.

Overall, this example highlights the limitations and challenges of both ABC and R-BSL under model misspecification, thereby motivating the need for our proposed methodology.

\section{Robust ABC}
\label{rabc}

%\dtf{You have to cut this down even more. Move Section 4.1 in its entirety to the supplementary appendix. Reference in the introductory section that this is there. Further, you have to cut down the introductory material and Section 4.2. There are some easy ways to do this in terms of formatting; e.g., make the plots smaller and use better spacing for your tables which are too long in general. You can also make the font of the tables and figures small using the .tex configurations.}

In this section, we propose a strategy for conducting ABC inference that takes into account model misspecification, and allows the researcher to detect incompatible summaries.

\subsection{A misspecification robust ABC approach}
\label{RABC}
Even in misspecified models certain summaries can be matched under the assumed model, while misspecification will prevent others from being matched. Thus, we assume it is feasible to partition the observed summaries into two components: $$\eta(\y)=(\psi(\y)',\varphi(\y)')',$$ where $\psi(\y)$ represents those summaries which can be matched under $P_\theta$, while $\varphi(\y)$ are those that may or may not be able to be matched under $P_\theta$. The investigation of this partition can always be produced in a preliminary step using simulations from the prior-predictive distribution, and so the existence of such a partition is not an onerous assumption.

The key idea behind our proposed R-ABC approach is to condition our inferences on summaries that we know can be matched, while ensuring that additional summaries, for which no match may be possible, do not adversely impact our inferences. To this end, our approach localizes inferences around the summaries we can match, $\psi(\y)$, and robustifies our inferences for those we may not be able to match, i.e., $\varphi(\y)$.  

To `robustify' our inferences, we follow the idea of mean adjustment proposed in \cite{frazier2021robust} (see also, \citealp{kelly2024misspecificationrobust}) and consider adding to $\varphi(\y)$ adjustment components $\Gamma=(\gamma_{1}, \gamma_{2}, ...,  \gamma_{d_{\varphi}})^{'} \in \mathcal{G} \subseteq \mathbb{R}^{d_{\varphi}}$, where $d_{\varphi} = \mathrm{dim}(\varphi)$, to create a new vector of simulated summaries:
$$\varphi(\z, \Gamma):= \varphi(\z) +\Gamma.$$ Given that $\Gamma$ is a random parameter, the joint vector of unknowns becomes $\zeta:=(\theta^{'}, \Gamma^{'})^{'}$. Following \cite{frazier2021robust} we choose a prior distribution for $\Gamma$ that favors compatibility as it has most of its mass around zero, but has thick enough tails to capture incompatibility. Unlike the original summaries $\varphi(\z)$, so long as $\Gamma$ has sufficiently wide support, it will always be possible for $\varphi(\z,\Gamma)$ to match $\varphi(\y)$. A detailed discussion on priors for $\Gamma$ is given in Section \ref{adjust_prior_section}.

To implement this R-ABC approach we first conduct ABC inference on $\theta$ using only $\psi(\y)$, and select an initial set of values for $\theta$ that satisfy
$$
d\{\psi(\z),\psi(\y)\}\le \epsilon_1,
$$ 
for some tolerance $\epsilon_1$. Using these draws as the basis of a proposal distribution, we then implement ABC using the adjusted summaries $\varphi(\z,\Gamma)$, and select values of $\theta$ if they satisfy the following joint selection condition:
\begin{equation}
	\text{select }\theta^i\text{ if } d\{\psi(\widetilde\z),\psi(\y)\}\le \epsilon_1\text{ and }d\{\varphi(\widetilde\z,\Gamma), \varphi(\y)\}\le \epsilon_2 \quad \text{where }\widetilde\z\sim P_{\theta},
	\label{joint_cond}
\end{equation}
where the notation $\widetilde{\z}$ makes clear that the dataset we use at the second step is distinct from that used in the first step (which was denoted by $\z$).

Our proposed R-ABC approach ultimately produces draws from the posterior for  $\zeta=(\theta',\Gamma')'$:
\begin{equation}
	\pi[\zeta|\eta(\y)] \propto \pi(\zeta) \int p(\z|\theta) \1[d\{\psi(\z), \psi(\y)\} \leq \epsilon_{1}]\1[d\{\varphi(\z, \Gamma), \varphi(\y)\} \leq \epsilon_{2}] d\z,
\end{equation}
where $\epsilon_{1}$ and $\epsilon_{2}$ represent the tolerances for step one and step two, respectively, of the algorithm. By retaining draws only if they meet the tolerance specified by both components, $\epsilon_{1}$ and $\epsilon_2$, we use information across all summaries but in an asymmetric fashion: the requirement that $d\{\psi(\z),\psi(\y)\}\le\epsilon_1$ restricts inferences from being worse than those based on $\psi(\y)$ alone.

A second important feature of this approach is that, by decomposing $\eta(\y)$ into $(\psi(\y)',\varphi(\y)')'$, we treat two lower-dimensional vectors of summary statistics in turn, rather than pay the price for the entire dimension of $\eta(\y)$ directly. That is, this approach partly mitigates the ``curse of dimensionality" associated with ABC. Indeed, as suggested in \cite{drovandi2024improving}, it is often easier to produce ABC inferences based on two groupings of low-dimensional components rather than one large dimensional vector. 

The localization step of our R-ABC approach bears some resemblance to the localization approach suggested by \cite{drovandi2024improving} to increase the computational efficiency of ABC methods. The idea suggested in \cite{drovandi2024improving} was to first crudely localize the posterior approximation using all summary statistics, and then refine it to an accurate low-dimensional approximation using a low-dimensional summary statistic. In contrast, our localization is reversed: we first use low-dimensional informative summary statistics $\psi(\y)$ to target a useful posterior approximation for certain components of $\theta$, and in the second step we add summaries that may be informative, but possibly misspecified, to refine this approximation. Similar to \cite{drovandi2024improving}, we ensure that the resulting inferences do not become less accurate at the second stage through the imposition of a joint selection step.  %That is, in a sense, our approach is based on converse logic in \cite{drovandi2024improving}: rather than start with all the summaries, we start with a select few, and instead of obtaining accurate low-dimensional approximations based on a few summaries, we seek to obtain a robust approximation based on additional summaries. 

\subsection{Adjustment components and prior specifications}
\label{adjust_prior_section}

Since the adjustment components $\Gamma$ are random parameters, a suitable prior is required. Since there is no reason to believe a \textit{priori} that $\theta$ and $\Gamma$ are related, we take as our joint prior for $\zeta=(\theta^{'}, \Gamma^{'})^{'}$, $\pi(\zeta):= \pi(\theta) \pi(\Gamma)$ and we restrict the prior on $\Gamma$ to have independent elements: $ \pi(\Gamma):= \prod_{j=1}^{d_{\varphi}} \pi(\gamma_{j}).$

In step two of our proposed R-ABC approach, some components of the original $\varphi(\z)$ are likely compatible, and we therefore want to ensure that the adjustment does not unduly perturb the components that are compatible. Hence, we select a prior that concentrates the vast majority of its mass near the origin. This choice induces ``shrinkage" in the adjustment components $\Gamma$, that is, the adjustment components corresponding to incompatible summaries should receive significant posterior probability away from the origin, while those corresponding to compatible summaries must have the majority of their posterior mass near the origin. With these requirements in mind, we propose using Laplace and spike-and-slab priors.

\subsubsection{Laplace prior}
Following \cite{frazier2021robust}, we consider independent Laplace (i.e., double-exponential) priors for each component of $\Gamma$, with fixed location 0 and common scale $\lambda>0$:
\begin{equation}
	\pi(\Gamma) = \prod_{j=1}^{d_{}\varphi} \frac{1}{2\lambda} e^{-\frac{|\gamma_{j}|}{\lambda}} = \bigg(\frac{1}{2\lambda} \bigg)^{d_{\varphi}} e^{-\frac{1}{\lambda} \sum_{j=1}^{d_{\varphi}} |\gamma_{j}|}.
\end{equation}
We denote this prior by La(0, $\lambda$). This prior guarantees that the bulk of the prior mass for $\gamma_{j}$ is near the origin, but has thick enough tails so that $\varphi(\z, \Gamma)$ is compatible with virtually any $\varphi(\y)$ that would be used in practice.

Additionally, the hyperparameter $\lambda$ should be chosen in a manner that allows the parameters $\Gamma$ to correct for the existence of incompatible summaries, when they are in evidence. That is, $\lambda$ should be chosen so as not to over-inflate the variance of the simulated summaries that are compatible, but also to ensure that there is enough mass away from the origin to allow us to meaningfully distinguish between large and small differences between $\varphi(\y)$ and $\varphi(\z)$. As a default choice of prior, we suggest selecting $\lambda=0.125$, as this places the prior support between $\pm1.$ The default prior La(0, 0.125) is used throughout all numerical experiments conducted in the paper.\footnote{Different values of $\lambda$ ($\lambda=0.5, \lambda=0.25$ and $\lambda=0.125$) were tested across all numerical and empirical examples in the paper and with $\lambda=0.125$ consistently producing the best results. However, $\lambda$ may depend on the scale and variability of the model summaries, and further investigation is needed to determine the most appropriate value for broader applications.} 

\subsubsection{Spike-and-slab prior}

While the Laplace prior provides a mechanism for regularization, it does not allow for genuine shrinkage in the sense of setting parameters exactly to zero -- that is, for non-zero $\lambda$ the probability that $\gamma_j=0$ is zero for all $j$. This is a fundamental difference with the spike-and-slab prior, which can produce a positive probability to $\gamma_j=0$.  Hence, throughout the remainder we do not refer to the Laplace prior as a ``shrinkage prior". In this context, the use of genuine shrinkage priors would be highly useful as their use would entirely eliminate unnecessary adjustment components. Genuine shrinkage priors were not considered in other adjustment approaches like R-BSL since the mixed nature of these priors complicates posterior sampling. However, in our context, since R-ABC relies on an accept-reject mechanism, we only need to be able to simulate from the prior.

Let \text{Ber}($p$) denote a Bernoulli random variable with success probability $p$ and, let Lap(0, $\lambda$) denote a Laplace random variable with location 0 and scale $\lambda > 0.$ Our prior for each component of $\gamma_{j}$ will be a spike-and-slab prior with spike at $\gamma_{j}=0$, which occurs with probability $p$, and continuous slab distribution given by Lap(0, $\lambda$). Conditional on $p$ and $\lambda$, this yields the conditional prior specification: $$q_{j} \sim \text{Ber}(p), \quad \gamma_{j}|q_{j}=0 \sim \delta(\gamma_{j}),  \quad \gamma_{j}|q_{j}=1 \sim \text{Lap}(0, \lambda),$$
where $\delta(\cdot)$ is the Dirac delta function centered at zero, representing the spike. Throughout all numerical experiments, we consider $p=0.5$ and $\lambda=0.125$ as our default choice for the hyperparameters in the spike-and-slab prior.

\subsection{Algorithm and detection of model misspecification} 
\label{R-ABC_alg}

\subsubsection{R-ABC Algorithm}
Our proposed R-ABC approach could be implemented by using the standard accept/reject ABC algorithm; however, this is computationally inefficient. Instead, we implement the second step using the ABC-SMC replenishment algorithm of \cite{drovandi2011estimation} and under two types of prior choices for $\Gamma$, Laplace and spike-and-slab. The implementation of R-ABC must change depending on the prior choice for $\Gamma$, hence, we have formulated our proposed R-ABC approach separately in Algorithm \ref{R-ABC-SMC-lap} (Laplace) and Algorithm \ref{R-ABC-SMC-spk} (Spike-and-slab), respectively, in Appendix \ref{appendix_A}.

In Algorithm \ref{R-ABC-SMC-lap}, where the Laplace prior is used for the adjustment components, both proposed $\theta$ and $\Gamma$ are sampled jointly from a multivariate Gaussian distribution, with a covariance tuned based on the population of SMC particles of $\theta$ and $\Gamma$. 

In contrast, in Algorithm \ref{R-ABC-SMC-spk}, where the spike-and-slab prior is used for the adjustment components, $\theta$ and $\Gamma$ are proposed independently. A new value for $\theta$ is proposed via a Gaussian distribution, with a covariance that is tuned based on the population of SMC particles $\theta$. The novelty of this algorithm lies in the way that it proposes a new value for $\Gamma$. A new value for $\Gamma$ is drawn from a mixture distribution, combining a point mass at zero and a continuous component represented by a mixture of normal distributions for the non-zero values. In this algorithm, we employ an adaptive proposal mechanism for $\Gamma$, where the probability of each $\gamma_{j}$ component being zero is dynamically updated based on the retained samples. Specifically, the algorithm tracks the proportion of retained $\Gamma$ values that are zero, and this information is used to adjust the probability of proposing a zero value in subsequent iterations. By doing so, the algorithm adaptively learns and refines the posterior distribution over $\Gamma$, improving its ability to distinguish between components that should be set to zero and those that should take non-zero values.

As noted, it is known that the mixed discrete-continuous nature of these priors may complicate posterior sampling. However, our approach effectively manages the proposal for $\Gamma$ by separating the spike (zero) and slab (non-zero) components, thus reflecting the discrete nature inherent in the original spike-and-slab prior.

\subsubsection{Detecting model incompatibility}
By examining the posterior distribution of the adjustment components, R-ABC allows researchers to detect incompatible summary statistics. These adjustment components contain information about the nature, and severity of the misspecification. The priors for the adjustment components have the majority of their mass around zero, and when the posteriors have significant mass away from the origin, this is strong evidence that the corresponding summary statistic is incompatible. Conversely, if the posterior distribution of an adjustment component remains concentrated around the origin, the adjustment components are not necessary, i.e., the observed summary statistic can be matched by the assumed DGP.

If one wishes to conduct a formal hypothesis test on the difference between the prior and the posterior, there are a wide number of techniques that can be used. In our numerical and empirical examples, we apply the `two-sample randomization test for location' (\citealp{millard2013envstats}; \citealp{millard2018envstats}) to examine whether there is a difference between the locations of each distribution. This involves testing the null hypothesis that the difference in location parameters is equal to 0:
$$
H_{0}: \mu_{\gamma_{j}} - \mu_{0} = 0,
$$
where $\mu_{\gamma_{j}}$ and $\mu_{0}$ respectively represent the posterior mean of the adjustment parameter $\gamma_{j}$ and the prior mean for $\gamma_{j}$.

If the p-value of the test is greater than a chosen significance level, there is not enough evidence to conclude that a significant difference exists between the prior and posterior distributions of the adjustment components at the chosen significance level.

\subsection{R-ABC vs R-BSL} 
\label{RABC-vs-RBSL}
R-ABC can be considered as an alternative to R-BSL. While both can be applied in similar contexts, R-BSL faces challenges when the summary statistics deviate from a Gaussian distribution. In such cases, R-BSL may struggle to sample effectively from the posterior distribution. Moreover, R-BSL cannot reliably differentiate between model misspecification and a strong violation of the underlying Gaussian assumption for the likelihood of the summaries (\citealp{frazier2021robust}). The acceptance rate of R-BSL is also known to decrease as model misspecification worsens, which can require running many iterations to achieve reliable inferences. Therefore, R-BSL is better suited to cases where the summary statistics exhibit ``thin tails", allowing the Gaussian approximation to remain valid.

In contrast, R-ABC relies on a non-parametric estimate of the likelihood of the summaries and therefore does not make any distributional assumptions about the summary statistics. \cite{frazier2018asymptotic} demonstrate that ABC inference remains valid even when dealing with summary statistics with ``heavy tails". Hence, our R-ABC approach is particularly useful when the summary statistics are non-Gaussian or have heavy tails.

Furthermore, using the genuine shrinkage priors such as spike-and-slab for the adjustment components in R-BSL could complicate posterior sampling when using  Metropolis-Hastings (MH)-MCMC. Even though SMC-based versions may be able to handle this more effectively, the standard implementation of BSL would be MH-MCMC. Due to the mixed discrete-continuous nature of such prior, incorporating such components within R-BSL directly violates the underlying Gaussian assumption for the likelihood. However, this limitation does not apply to R-ABC, which only requires the ability to simulate from the prior distribution.

Given these differences between the two methods, it is worthwhile to compare their performance, particularly under model misspecification. Through the following simulation examples, we demonstrate that there are clear scenarios where R-ABC is preferable to R-BSL.

\section{Simulation exercises}
\label{simulation_ex}
In this section, we conduct a set of simulation experiments to demonstrate the performance of R-ABC compared to standard ABC approaches and R-BSL. These examples demonstrate that R-ABC provides more reliable statistical inferences than ABC-SMC and ABC-SMC-Reg when the model is misspecified. Moreover, the simulations results reveal scenarios where R-ABC may be advantageous over R-BSL. We also show that the $\Gamma$ components of the R-ABC approach can be used to reliably detect the summaries that may be incompatible with the assumed model. Note that in Appendix \ref{appendix_B}, we present an example where a moving average (MA) model is used as the assumed model, while a SV model serves as the true DGP. In this case, both R-ABC and R-BSL yield closely similar inferences, given that the Gaussian assumption of the summary statistics holds reasonable in this setting. Furthermore, an additional simulation example using a normal model to demonstrate the performance of R-ABC is provided in Appendix \ref{appendix_E}.

\subsection{Misspecified g-and-k model} \label{gnk_example}

\subsubsection{Models and computational details}
To illustrate the performance of R-ABC against R-BSL in another example, we use the g-and-k model (\citealp{drovandi2011estimation}; \citealp{fearnhead2012constructing}; \citealp{frazier2020model}) which is most commonly stated through its quantile function:
\begin{equation*}
	q \in (0,1) \mapsto a + b \bigg(1 + 0.8 \frac{1-exp\big(-g z(q)\big)}{1+exp\big(-g z(q)\big)}\bigg) \big[1 + z(q)^{2}\big]^{k} z(q),
\end{equation*}
where $-\infty < a < \infty, b>0, -\infty < g < \infty, k>-0.5$ and $z(q)$ refers to the $q$-th quantile of the standard normal distribution. The parameters $\theta=(a, b, g, k)^{\prime}$ control the location, scale, skewness and kurtosis, respectively. As with previous studies, we set the prior on each component of $\theta$ as $U(0,10)$, where $U(a,b)$ denotes the uniform distribution on $[a,b]$. 

Following \cite{drovandi2011likelihood},  we use robust measures of location, scale, skewness, and kurtosis as summary statistics, denoted as, $\eta(\y)=(S_{1}, S_{2}, S_{3},  S _{4})^\prime$, where
$$
S_{1}=L_{2}, \quad S_{2}=L_{3}-L_{1}, \quad S_{3}=\frac{L_{3}+ L_{1} - 2L_{2}}{S_{2}}, \quad S_{4}=\frac{E_{7} - E_{5} + E_{3} - E_{1}}{S_{2}},
$$
with $L_{i}$ representing the $i$-th quartile and $E_{i}$ representing the $i$-th octile.

The g-and-k distribution is recognized for its flexibility, but despite this flexibility the distribution is unimodal and, hence, cannot capture multi-modality in the data. Here, we assume that the data comes from a g-and-k distribution, but that the actual observed data is generated from a distribution with bi-modality. Specifically, the observed data is generated as iid draws from a Gaussian mixture
\begin{equation}
	y_{t}  \sim w N( \mu_{1},\sigma_{1}^{2})  + (1-w) N( \mu_{2},\sigma_{2}^{2}),
	\label{gaus_mix}
\end{equation}
where $\mu_{1}, \mu_{2}$ denote means, $\sigma_{1}, \sigma_{2}$ denote the respective standard deviations, and $w$ is the mixing proportion of the two normal distributions.

We now compare the performance of the R-ABC and R-BSL approaches using a repeated sampling experiment: From the DGP in (\ref{gaus_mix}), we simulate 50 replications of the observed data $\y$ of size $n = 5000$, with parameter values:
$$
(\mu_{1}, \sigma_{1}^{2}) = (1,2), \quad (\mu_{2}, \sigma_{2}^{2}) = (7,2), \quad w=0.6.
$$
This DGP produces observed data with significant bi-modality, as shown in Figure \ref{gnk_obs_data}, and can be verified to produce incompatible summaries (see Figure \ref{gnk_post_pred_sum} in Appendix \ref{appendix_C} for further details). Both ABC-SMC and ABC-SMC-Reg are implemented using the ABC-SMC replenishment. We terminate the algorithm when the acceptance rate in the MCMC step drops below 1\%.

\begin{figure}[h!]
	\centering 
	\includegraphics[width=0.5\linewidth]{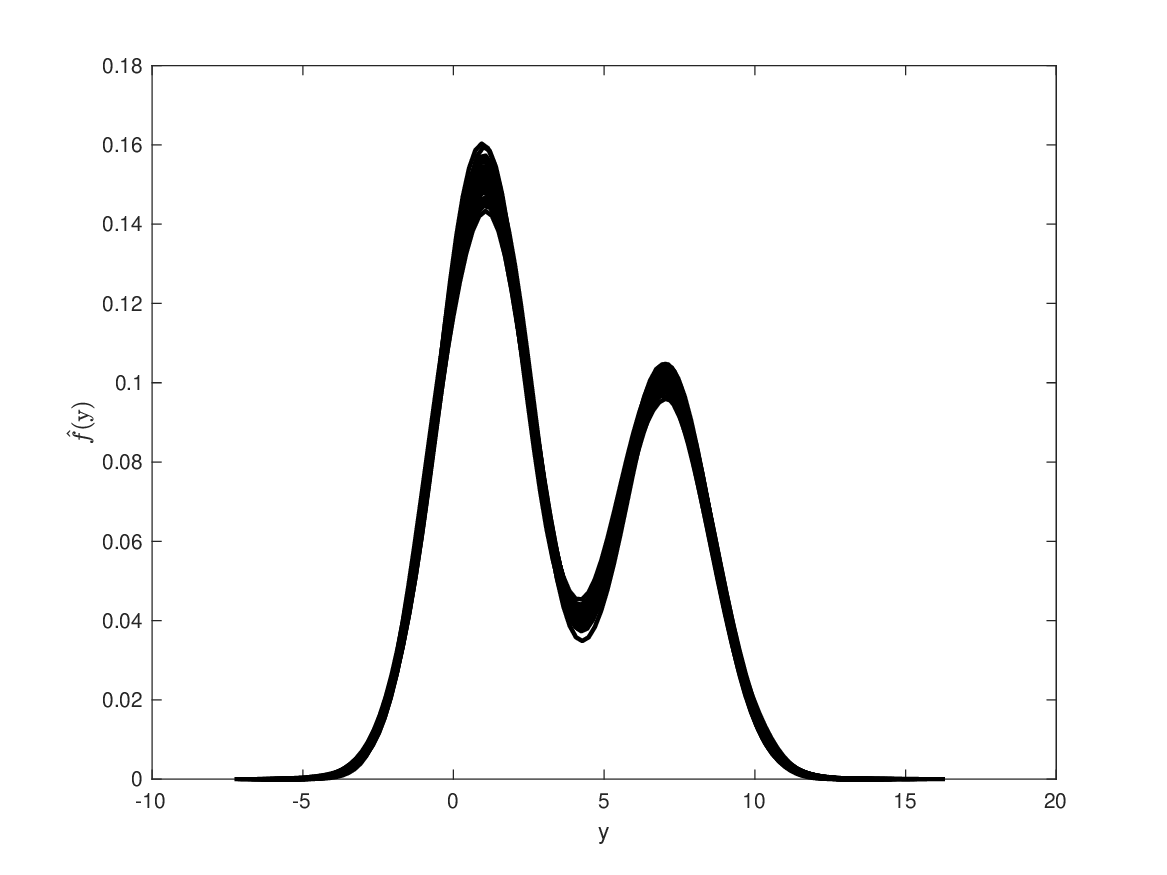} 
	\caption{Kernel density estimates of 50 replications of data simulated from the Gaussian mixture model in (\ref{gaus_mix}).} 
	\label{gnk_obs_data} 
\end{figure}

Since each component of $\theta$ corresponds to a robust summary statistic in $\eta(\y)$, one might use any of the summary statistics in $\eta(\y)$ as $\psi(\y)$ when performing R-ABC. However, \cite{drovandi2024improving} observe poor performance of the marginal approximation for the parameter $a$ based on only $S_{1}$. Upon further investigation, they discover that this is due to the distribution of $S_{1}$ also being influenced by the scale $b$, and that there is a dependency between $a$ and $b$ in the posterior given $S_{1}$. Hence, we first analyze how each of the summaries $S_{1}, S_{2}, S_{3}$, and $S_{4}$ relate to the parameters $a, b, g$, and $k$. These findings are discussed in detail in Appendix \ref{appendix_C}. According to the results in Appendix \ref{appendix_C}, since the parameter $g$ depends only on $S_{3}$, we choose the following partition of the summaries when performing the R-ABC:
$$\psi(\y)=S_{3}, \quad \varphi(\y)=(S_{1}, S_{2}, S_{4})^{\prime}.
$$

Step one of the R-ABC approach is conducted using the ABC accept/reject algorithm using only $\psi(\y)$. We simulate $N_{1}=25000$ draws from the prior and retain those leading to the smallest 5\% of the overall simulated distances. Then we implement step two of the R-ABC approach using the adjusted summaries, where three adjustment components, $\Gamma_{\text{R-ABC}} = (\gamma_{(rabc,1)}, \gamma_{(rabc,2)}, \gamma_{(rabc,3)})^{\prime}$, are introduced for $S_1$, $S_2$, and $S_4$, respectively. This step employs the modified ABC-SMC replenishment algorithm, described in Algorithms \ref{R-ABC-SMC-lap} and \ref{R-ABC-SMC-spk}, which respectively use Laplace and spike-and-slab priors for the adjustment components. Both algorithms are initialized with the ABC samples retained from step one and continue running until the MCMC acceptance rate falls below 1\%.

In the implementation of R-BSL, both versions use $m=50$ simulated datasets. For each replication, we run the MCMC sampler for 10000 iterations, discard the first 5000 as burn-in, and thin the remaining sample by selecting every fifth draw. All four summaries are adjusted in R-BSL, hence four adjustment components are needed. That is, we use $\Gamma_{\text{R-BSL}}=(\gamma_{(rbsl,1)}, \gamma_{(rbsl,2)}, \gamma_{(rbsl,3)}, \gamma_{(rbsl,4)})^{\prime}$ for $S_{1}, S_{2}$, $S_{3}$ and $S_{4}$, respectively.

The following section discusses the repeated sampling behavior of the R-ABC and R-BSL approaches.\footnote{In this Monte Carlo design, the pseudo-true value is given by  $\theta^{*} = (a^{*}, b^{*}, g^{*}, k^{*})^{'} =(2.3663, 4.1757, 1.7850, 0.1001)^{\prime}.$ For a detailed explanation of how the pseudo-true value is determined, refer to Appendix \ref{appendix_C}.} Note that we discuss only the results corresponding to the parameter $k$ in the main text, while the discussion of the remaining parameters is provided in Appendix \ref{appendix_C}.

\subsubsection{Comparison: R-ABC vs R-BSL}

\begin{figure}[h!]
	\centering 
	\includegraphics[width=1.1\linewidth]{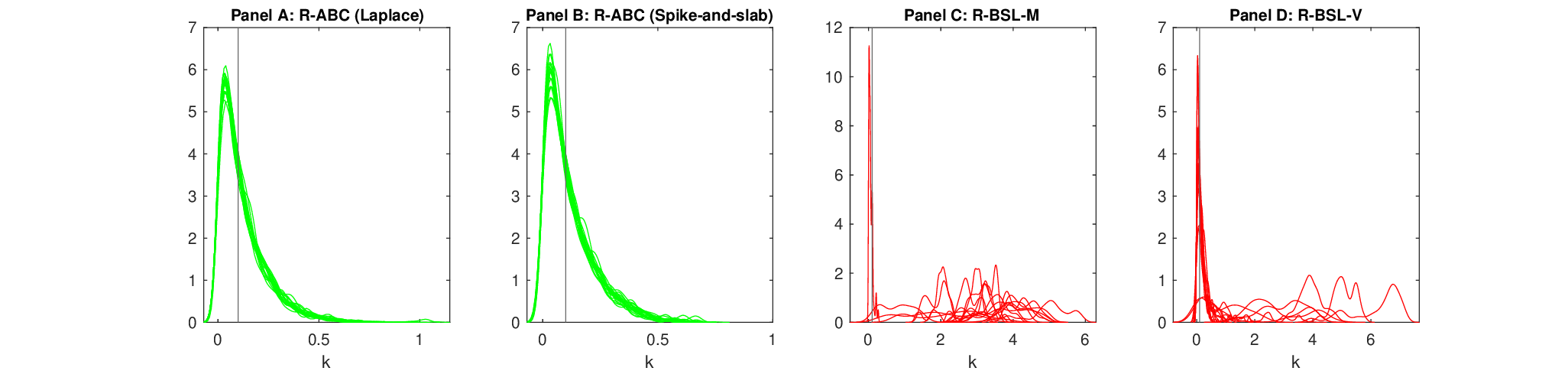} 
	\caption{Panels A and B respectively provide the posteriors for $k$ using the R-ABC approach when using Laplace and spike-and-slab priors for the adjustment components. Panels C and D respectively provide the same information for R-BSL-M and R-BSL-V. The vertical line represents the pseudo-true value. The hyperparameter for the priors associated with the adjustment components of R-ABC is $\lambda=0.125.$} 
	\label{gnk_theta_rabc_rbsl} 
	
\end{figure}

Figure \ref{gnk_theta_rabc_rbsl} displays the posterior distributions from the R-ABC and R-BSL approaches for the parameter $k$ of the g-and-k distribution across different data replications. Note that for interpretability, only the results from the first 20 data replications are shown in Figure \ref{gnk_theta_rabc_rbsl}. Results for the other parameters are presented in Figure \ref{gnk_theta_rabc_rbsl_abg} in Appendix \ref{appendix_C}.

The R-BSL MCMC sampler is initialized with random values. In this instance, the average acceptance rates across the replicated datasets for R-BSL-M and R-BSL-V are 14\% and 49\%, respectively. The corresponding R-BSL results appear in Panels C and D of Figure \ref{gnk_theta_rabc_rbsl}. From the posterior distributions shown in Figure \ref{gnk_theta_rabc_rbsl}, it is evident that, across each data replication, R-ABC $\big(\text{Panels A and B}\big)$ concentrate closely around the pseudo-true value. In contrast, R-BSL-M and R-BSL-V $\big(\text{Panels C and D}\big)$ exhibit considerable variability across replications, and cannot capture the pseudo-true value in some cases. Note that the qualitatively similar conclusions can be drawn for the other parameters as observed in Figure \ref{gnk_theta_rabc_rbsl_abg} in Appendix \ref{appendix_C}. An animation plot of Figures \ref{gnk_theta_rabc_rbsl} and \ref{gnk_theta_rabc_rbsl_abg} is available in Appendix \ref{appendix_C} (see Figure \ref{anim_gnk_1}), further illustrating the higher variability of the R-BSL posteriors across replications compared to the R-ABC posteriors.

Given this unexpected behavior of the R-BSL posteriors, we conducted an additional implementation of R-BSL, this time initializing the MCMC sampler at the infeasible pseudo-true value. The resulting posteriors are displayed in Figure \ref{gnk_theta_rbsl_pstv} in Appendix \ref{appendix_C}, and show that even with this infeasible initialization, R-BSL delivers qualitatively similar results to those based on the random initialization, suggesting that the variability of the posterior across the replications is intrinsic to R-BSL.\footnote{In this case, the average acceptance rates across replicated datasets are 8\% for R-BSL-M and 55\% for R-BSL-V.}

{\renewcommand{\arraystretch}{0.8}
\begin{table}[h!]
	\caption{\scriptsize Monte Carlo coverage (Cov), bias of the posterior mean (Bias), and average posterior standard deviation (Std) for the parameter $k$ in the g-and-k example. Cov is the percentage of times that the marginal 95\% credible set contains $k^*$. Std is the average posterior standard deviation across the Monte Carlo trials. The rows refer to either the ABC or BSL approach. The last four rows refer to the R-BSL approach, where R-BSL-M, R-BSL-V refer to the case where random starting values are used for the MCMC sampler, while R-BSL-M (PSTV), R-BSL-V (PSTV) refer to the case where pseudo-true values are used as starting values of MCMC sampler.}
	\centering%
	\small
	\begin{tabular}{lrrrrrrr}
		\hline\hline
		&   \multicolumn{3}{c}{$k$}  \\
		\cline{2-4} 
		& \multicolumn{1}{c}{Cov} & \multicolumn{1}{c}{Bias} & \multicolumn{1}{c}{Std} \\
		{\textbf{Method}} \\
		ABC-SMC	&	0\%	&	-0.0836	&	0.0150	\\
		ABC-SMC-Reg	& 0\%	&	-0.1398	&	0.0147	\\
		R-ABC (Laplace)	&	100\%	&	0.0201	&	0.1144	\\
		R-ABC (Spike-and-slab) 	&	100\%	&	0.0174	&	0.1100	\\
		R-BSL-M	&	18\%	&	2.2919	&	0.3963	\\
		R-BSL-V	&    62\%	&	1.8305	&	0.5523	\\
		R-BSL-M (PSTV)	&	92\%	&	-0.0287	&	0.0604	\\
		R-BSL-V (PSTV)	&	92\%	&	0.6027	&	0.3361	\\
		\hline\hline
	\end{tabular}%
	\label{tab_gnk_k}%
\end{table}	
}

When examining the posteriors for the adjustment components of R-ABC and R-BSL across 50 replications (Figures \ref{fig_gnk_gamRABC_random} and \ref{fig_gnk_gamRBSL_random} in Appendix \ref{appendix_C}), it is evident that the posterior of the $\gamma$ component associated with $S_{4}$ deviates significantly from its prior across all methods. In contrast, posteriors of the other adjustment components closely align with their respective priors. That is, both R-ABC and R-BSL reveal the inability of the model in matching the robust summary related to kurtosis. A detailed discussion on the adjustment components is provided in Appendix \ref{appendix_C}.

To further compare the two approaches, we assess the posterior bias, average posterior standard deviation, and Monte Carlo coverage for each method across replications. Note that here we report the results for R-BSL under two initialization conditions of the MCMC sampler: random starting values (R-BSL-M and R-BSL-V) and pseudo-true values (R-BSL-M (PSTV) and R-BSL-V (PSTV)). For parameter $k$, these comparative results are summarized in Table \ref{tab_gnk_k}, while the corresponding results for other parameters are provided in Table \ref{tab_gnk_abg} in Appendix \ref{appendix_C}.

Regarding the bias of the posterior mean for parameter $k$, R-BSL-M and R-BSL-V, which initialized the MCMC sampler with random starting values, exhibit a high bias while R-ABC (spike-and-slab) achieves the lowest bias. In terms of posterior variability, ABC-SMC and ABC-SMC-Reg exhibit smallest posterior uncertainty, whereas R-BSL-M and R-BSL-V show the higher posterior uncertainty for parameter $k$. In contrast, R-ABC yields comparatively smaller posterior uncertainties, although these remain slightly higher than those of ABC-SMC, ABC-SMC-Reg and R-BSL-M (PSTV). Further, both R-BSL approaches, which initialized the MCMC sampler with random starting values, display very poor coverage for parameter $k$. Most notably, both ABC-SMC and ABC-SMC-Reg produce credible sets that do not contain the pseudo-true value in any of the Monte Carlo replications, resulting in 0\% coverage. In comparison to all methods, R-ABC produces a reasonable coverage for the parameter $k$. Detailed discussions of the results for other parameters are included in Appendix \ref{appendix_C}.

These findings clearly demonstrate a situation where R-ABC is preferable to R-BSL. However, as previously mentioned, we do not view these two methodologies as direct competitors. R-ABC is applicable regardless of whether the summaries follow an approximate Gaussian distribution. On the other hand, if the summaries are indeed approximately Gaussian, R-BSL should provide more precise inferences. It is ultimately up to the researcher to determine the appropriate context for employing each method. For instance, one might first use R-ABC to assess model misspecification and better understand the distribution of the summaries. If it is found that the summaries are approximately Gaussian, applying R-BSL could yield more precise inferences.

\section{Empirical illustration} 
\label{empirical_exercise}

In this section, we analyze the performance of R-ABC in the motivating empirical example introduced in Section \ref{motivating_ex}. 

Several empirical studies have demonstrated the use of $\alpha$-stable processes to capture the non-Gaussian features of financial returns (\citealp{carr2003finite}; \citealp{peters2012likelihood};
\citealp{lombardi2009indirect}; \citealp{martin2019auxiliary}), making it an ideal choice for analyzing volatile assets like S\&P500 index. In this spirit, we define the following $\alpha$-stable SV for the continuously-compounded return, $r_{t}$, on the S\&P500 index:
\begin{equation*}
	r_{t}=\sigma_{t}w_{t}\quad\text{;}\quad w_{t}\overset{iid}{\sim}
	\mathcal{S}(\theta_{4}, \theta_{5},\mu,\sigma)
\end{equation*}
\begin{equation*}
	\ln \sigma_{t}^{2} = \theta_{1} + \theta_{2}\ln \sigma_{t-1}^{2} + \theta_{3} \nu_{t}\quad\text{;}\quad \nu_{t}\overset{iid}{\sim} N(0,1),
\end{equation*}
where  $\sigma_{t}^{2}$ is the variance at time $t$, and $\mathcal{S}(\theta_{4}, \theta_{5},\mu,\sigma)$ denotes an $\alpha$-stable L\'{e}vy process with location parameter $\mu$, scale parameter $\sigma$, tail index $\theta_{4} \in(1,2)$, and skewness parameter $\theta_{5} \in [-1,1]$. The values of $\theta_{4}$ and $\theta_{5}$ control the degree of leptokurtosis and skewness (respectively) in the innovations of the returns. In particular, $\theta_{4}$ can be identified as the tail thickness parameter, with tails becoming heavier as $\theta_{4}$ decreases. The parameter $\theta_{2}$ represents the persistence or autocorrelation of the volatility process over time. A higher value of $\theta_{2}$ (close to 1) implies that the current volatility ($\ln \sigma_{t}^{2}$) depends strongly on its past value ($\ln \sigma_{t-1}^{2}$). This means volatility shocks have long-lasting effects, and this is a common feature observed in financial time series, where periods of high or low volatility tend to persist. An $\alpha$-stable stochastic volatility model is thus particularly well-suited for capturing the extreme return volatility observed in this series.

We choose to fix the skewness by setting $\theta_{5}=0$ and simplify the analysis further by fixing $\mu=0, \sigma=1$, and the volatility location parameter $\theta_{1}=0$. For the remaining parameters, we employ uniform priors as follows:
$$
\theta_2\sim U(0.7,1),\quad\theta_3\sim U(0.01,1),\quad\theta_4\sim U(1,2).
$$
%These specific priors, also used in \cite{martin2019auxiliary}, are informed by empirical findings regarding the plausible range of these parameters from previous studies on daily S\&P500 returns.
We generate summary statistics for ABC inference using an auxiliary model that accommodates the heavy-tailed nature of the data (\citealp{martin2019auxiliary}). Specifically, we employ a first-order generalized autoregressive conditional heteroscedastic (GARCH(1,1)) model:
\begin{flalign}
	r_{t}&=x_{t} \epsilon_{t} 	\label{aux_model_garch}\\ x_{t}&=\beta_{1}+\beta_{2} x_{t-1}\left|\epsilon_{t-1}\right|+\beta_{3} x_{t-1},
	\label{aux_model_garch_v}
\end{flalign}
where $\epsilon_{t}\stackrel{iid}{\sim} \text{t}_{\beta_4}$, and $\text{t}_{\beta_4}$ denotes a standardized Student-t random variable with ${\beta_4}$ degrees of freedom.\footnote{See \citealp{lombardi2009indirect}; \citealp{garcia2011estimation} for details on the use of indirect inference in similar model contexts.} We use the absolute value of $\epsilon_{t-1}$ in the volatility equation to reduce numerical instabilities that may arise from extreme values in the $\alpha$-stable distribution.

The model in (\ref{aux_model_garch}) and (\ref{aux_model_garch_v}) provides an auxiliary likelihood, allowing for straightforward calculation of the auxiliary scores. As discussed in \cite{martin2019auxiliary}, in state space models like the $\alpha$-stable volatility model, using the scores of auxiliary likelihoods as summary statistics facilitates ABC-based inference. Accordingly, we use the scores of this auxiliary GARCH model as our summary statistics for ABC, denoted by $\eta(\mathbf{y})=(S_{1}, S_{2}, S_{3},  S_{4})^\prime$ where
\begin{flalign}
	S_{1} =T^{-1}\frac{\partial L_{a}}{\partial \beta_{1}}, \quad S_{2} =T^{-1}\frac{\partial L_{a}}{\partial \beta_{2}}, \quad S_{3} =T^{-1}\frac{\partial L_{a}}{\partial \beta_{3}}, \quad S_{4} =T^{-1}\frac{\partial L_{a}}{\partial \beta_{4}}.
	\label{summary_emp}
\end{flalign}
Here, $L_{a}$ is the log-likelihood of the auxiliary GARCH model, and $T$ is the total number of observations.

Following the analysis in Section \ref{motivating_ex}, we use the same models and precisely the same observed data and apply R-ABC. Note that, for each ABC approach implemented in Section \ref{motivating_ex}, we stop the SMC-ABC algorithm when the acceptance rate in the MCMC step drops below 1\%, as flagged earlier. The proposal distribution in the MCMC ABC kernel is a Gaussian random walk, with a covariance tuned based on the population of SMC particles.

In the R-ABC implementation, $S_{4}$ is less favored for use in step one because it is directly influenced by the degrees of freedom parameter $\beta_{4}$ in the auxiliary model. Including $S_{4}$ in step one could skew the retained draws toward parameter values influenced by $\beta_{4}$, potentially pushing the tail index $\theta_{4}$ closer to 2, and suggesting lighter tails, which do not align with the heavy-tailed characteristics of the observed data. To ensure robust inference about $\theta_{4}$, the summaries in step one are restricted to those not affected by $\beta_{4}$. Accordingly, we apply R-ABC with the following summary partition:
$$
\psi(\y)=S_{1}, \quad \varphi(\y)=(S_{2}, S_{3}, S_{4})^{\prime}.\\
$$

Using the summary partition outlined above, we implement the R-ABC approach starting with step one, which employs the ABC accept/reject algorithm. Here, we generate $N=50000$ simulated datasets, setting the tolerance level to the 5\% quantile of the simulated distances. Step two of R-ABC is then carried out using a modified ABC-SMC replenishment algorithm, detailed in Algorithms \ref{R-ABC-SMC-lap} and \ref{R-ABC-SMC-spk}, which employ Laplace and spike-and-slab priors for the adjustment components, respectively. These algorithms are initialized with the ABC samples retained from step one and continue running until the MCMC acceptance rate falls below 1\%. Three adjustment parameters are needed in the second step $\Gamma_{\text{R-ABC}}=(\gamma_{(rabc,1)}, \gamma_{(rabc,2)}, \gamma_{(rabc,3)})^{\prime}$ respectively for $S_{2}, S_{3}$ and $ S_{4}$. The hyperparameter value of the adjustment priors is set to the default value, $\lambda=0.125.$ 

In the implementation of R-BSL in Section \ref{motivating_ex}, both versions rely on $m=50$ simulated datasets and are based on $100000$ MCMC iterations, with the first $10000$ discarded as burn-in and the remaining samples thinned by selecting every $100^{th}$ draw. The MCMC sampler is initialized with $\theta_{2}=0.8, \theta_{3}=0.2$ and $\theta_{4}=1.3$. Since all four summary statistics are adjusted in R-BSL, four adjustment components are required, denoted by $\Gamma_{\text{R-BSL}}=(\gamma_{(rbsl,1)}, \gamma_{(rbsl,2)}, \gamma_{(rbsl,3)}, \gamma_{(rbsl,4)})^{\prime}$.

%The next section presents results for both R-ABC and R-BSL, assessing their effectiveness in modeling the S\&P500 return series.\color{black}

%\subsection{Empirical results}

\begin{figure}[h!]
	\centering 
	\includegraphics[width=0.6\linewidth]{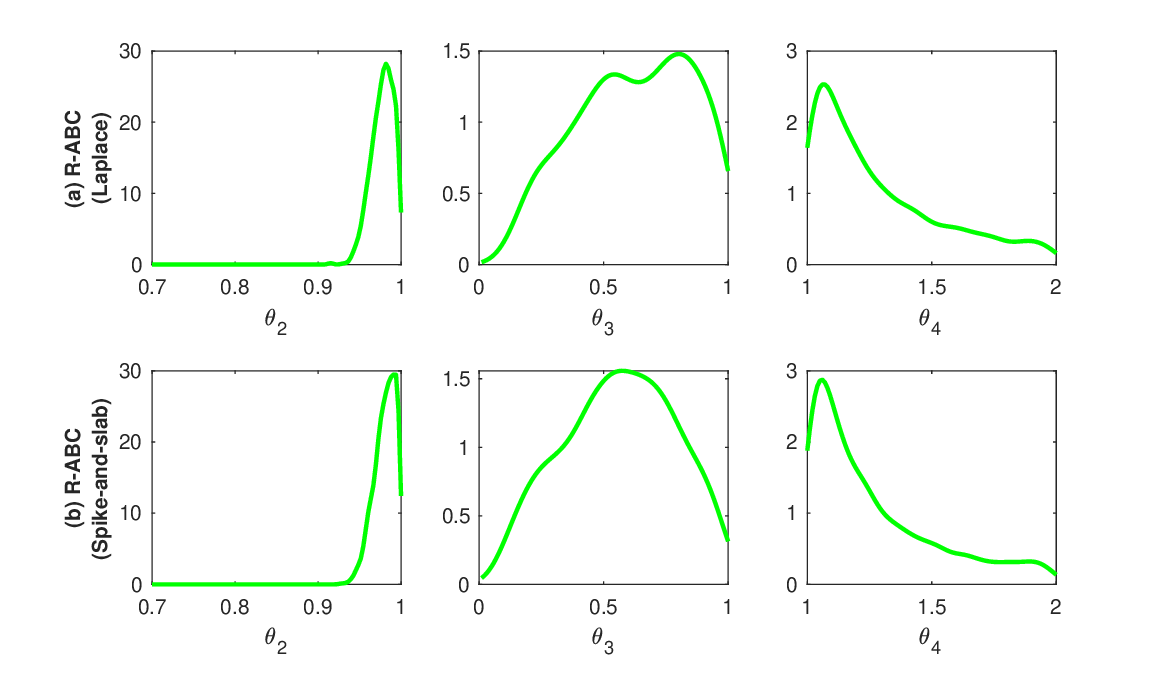} 
	\caption{Rows (a) and (b) respectively provide the posteriors for $\theta_{2}, \theta_{3}$ and $\theta_{4}$ using the R-ABC approach when using Laplace and spike-and-slab priors for the adjustment components. The hyperparameter for the priors associated with the adjustment components in R-ABC is $\lambda=0.125.$} 
	\label{emp_theta_rabc} 
\end{figure}

The posterior distributions of the unknown parameters $\theta_{2}, \theta_{3}$ and $\theta_{4}$ are illustrated in Figure \ref{emp_theta_rabc} for R-ABC when using Laplace and spike-and-slab priors for adjustment components. When compared to Panel A of Figure \ref{fig_emp_theta_gam}, the results in Figure \ref{emp_theta_rabc} indicate that there is only a slight observable difference between ABC-SMC and R-ABC, with both methods concentrating around empirically plausible values for each parameter. However, the R-ABC posteriors for the tail index parameter $\theta_{4}$ differs notably from those obtained using ABC-SMC-Reg and R-BSL. Specifically, unlike ABC-SMC-Reg and R-BSL, the R-ABC posteriors for $\theta_{4}$ in Figure \ref{emp_theta_rabc} are centered near unity reflecting the extremely heavy-tailed nature of the returns data.

We present the adjustment components for R-ABC and R-BSL in Appendix \ref{appendix_D} (plotted in Figures \ref{emp_gam_rabc} and \ref{emp_gam_rbsl}). As shown in Figure \ref{emp_gam_rabc}, both R-ABC (Laplace) and R-ABC (Spike-and-slab) exhibit slight deviations of the adjustment components $\gamma_{(rabc,1)}, \gamma_{(rabc,2)}$ and $\gamma_{(rabc,3)}$ from the origin.\footnote{These deviations are confirmed by results from the two-sample randomization test for location at the 5\% significance level. For the R-ABC (Laplace), the test rejects $H_{0}$ for $\gamma_{(rabc,1)}, \gamma_{(rabc,2)}$ and $\gamma_{(rabc,3)}$ with p-values of 0, 0.0432 and 0 respectively. For R-ABC (Spike-and-slab), the test rejects $H_{0}$ for $\gamma_{(rabc,1)}, \gamma_{(rabc,2)}$ (p-values 0 and 0.0068) but does not reject $H_{0}$ for $\gamma_{(rabc,3)}$ in this case (p-value=0.0536).} Adjustment components for R-BSL are shown in Figure \ref{emp_gam_rbsl}. According to the top row, nearly all adjustment components exhibit slight deviations from the prior under R-BSL-M. However, as seen in the bottom row, there are slight deviations in $\gamma_{(rbsl,2)}$, $\gamma_{(rbsl,3)}$ and $\gamma_{(rbsl,4)}$, but those are minimal.

In summary, results of the adjustment components across both methods suggest that the assumed model cannot reliably match the summaries $S_{2}, S_{3}$ and $S_{4}$. This implies that the model appears to be significantly misspecified in this setting.

\section{Discussion}
\label{discussion}

Approximate Bayesian computation (ABC) is a powerful tool used to conduct inference in complex models when traditional likelihood-based methods are challenging or infeasible to apply. However, using ABC with misspecified models can result in unreliable inferences. Particularly, post-processing methods like the local linear regression adjustment, often employed in addressing the curse of dimensionality in ABC inference, can lead to unintended consequences, and deliver very poor inferential outcomes. To circumvent the poor behavior of ABC in these settings, in this article we propose a modification of ABC that is robust to model misspecification.

Our approach, called robust ABC (R-ABC), begins by partitioning summary statistics into two groups: those summaries that can be matched under the assumed model and those that may or may not be able to be matched under the assumed model. By initially conducting ABC-based inference using the summaries we can match, we lay the groundwork for more reliable parameter estimation. In the second step, we adjust the uncertain summaries by introducing additional free parameters that are designed to account for possible incompatibilities. This adjustment is inspired by the robust Bayesian synthetic likelihood (R-BSL) method introduced in \cite{frazier2021robust}, which applies a similar strategy to produce BSL posteriors that are robust to model misspecification. Through various simulation exercises and an empirical application, we illustrate how this new method outperforms standard ABC approaches, showcasing its statistical benefits when the model is misspecified.

This new approach not only improves statistical inferences on model parameters under model misspecification but also enables users to identify summary statistics that are potentially incompatible with the assumed DGP. This identification is achieved by comparing the posterior of the adjustment components to their prior distribution; if the posterior of an adjustment component deviates from the prior, it suggests that this summary statistic cannot be aligned with the assumed model. Notably, using a true shrinkage prior, like the spike-and-slab, for the adjustment components, helps to fully eliminate unnecessary adjustment components by setting them to 0. Overall, the two-step structure of R-ABC allows us to separate out the influence of compatible and incompatible summaries more effectively, reducing dimensionality and the associated computational burden without compromising accuracy. While our R-ABC method has similarities with the R-BSL approach proposed in \cite{frazier2021robust}, we identify that there are instances where R-ABC yields reliable inferences compared to R-BSL, especially when the summary statistics deviate from Gaussian assumptions. Given these advantages and the more prevalent application of ABC across various fields for conducting inference in complex models, we believe this proposed R-ABC approach will serve as a valuable tool for practitioners, particularly in empirical applications where model misspecification is a significant challenge. 

While the current R-ABC approach focuses on adjusting the second set of summaries by introducing additional free parameters, future research could investigate alternative robustification methods to further enhance the flexibility of the approach and extend its applicability across more diverse applications. Additionally, further exploration into optimal hyperparameter selection for the adjustment priors may be valuable, although our default choices have shown to be reasonable and effective in yielding good simulation and empirical results in our examples.

\bigskip

\bibliographystyle{apalike} 
\bibliography{ref}

\newpage
\section*{Appendix}
\appendix
\vspace{-1.2cm}

\def\spacingset#1{\renewcommand{\baselinestretch}%
	{#1}\small\normalsize} \spacingset{1}

%\spacingset{1.75} % DON'T change the spacing! -- jcgs site

\spacingset{1.45} %overleaf template

\section*{}
% Define appendix-specific formatting

\titleformat{\subsection}[block]{\large\bfseries}{\Alph{subsection}}{1em}{}

This file includes the following sections:

\begin{enumerate}
	\item Appendix \ref{appendix_A}: ABC, ABC-Reg and the R-ABC (with steps one and two documented separately) algorithms.
	
	\item Appendix \ref{appendix_B}: Simulation example to demonstrate the performance of R-ABC and R-BSL (MA(2) example) which supplements the main text in Section \ref{simulation_ex}.
	
	\item Appendix \ref{appendix_C}: Additional results for the misspecified g-and-k example which supplements the main text in Section \ref{gnk_example}.
	
	\item Appendix \ref{appendix_D}: Additional results for the empirical illustration which supplements the main text in Section \ref{empirical_exercise}.
	
	\item Appendix \ref{appendix_E}: Additional simulation example to demonstrate the performance of R-ABC.
	
\end{enumerate}

\titleformat{\section}[block]{\large\bfseries}{\Alph{section}}{1em}{}

\section{ABC, ABC-Reg and R-ABC algorithms}
\label{appendix_A}

\medskip\medskip
\subsection*{A.1 ABC and  ABC-Reg algorithms}
\medskip

\begin{algorithm}
	\caption{ABC accept/reject algorithm} \label{ABC}
	\begin{algorithmic}[1]
		\State Simulate ${\theta }^{i}$, $i=1,2,...,N$, from $\pi({\theta }),$
		\State Simulate $\mathbf{z}^{i}=(z_{1}^{i},z_{2}^{i},...,z_{n}^{i})^{\prime }$, $i=1,2,...,N$, from $p(\cdot|\theta^{i})$;
		\State For each $i=1,...,N$, select ${\theta }^{i}$ such that: $d\{\eta(\mathbf{z}^{i}),\eta(\mathbf{y})\}\leq \epsilon.$
	\end{algorithmic}
\end{algorithm}

\begin{algorithm}
	\caption{ABC-Reg Algorithm}\label{ABC-Reg}
	\begin{algorithmic}[1]
		\State Let $\{\theta^i,\z^i \}_{i=1}^{N}$ be $N$ draws from an initial ABC posterior approximation.
		\State For each $i$, calculate weights $w^i\propto K_\epsilon[d\{\eta(\z^i),\eta(\y)\}]$, with $K_\epsilon(\cdot)$ a bounded kernel function with bandwidth $\epsilon$. 
		
		\State Fit the regression model $\theta^i=\beta_0+\beta_1^{\prime} \eta(\mathbf{z}^i)+\nu_i,$ where $\nu_i$ denotes the model residual, to obtain $\hat\beta_0,\;\hat\beta_1$; i.e., minimize (in $\beta_0,\beta_1$)
		$
		\sum_{i=1}^{N}\left[\theta^i-\beta_0-\beta_1'\eta(\z^i)\right]^2w_i.
		$ 
		\State Adjust $\theta^i$ according to
		$
		\tilde\theta^i=\hat\beta_1'\eta(\y)+\hat\nu^{i}=\hat\beta_1'\eta(\y)+[\theta^i-\hat\beta_1'\eta(\z^i)].
		$
	\end{algorithmic}
\end{algorithm}

 In Algorithm 1, $N$ is the number of draws from $\pi(\mathbf{\theta})$, and the tolerance level $\epsilon$ is chosen to be small. In practice, most researchers use a modified version of Algorithm 1, known as ``reference-table acceptance-rejection ABC" (\citealp{cornuet2008inferring}), where the acceptance step is replaced with a nearest-neighbour selection step (\citealp{biau2015new}). In this version, the accepted draws of $\theta$ correspond to an empirical quantile over the simulated distances $d\{\eta(\z),\eta(\y)\}$. Specifically, Step 3 in Algorithm 1 is replaced with: Select all $\theta^{i}$ associated with the $q=\delta/N$ smallest distances $d\{\eta(\z),\eta(\y)\}$ for some $\delta$.

\medskip\medskip
\subsection*{A.2 Standard ABC accept/reject algorithm for performing step one of R-ABC approach}
\medskip\medskip
\begin{algorithm}[H]
	\label{R-ABC-SMC-step1}
	\caption{Step one of R-ABC Approach}
	\medskip
	\textbf{Step one: Standard ABC accept/reject algorithm}\\
	\medskip
	\textit{Inputs}: Observed data $\y$, prior distribution $\pi(\theta)$, distance metric $d$, number of draws from prior $N_{1}$, first set of summaries $\psi(\y)$, ABC tolerance $\epsilon_{1}$.\\
	\medskip
	\For{ i in 1 to $N_{1}$}{
		\medskip
		\Repeat{$d^{i} \leq \epsilon_{1}$}{
			\medskip
			1) $\theta^{i} \sim \pi(\theta)$
			\medskip \\
			2) $\mathbf{z}^{i}=(z^{i}_{1},z^{i}_{2}, ..., z^{i}_{n} ) \sim p(\cdot |\theta^{i})$
			\medskip \\
			3) $d^{i}= d \{ \psi(\mathbf{z}^{i}), \psi(\mathbf{y})\} $
			\medskip
		} 
	}
	
	\medskip \medskip
	\textit{Final output}: Samples $\{\theta^{i}, d^{i}, \psi(\mathbf{z}^{i})\}_{i=1}^{M}$ based on the approximate posterior with kernel $ \1[d\{\psi(\z), \psi(\y)\} \leq \epsilon_{1}]$
\end{algorithm} 

\medskip\medskip

\subsection*{A.3 ABC-SMC replenishment algorithm adapted for performing step two of R-ABC approach using Laplace prior for $\Gamma$}
\spacingset{0.9} 
{\centering
	\begin{minipage}{1\linewidth}
		\begin{algorithm}[H]
			\label{R-ABC-SMC-lap}
			\tiny
			\textit{Inputs}: Observed data $\y$, prior distribution of adjustment components $\pi(\Gamma)$, distance metric $d$, proportion of particles to drop at each iteration $\alpha=0.5$, number of particles $N$, first set of summaries $\psi(\y)$, second set of summaries $\varphi(\y)$, minimum acceptance probability $p_{acc\_min}$, retained $\theta$ from step one, ABC tolerance of step one as $\epsilon_{1}$, proposal distribution $q$, integer part of $\alpha N$ is $N_{a}$.\\
			\caption{ABC-SMC replenishment algorithm for performing step two of R-ABC}
			\medskip
			\small
			\For{ i in 1 to N}{
				\medskip
				\Repeat{$d^{i} \leq \epsilon_{0}$}{
					\medskip
					1) $\Gamma^{i} \sim \pi(\Gamma)$; Adjustment component sampling from Laplace(0, $\lambda$)
					\medskip\\
					2) $\mathbf{z}^{i}=(z^{i}_{1},z^{i}_{2}, ..., z^{i}_{n} ) \sim p(\cdot | \theta^{i} )$; $\theta^{i}$ from step one
					\medskip \\
					3) $d^{i}=d\big( \varphi(\mathbf{z}^{i}) + \Gamma^{i}, \varphi(\mathbf{y}) \big)$; Summary adjustment
					\medskip
				} 
			}
			\medskip
			sort the particle set $(\theta^{i}, d^{i}, \Gamma^{i})$ by $d^{i}$; compute max distance $\epsilon_{MAX} = d^{N}$;\\
			set $p_{acc}=1;$\\
			\medskip
			\While{$p_{acc}>p_{acc\_min}$}{
				\medskip
				drop $N_{a}$ particles with largest $d$ and compute $\epsilon$ for the next target: $\epsilon_{NEXT}=d^{N-N_{a}}$;\\
				set the acceptance counter $i_{acc}=0$;\\
				\medskip
				\For{j in 1 to $N_{a}$}{
					resample: $\theta^{N-N_{a}+j}$ from $\bigl\{ \theta^{i}\bigr\}_{i=1}^{N-N_{a}}$ and select $\Gamma^{N-N_{a}+j}$ corresponding to the random sampled index;\\
					\medskip
					\For{k in 1 to R}{
					propose move $(\theta^{**}, \Gamma^{**}) \sim q(\cdot| \theta^{N-N_{a}+j}, \Gamma^{N-N_{a}+j})$;\\
					\medskip
					$\mathbf{z} \sim p(\cdot | \theta^{**})$; $d_{2}= d\big( \varphi(\mathbf{z}) +  \Gamma^{**}, \varphi(\mathbf{y}) \big)$; $d_{1}= d\big( \psi(\mathbf{z}), \psi(\mathbf{y}) \big)$ (Step one related ABC distances)\\
					\medskip
					compute acceptance ratio:
					$$
					\tiny
					MH = min \biggl\{ 1, \frac{\pi(\theta^{**}) \pi(\Gamma^{**})}{\pi(\theta^{N-N_{a}+j})\pi(\Gamma^{N-N_{a}+j})} \1 \big[d\big(  \varphi(\mathbf{z}) +  \Gamma^{**}, \varphi( \mathbf{y})\big) \leq \epsilon_{NEXT}\big] . \1 \big[d\big( \psi(\mathbf{z}), \psi(\mathbf{y}) \big) \leq \epsilon_{1}\big]\biggr\}
					$$
					\medskip
					\If{$u \sim U(0,1) < $  MH}{
						set $\theta^{N-N_{a}+j} = \theta^{**}$; set $\Gamma^{N-N_{a}+j} = \Gamma^{**}$; set $d^{N-N_{a}+j} = d\big(  \varphi(\mathbf{z}) +  \Gamma^{**}, \varphi( \mathbf{y})\big)$;\\
						set $i_{acc}=i_{acc}+1$;\\
					}
				}
				}
				\medskip
				compute the acceptance probability $p_{acc} = i_{acc}/(RN_{a})$;\\
				sort the particle set $(\theta^{i}, d^{i}, \Gamma^{i})$ by $d^{i}$; compute max distance $\epsilon_{MAX} = d^{N}$;\\
				\medskip
			}
			\medskip
			\textit{Final output}: Samples $\{\theta^{i}, \Gamma^{i}, \}_{i=1}^{N}$ based on the approximate posterior with kernel $ \1[d\{\psi(\z), \psi(\y)\} \leq \epsilon_{1}]\1[d\{\phi(\z, \Gamma), \varphi(\y)\} \leq \epsilon_{NEXT}]$, where $\epsilon_{NEXT}$ is the ABC tolerance of step two.
		\end{algorithm}
	\end{minipage}
	\par
}

\subsection*{A.4 ABC-SMC replenishment algorithm adapted for performing step two of R-ABC approach using spike-and-slab prior for $\Gamma$}
\spacingset{1} 
{\centering
	\begin{minipage}{1\linewidth}
		\begin{algorithm}[H]
			\label{R-ABC-SMC-spk}
			\tiny
			\textit{Inputs}: Observed data $\y$, prior distribution of adjustment components $\pi(\Gamma)$, distance metric $d$, proportion of particles to drop at each iteration $\alpha=0.5$, number of particles $N$, first set of summaries $\psi(\y)$, second set of summaries $\varphi(\y)$, minimum acceptance probability $p_{acc\_min}$, retained $\theta$ from step one, ABC tolerance of step one as $\epsilon_{1}$, proposal distribution for $\theta$ as $q_{\theta}$, proposal distribution for $\Gamma$ as $q_{\Gamma}$, integer part of $\alpha N$ is $N_{a}$.\\
			\caption{ABC-SMC replenishment algorithm for performing step two of R-ABC}
			\medskip
			\small
			\For{ i in 1 to N}{
				\Repeat{$d^{i} \leq \epsilon_{0}$}{
					1) $\Gamma^{i} \sim \pi(\Gamma)$; Adjustment component sampling from spike-and-slab with slab of Laplace(0, $\lambda$)
					\medskip\\
					2) $\mathbf{z}^{i}=(z^{i}_{1},z^{i}_{2}, ..., z^{i}_{n} ) \sim p(\cdot | \theta^{i} )$; $\theta^{i}$ from step one
					\medskip \\
					3) $d^{i}=d\big( \varphi(\mathbf{z}^{i}) + \Gamma^{i}, \varphi(\mathbf{y}) \big)$; Summary adjustment
				} 
			}
			\medskip
			sort the particle set $(\theta^{i}, d^{i}, \Gamma^{i})$ by $d^{i}$; compute max distance $\epsilon_{MAX} = d^{N}$;\\
			set $p_{acc}=1;$\\
			\medskip
			\While{$p_{acc}>p_{acc\_min}$}{
				drop $N_{a}$ particles with largest $d$ and compute $\epsilon$ for the next target: $\epsilon_{NEXT}=d^{N-N_{a}}$;\\
				\medskip
				\For{j in 1 to $N_{a}$}{
					resample: $\theta^{N-N_{a}+j}$ from $\bigl\{ \theta^{i}\bigr\}_{i=1}^{N-N_{a}}$ and select $\Gamma^{N-N_{a}+j}$ corresponding to the random sampled index;\\
					\medskip
					\For{k in 1 to R}{
					propose move $\theta^{**} \sim q_{\theta}(\cdot| \theta^{N-N_{a}+j})$;\\
					propose move $\Gamma^{**} \sim q_{\Gamma}(\cdot|\Gamma^{N-N_{a}+j})$;\\
					\medskip
					$\mathbf{z} \sim p(\cdot | \theta^{**})$; $d_{2}= d\big( \varphi(\mathbf{z}) +  \Gamma^{**}, \varphi(\mathbf{y}) \big)$; $d_{1}= d\big( \psi(\mathbf{z}), \psi(\mathbf{y}) \big)$ (Step one related ABC distances)\\
					\medskip
					compute acceptance ratio:
					\vspace{0.5cm}
					\scalebox{0.95}{$
					\tiny
					MH = min \biggl\{ 1, \frac{\pi(\theta^{**}) \pi(\Gamma^{**})q(\Gamma^{N-N_{a}+j}|\Gamma^{**})}{\pi(\theta^{N-N_{a}+j})\pi(\Gamma^{N-N_{a}+j})q(\Gamma^{**}|\Gamma^{N-N_{a}+j})} \1 \big[d\big(  \varphi(\mathbf{z}) +  \Gamma^{**}, \varphi( \mathbf{y})\big) \leq \epsilon_{NEXT}\big] . \1 \big[d\big( \psi(\mathbf{z}), \psi(\mathbf{y}) \big) \leq \epsilon_{1}\big]\biggr\}
					$}
					\If{$u \sim U(0,1) < $  MH}{
						set $\theta^{N-N_{a}+j} = \theta^{**}$; set $\Gamma^{N-N_{a}+j} = \Gamma^{**}$; set $d^{N-N_{a}+j} = d\big(  \varphi(\mathbf{z}) +  \Gamma^{**}, \varphi( \mathbf{y})\big)$;\\
						set $i_{acc}=i_{acc}+1$;\\
					}
				}
				}
				\medskip
				compute the acceptance probability $p_{acc} = i_{acc}/(RN_{a})$;\\
				sort the particle set $(\theta^{i}, d^{i}, \Gamma^{i})$ by $d^{i}$; compute max distance $\epsilon_{MAX} = d^{N}$;\\
			}
			\medskip
			\textit{Final output}: Samples $\{\theta^{i}, \Gamma^{i}, \}_{i=1}^{N}$ based on the approximate posterior with kernel $ \1[d\{\psi(\z), \psi(\y)\} \leq \epsilon_{1}]\1[d\{\phi(\z, \Gamma), \varphi(\y)\} \leq \epsilon_{NEXT}]$, where $\epsilon_{NEXT}$ is the ABC tolerance of step two.
		\end{algorithm}
	\end{minipage}
	\par
}

\spacingset{1.45} %overleaf template
\section{Moving average example}
\label{appendix_B}

\label{ma2_example}
Here we demonstrate R-ABC in a common example in the approximate inference literature, the moving average (MA) model. The researcher believes data $\y$ is generated according to the following MA model of order 2 (MA(2)):
\begin{equation}
	z_{t}=e_{t}+\theta _{1}e_{t-1}+\theta _{2}e_{t-2},  \label{MA2}
\end{equation}
where $e_{t}\stackrel{iid}{\sim} \mathcal{N}(0,1)$ and the unknown parameters $\theta=(\theta_{1},\theta_{2})^\prime$ are assumed to obey
\begin{equation}
	-2<\theta _{1}<2,\;\theta _{1}+\theta _{2}>-1,\theta _{1}-\theta _{2}<1.
	\label{const1}
\end{equation}
Our prior information on $\theta=(\theta_{1},\theta_{2})^\prime$ is uniform over the invertibility region in \eqref{const1}. A useful choice of summary statistics for the MA(2) model are the
sample autocovariances $\eta _{j}(\z)=\frac{1}{T}%
\sum_{t=1+j}^{T}z_{t}z_{t-j}$, for $j=0,1,2$. Let $\eta(\z)$ denote the summaries $\eta(\z)=(\eta _{0}(\z), \eta _{1}(\z), \eta _{2}(\z))^\prime$. Then, under the DGP in (\ref{MA2}), it can be shown that the summaries $\eta(\z)$ satisfy
\begin{equation}
	\label{ma_btheta}
	\text{plim}_n\eta(\z) =b(\theta):=
	\begin{pmatrix}
		1+\theta^2_{1}+\theta^2_{2},&\theta_{1}(1+\theta_{2}),&\theta_{2}
	\end{pmatrix}^{\prime}.
\end{equation}

While the researcher believes the data is generated according to an MA(2) model, the true DGP for $\y$ evolves according to a stochastic volatility (SV) model,
\begin{flalign}
	y_{t}&=\exp(h_{t}/2)u_{t}\nonumber\\h_{t}&=\omega+\rho h_{t-1}+v_{t}\sigma_{v},\label{SV}
\end{flalign} where $0<\rho<1$, $0<\sigma_{v}<1$, and $u_{t}$ and  $v_{t}$ are both iid standard Gaussian. In this case, if one takes $\mathbf{\eta }\left( \mathbf{y}\right) :=(\eta _{0}\mathbf{(y)}, \eta_{1}\mathbf{(y)}, \eta _{2}\mathbf{(y)})^\prime$ it follows that, under the DGP in (\ref{SV}),
\begin{equation}
	\label{ma_b0}
	\text{plim}_n\eta(\mathbf{y}) =b_{0}:=
	\begin{pmatrix}\exp\left( \frac{\omega}{1-\rho}+\frac{1}{2}\frac{\sigma_v^2}{1-\rho^2}\right)
		,&0,&0
	\end{pmatrix}^{\prime}.
\end{equation}
For any values of $\omega,\sigma_v$ and $\rho$ such that $$\exp\left( \frac{\omega}{1-\rho}+\frac{1}{2}\frac{\sigma_v^2}{1-\rho^2}\right)\neq1,$$ it follows that model is not compatible and hence misspecified in the sense of ABC. From the definition of $b(\theta)$ and $b_0$, it also follows that the value that minimizes $\|b(\theta)-b_0\|$ is $\theta^*=(0,0)^\prime$, and this is the value onto which one would expect the ABC posterior to concentrate in the limit.

To understand the performance of ABC, R-ABC and R-BSL in this misspecified setting, we consider the following Monte Carlo analysis: we generate 50 replications of the observed data of size $n$ = 1000, from the SV model in (\ref{SV}) with parameter values $\omega=-0.76, \rho=0.90$ and $\sigma_{v}=0.36$, and use ABC-SMC, ABC-SMC-Reg, R-ABC and R-BSL to conduct inference on $\theta$ in the misspecified MA(2) model. To implement ABC-SMC and ABC-SMC-Reg, we use all summary statistics and run the ABC-SMC replenishment algorithm until the MCMC acceptance rate drops below 1\%.

R-ABC is implemented using the following partition:
$$\psi(\y)= \eta_{2}(\y)=\frac{1}{T}\sum_{t=3}^{T}y_{t}y_{t-2} \quad \text{and} \quad \varphi(\y)=\left(\eta_{0}(\y), \eta_{1}(\y) \right)^{\prime} = \left(  \frac{1}{T}\sum_{t=1}^{T}y_{t}^{2},      \frac{1}{T}\sum_{t=2}^{T}y_{t}y_{t-1} \right)^{\prime}.
$$
Under the assumed MA(2) model, the second-order autocovariance depends solely on $\theta_{2}$ in the limit so that $\eta_{2}(\y)$ is highly informative about $\theta_{2}$. Given this, it is reasonable to assume that the second-order autocovariance, $\eta_{2}(\y)$,  can serve as the summary statistic to be matched under the assumed model, i.e., $\psi(\y)$. The remaining summary statistics, variance ($\eta_{0}(\y)$) and first-order autocovariance ($\eta_{1}(\y)$), are chosen for $\varphi(\y)$.

\begin{figure}[h!]
	\centering 
	\includegraphics[width=0.6\linewidth]{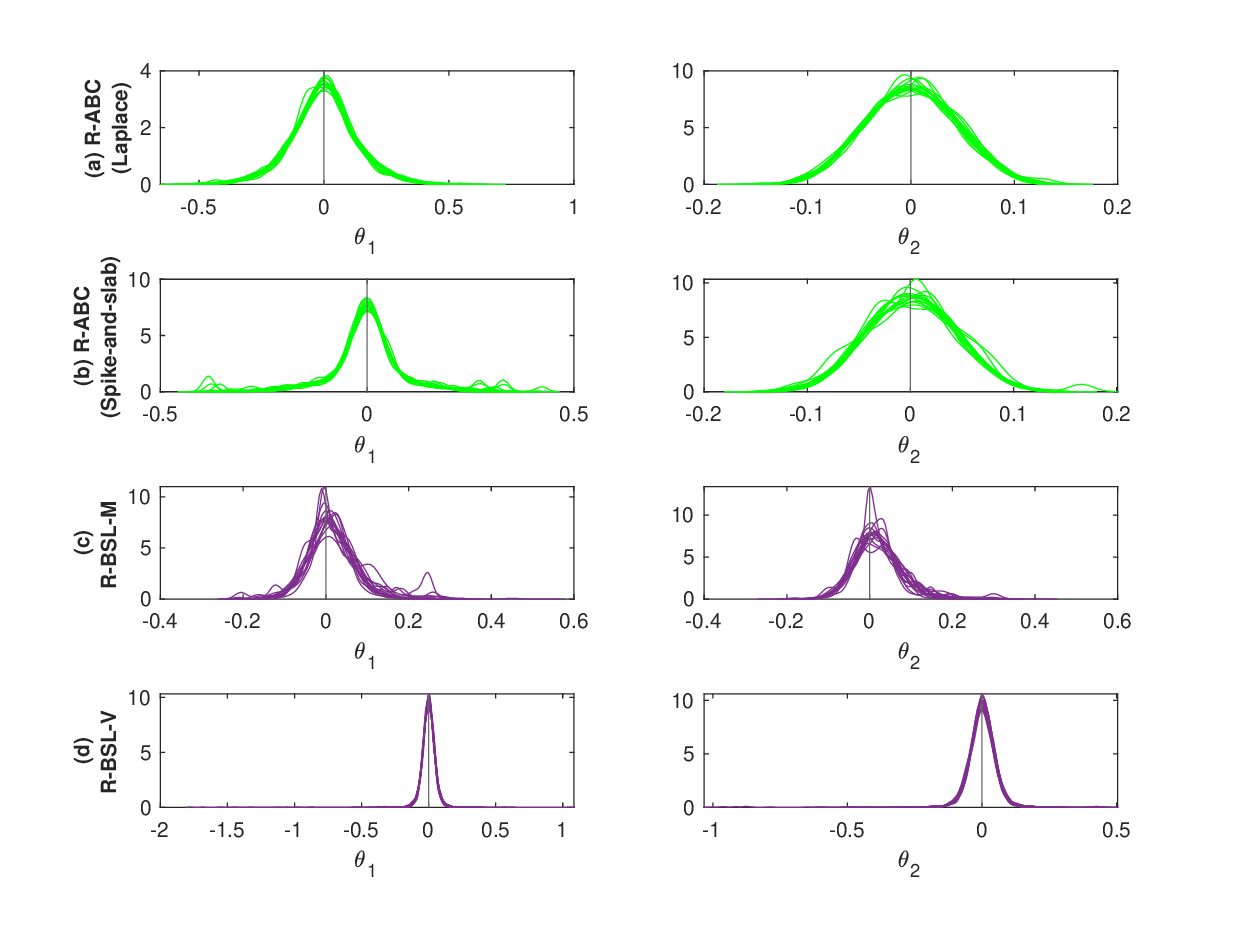} 
	\caption{Rows (a) and (b) respectively provide the posteriors for $\theta_{1}$ and $\theta_{2}$ using the R-ABC approach when using Laplace and spike-and-slab priors for the adjustment components. Rows (c) and (d) respectively provide the same information for R-BSL-M and R-BSL-V. The vertical line represents the pseudo-true values. The hyperparameter for the priors associated with the adjustment components of R-ABC is $\lambda=0.125.$} 
	\label{ma2_theta_robust} 
\end{figure}

\begin{figure}[h!]
	\centering 
	\includegraphics[width=0.8\linewidth]{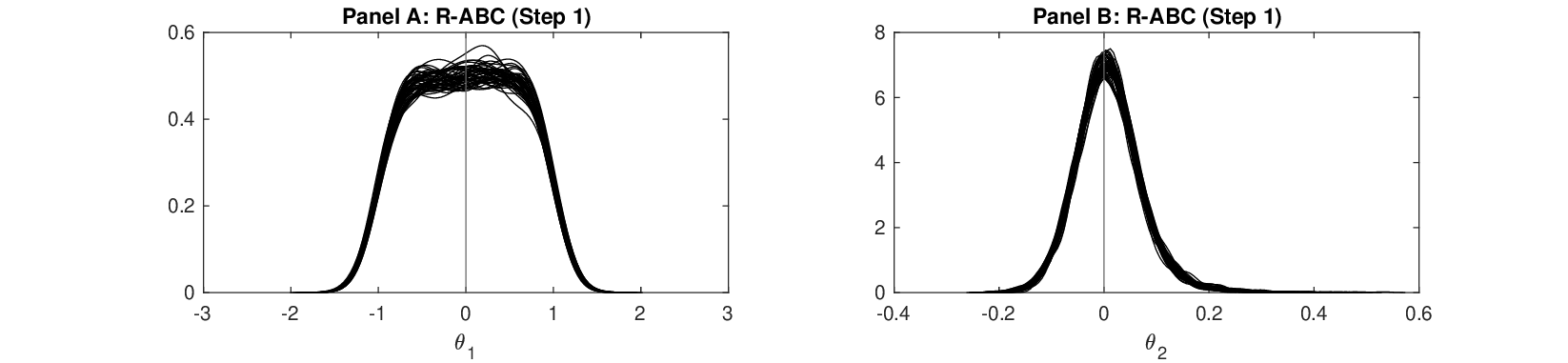} 
	\caption{Panels A and B respectively provide the posteriors for $\theta_{1}$ and $\theta_{2}$ using the step one of R-ABC approach. The vertical line represents the pseudo-true values.} 
	\label{ma2_theta_rabc_step1} 
\end{figure}
We conduct step one of the R-ABC approach using the ABC accept/reject algorithm using only $\psi(\y)$. We simulate $N_{1}=25000$ draws from the prior and retain those leading to the smallest 5\% of the overall simulated distances. Step two of the R-ABC approach is implemented using the adjusted summaries, where two adjustment components are now needed, for $\eta_{0}(\y)$ and $\eta_{1}(\y)$, in $\varphi(\y)$. We use Algorithms \ref{R-ABC-SMC-lap} and \ref{R-ABC-SMC-spk} when using the ABC-SMC replenishment algorithm in step two, and run it until the MCMC acceptance rate drops below 1\%. In Algorithms \ref{R-ABC-SMC-lap} and \ref{R-ABC-SMC-spk}, we respectively use Laplace and spike-and-slab priors for the adjustment parameters, $\Gamma_{\text{R-ABC}} = (\gamma_{(rabc,1)}, \gamma_{(rabc,2)})^{\prime}$.

Both versions of R-BSL are implemented using the default prior choices: a Laplace prior with $\lambda=0.5$ for R-BSL-M, and an exponential prior with $\lambda=0.5$ for R-BSL-V (\citealp{frazier2021robust}). Each R-BSL approach uses $m=50$ simulated datasets to estimate the mean and variance. For each replication, we run the MCMC sampler for 100000 iterations, discard the first 10000 as burn-in, and thin the remaining sample by selecting every 100th draw. In R-BSL, we adjust all three summaries and therefore three adjustment components are needed. That is, we use $\Gamma_{\text{R-BSL}}=(\gamma_{(rbsl,1)}, \gamma_{(rbsl,2)}, \gamma_{(rbsl,3)})^{\prime}$ for $\eta _{0}, \eta _{1}$ and $\eta _{2}$, respectively.

In Figure \ref{ma2_theta_robust}, we present the resulting posteriors for the $\theta$ components obtained from the R-ABC and R-BSL approaches across different data replications. We plot only the first 20 data replications in Figure \ref{ma2_theta_robust} for the interpretability. Additionally, the posteriors for the $\theta$ components from step 1 of the R-ABC approach are also provided in Figure \ref{ma2_theta_rabc_step1}, and an animation plot of Figure \ref{ma2_theta_robust} is represented in Figure \ref{anim_ma2}.

As expected, the posterior of $\theta_{2}$ resulting from step one of the R-ABC approach is highly concentrated around the pseudo-true value, whereas the posterior of $\theta_{1}$ resulting from step one of the R-ABC approach is similar to the prior (see Panels B and A respectively of Figure \ref{ma2_theta_rabc_step1}). 

Most notably, once the robustification is done, the resulting R-ABC posteriors obtained in step two (Rows (a) and (b) of Figure \ref{ma2_theta_robust}) are more closely centered around the pseudo-true value $\theta^*=(0,0)'$. Comparing the R-ABC posteriors for $\theta_{1}$, it is clearly visible that the posterior obtained when using the spike-and-slab prior for $\Gamma$ is more concentrated around the pseudo-true value than the posterior of $\theta_{1}$ based on the Laplace prior for $\Gamma$. This highlights the importance of using the original shrinkage-type priors for adjustment components in R-ABC in producing more reliable inferences, since these priors are capable of eliminating unnecessary adjustment components completely in the robustification step. As a result, they prevent the adjustment from unfairly affecting the compatible components, thereby leading to inferences that are more reliable.

Further, when comparing the two variants of R-BSL, it is evident that the R-BSL-V posteriors (Row (d) of Figure \ref{ma2_theta_robust}) exhibit less variability across replications and are concentrated closely around the pseudo true values. Interestingly, both the R-ABC (Spike-and-slab) and R-BSL-V posteriors demonstrate similar behaviour for both components of $\theta$. Across the replications, the variability of R-BSL-V posteriors is less compared to the R-ABC (Spike-and-slab) posteriors.

\begin{figure}[h!]
	\centering 
	\includegraphics[width=0.65\linewidth]{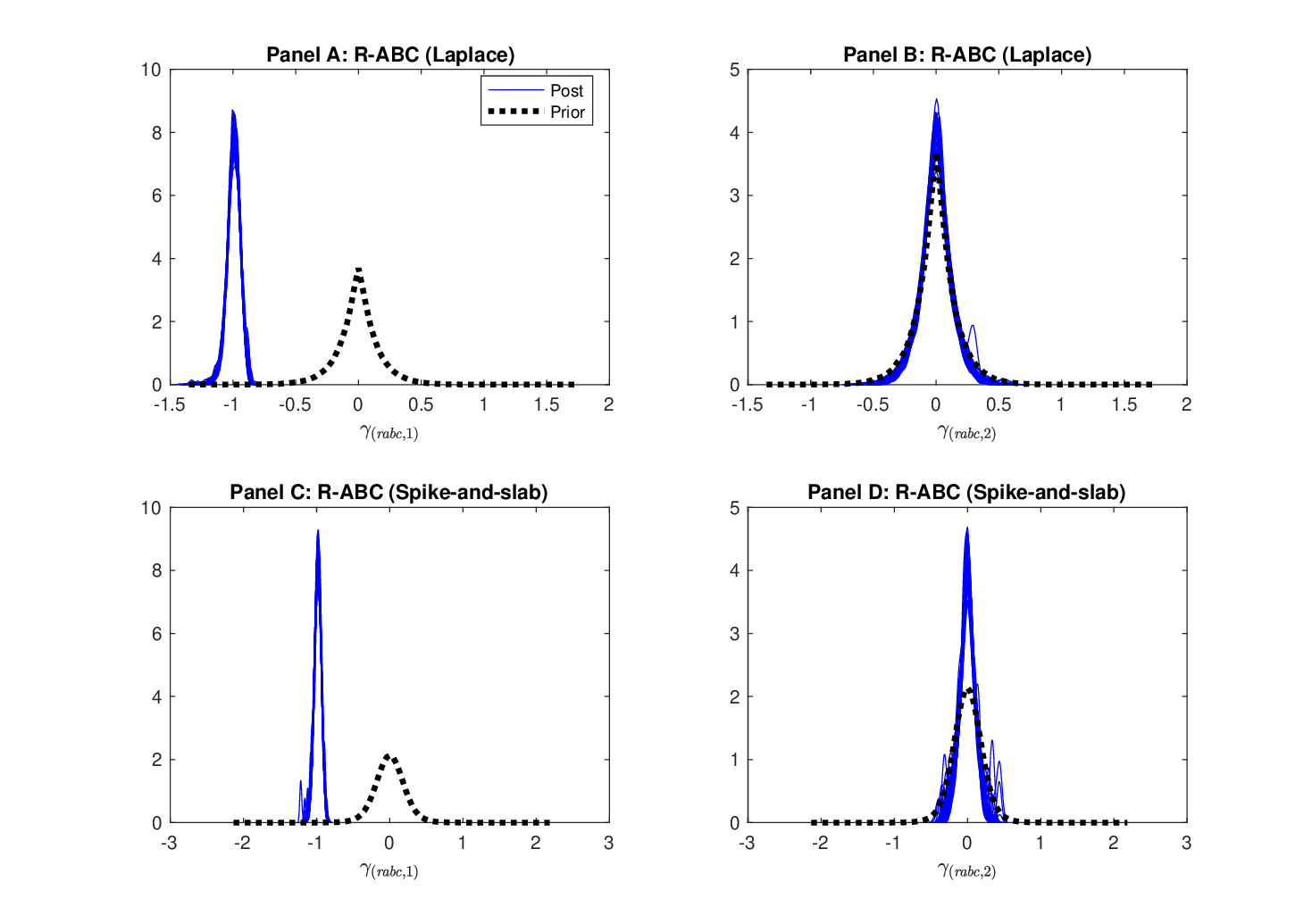} 
	\caption{Panels A and B respectively provide the posteriors for adjustment components, $\gamma_{1}$ and $\gamma_{2}$ of the R-ABC approach when using Laplace prior for $\Gamma$, while Panels C and D represent the same information when using spike-and-slab prior for $\Gamma$ in R-ABC approach. As indicated in the key, the marginal posterior of $\Gamma$ component (Post) is represented in a blue solid line. In Panels A and B, the black dotted lines correspond to the Laplace prior, whereas in Panels C and D, they represent the spike-and-slab prior. Both type of priors use the default hyperparameter $\lambda = 0.125$.} 
	\label{ma2_gam_rabc} 
\end{figure}

\begin{figure}[h!]
	\centering 
	\includegraphics[width=0.65\linewidth]{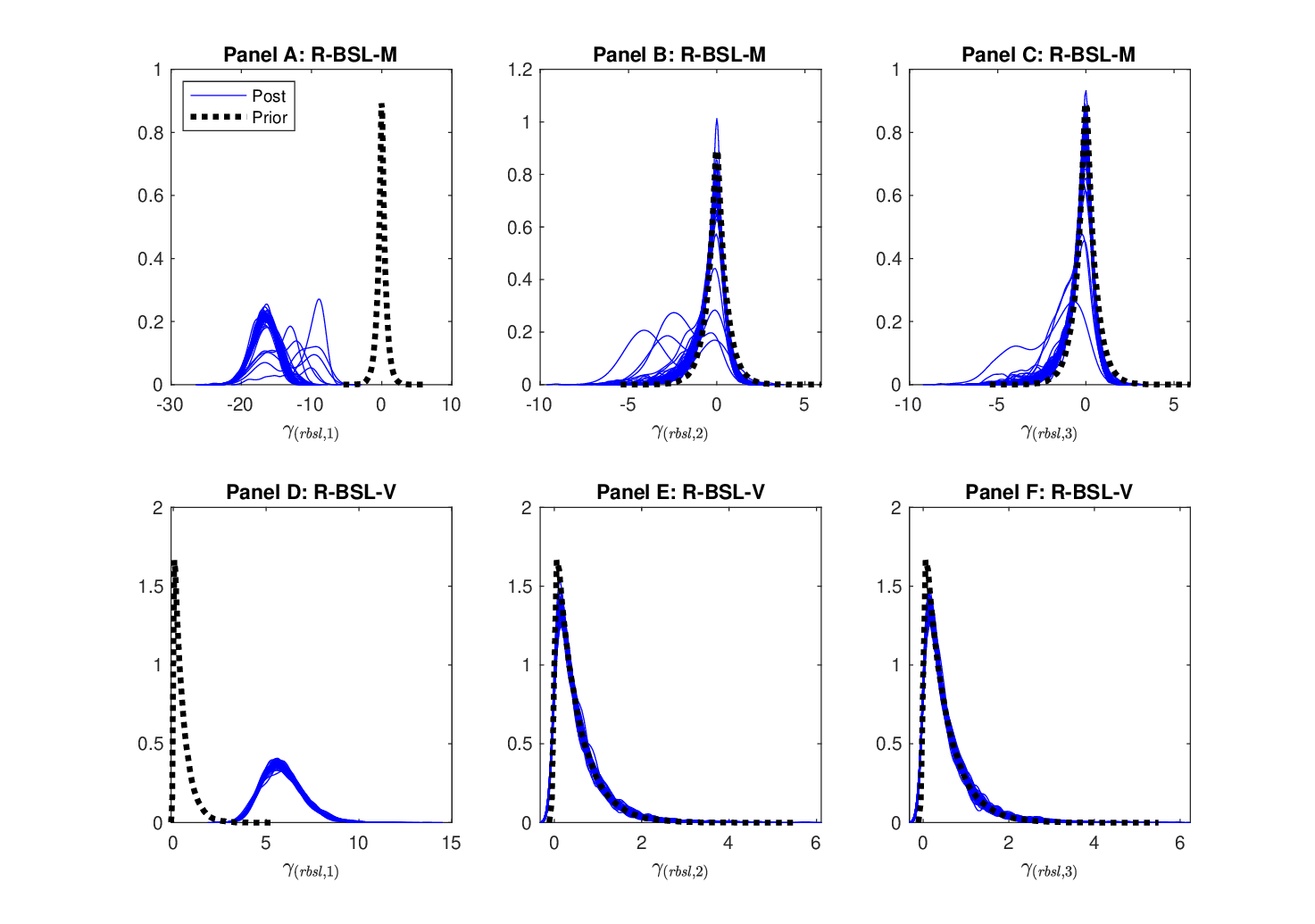} 
	\caption{Panels A, B, and C respectively provide the posteriors for adjustment components of the R-BSL-M approach while Panels D, E, and F provide the same information about R-BSL-V approach. As indicated in the key, the marginal posterior of $\Gamma_{\text{R-BSL}}$ component (Post) is represented in a blue solid line. The black dotted line in the first row represents the La(0,0.5) prior while the one in the bottom row corresponds to the exponential prior with $\lambda=0.5.$} 
	\label{ma2_gam_rbsl} 
\end{figure}

We now examine the marginal posteriors for the adjustment components of the R-ABC approach across 50 replications, which are given in Figure \ref{ma2_gam_rabc}. Panels A and B show the posteriors for $\Gamma$ when using the Laplace prior, while Panels C and D display the posteriors for $\Gamma$ when using the spike-and-slab prior. Referring back to the summaries used in the robustification step of R-ABC, under the true DGP in (\ref{SV}), the first-order autocovariance is zero for all values of $(\omega, \rho, \sigma_{v})^{\prime}$. Hence, we expect that R-ABC would detect incompatibility in the first summary statistic $\eta_{0}(\y)$, the sample variance, while the adjustment component for the first-order autocovariance would be indistinguishable from the prior. As observed in Panels A and C in Figure \ref{ma2_gam_rabc} under both priors and across all replications, the results are consistent with these expectations, with the R-ABC procedure clearly detecting that we cannot match the first summary statistic.\footnote{This is also confirmed by the results of the two-sample randomization test of location over the replicated datasets. Under R-ABC (Laplace), the proportion of times the test rejects $H_{0}$ for $\gamma_{(rabc,1)}$ and $\gamma_{(rabc,2)}$ respectively are 1 and 0.06. The same information under spike-and-slab prior are 1 and 0.18.} A similar conclusion can be drawn by examining the posteriors of the adjustment components for R-BSL as observed in Figure  \ref{ma2_gam_rbsl} .

To further compare the estimation methods, we record the bias of the posterior mean, and the average posterior standard deviation, and across the replications we calculate the Monte Carlo coverage of each procedure. We report the results for $\theta_{1}$ and $\theta_{2}$ in Table \ref{tab_ma2}.

{\renewcommand{\arraystretch}{0.8}
\begin{table}[h!]
	\caption{\scriptsize Monte Carlo coverage (Cov), bias of the posterior mean (Bias), and average posterior standard deviation (Std) for $\theta=(\theta_1,\theta_2)'$ in the MA(2) example. Cov is the percentage of times that the marginal 95\% credible set contains $\theta^*_j=0$, for $j=1,2$. Std is the average posterior standard deviation across the Monte Carlo trials. The rows refer to either the ABC or BSL approach.}
	\centering%
	\small  
	\begin{tabular}{lrrrrrrrrrr}
		\hline\hline
		&   \multicolumn{3}{c}{$\theta_{1}$}     & &  &   \multicolumn{1}{c}{$\theta_{2}$}  &  &\\
		\cline{2-4} \cline{6-8}
		& \multicolumn{1}{c}{Cov} & \multicolumn{1}{c}{Bias} & \multicolumn{1}{c}{Std} &       & \multicolumn{1}{c}{Cov} & \multicolumn{1}{c}{Bias} & \multicolumn{1}{c}{Std} \\
		{\textbf{ABC method}} \\
		ABC-SMC	&	100\%	&	-0.0006	&	0.0845	&&	100\%	&	-0.0006	&	0.0841	\\
		ABC-SMC-Reg	&	52\%	&	0.0183	&	0.0346	&&	56\%	&	0.0082	&	0.0342	\\
		R-ABC (Laplace)	&	100\%	&	-0.0007	&	0.1331	&&	100\%	&	0.0007	&	0.0439	\\
		R-ABC (Spike-and-slab) 	&	100\%	&	0.0000	&	0.0853	&&	100\%	&	0.0006	&	0.0434	\\
		R-BSL-M	&	100\%	&	0.0343	&	0.0722	&&	100\%	&	0.0331	&	0.0673	\\
		R-BSL-V	&	100\%	&	0.0000	&	0.0568	&&	100\%	&	0.0008	&	0.0540	\\
		\hline\hline
	\end{tabular}%
	\label{tab_ma2}%
\end{table}	
}
The results in Table \ref{tab_ma2} demonstrate that the bias for both parameters is relatively similar across the different estimation methods. Among these, R-ABC (Spike-and-slab) and R-BSL-V exhibit the smallest bias for $\theta_{1}$, while R-ABC (Spike-and-slab) and ABC-SMC achieve the smallest bias for $\theta_{2}$. ABC-SMC-Reg, however, demonstrates notably poor coverage for both parameters. In contrast, ABC-SMC, R-ABC, and R-BSL approaches achieve a coverage of 100\% for both parameters, ensuring the pseudo-true value is always captured. 

When comparing posterior variability, ABC-SMC-Reg displays the smallest posterior uncertainty. For $\theta_{1}$, the posterior uncertainties associated with both versions of R-ABC are higher than those of ABC-SMC and ABC-SMC-Reg. For $\theta_{2}$, the posterior uncertainties from both R-ABC approaches are smaller than ABC-SMC, but remain larger than those of ABC-SMC-Reg. In comparison to R-BSL, the posterior uncertainty of R-ABC (Spike-and-slab) is quite higher than those of R-BSL-V for $\theta_{1}$. However, for $\theta_{2}$, both R-BSL versions display higher posterior uncertainties than R-ABC approaches.

Given that the Gaussian assumption of the summary statistics is reasonable in this example, it clearly demonstrate that the R-BSL provides reliable inferences for unknown parameters. Additionally, both R-BSL-V and R-ABC (Spike-and-slab) yield closely similar results in this setting, providing accurate statistical inferences under model misspecification.

\begin{figure}[h!]
	\centering
	\includemedia[
	width=0.7\linewidth, % Adjust size
	height=0.35\linewidth, % Adjust size
	activate=onclick,
	addresource=Anim_ma2_theta_tau0_125.mp4,
	flashvars={
		source=Anim_ma2_theta_tau0_125.mp4   % Video source
		&autoPlay=true              % Automatically play when clicked
		&loop=true                  % Loop the video
	}
	]{}{VPlayer.swf} % Use the default video player
	\caption{Animation plot corresponding to Figure \ref{ma2_theta_robust} across the replications. Rows (a) and (b) respectively provide the posteriors for $\theta_{1}$ and $\theta_{2}$ using ABC-SMC and ABC-SMC-Reg, while the Rows (c) and (d) respectively represent the same information using the R-ABC approach when using Laplace and spike-and-slab priors for the adjustment components. Finally, Rows (e) and (f) respectively provide the same information for R-BSL-M and R-BSL-V. The hyperparameter for the priors associated with the adjustment components in R-ABC is $\lambda=0.125.$}%
	\label{anim_ma2} 
\end{figure}

\newpage
\section{Misspecified g-and-k example} 
\label{appendix_C}

\subsection*{C.1 Dependence of the distributional properties of robust summaries on the g-and-k parameters}

\begin{figure}[h!]
	\centering 
	\includegraphics[width=0.55\linewidth]{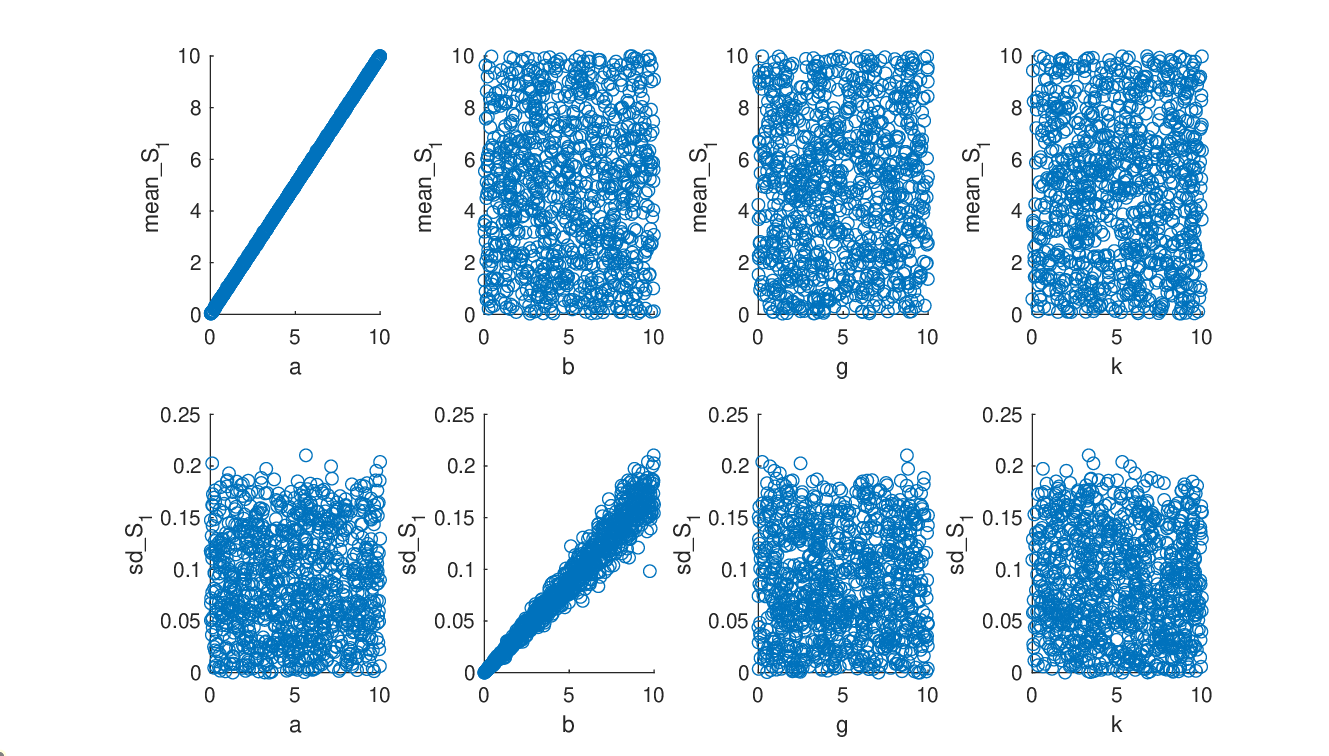} 
	\caption{Scatter plots of the g-and-k parameters against the estimated mean (top row) and standard deviation (bottom row) of $S_{1}$} 
	\label{fig_S1_gnk} 
\end{figure}

\begin{figure}[h!]
	\centering 
	\includegraphics[width=0.55\linewidth]{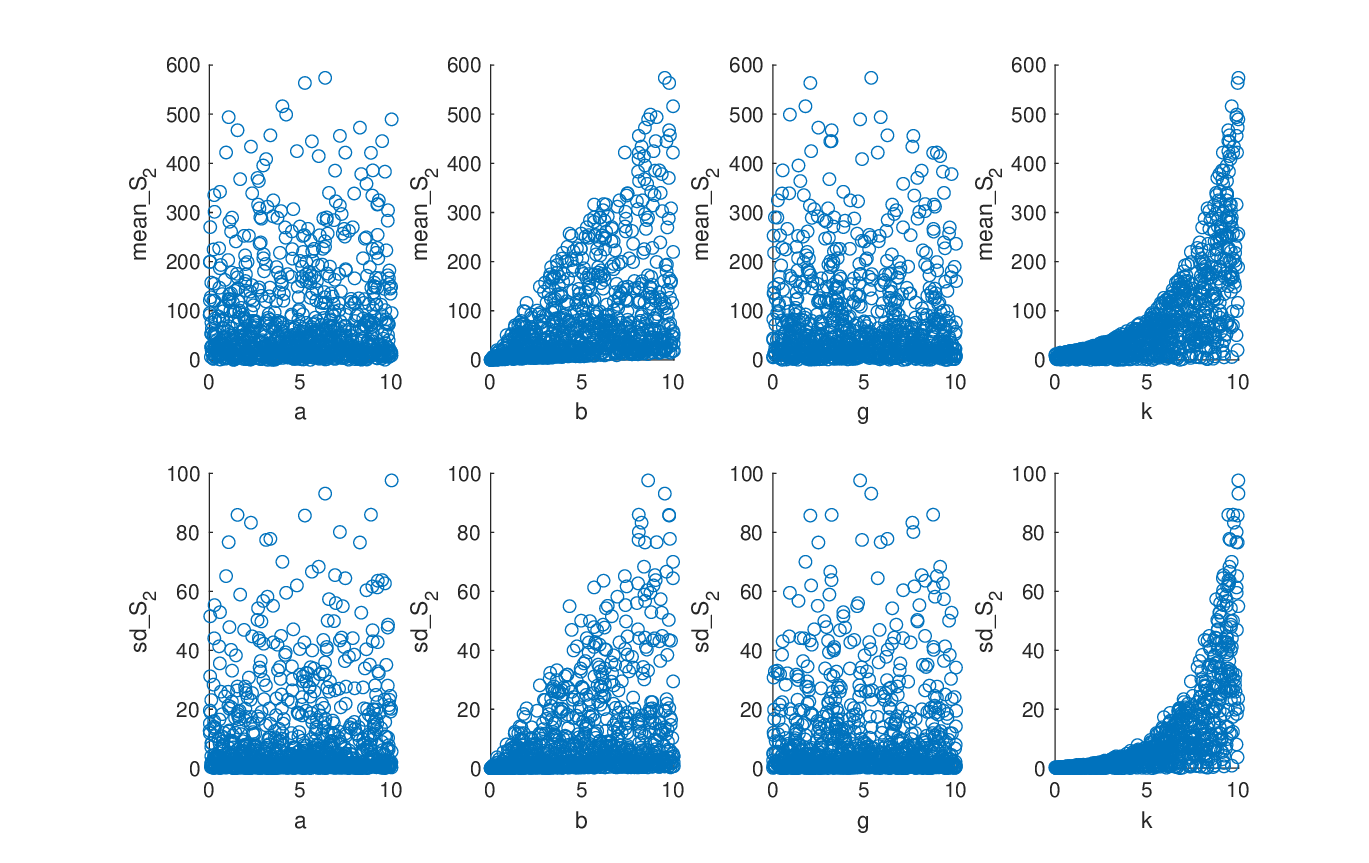} 
	\caption{Scatter plots of the g-and-k parameters against the estimated mean (top row) and standard deviation (bottom row) of $S_{2}$} 
	\label{fig_S2_gnk} 
\end{figure}

\begin{figure}[h!]
	\centering 
	\includegraphics[width=0.55\linewidth]{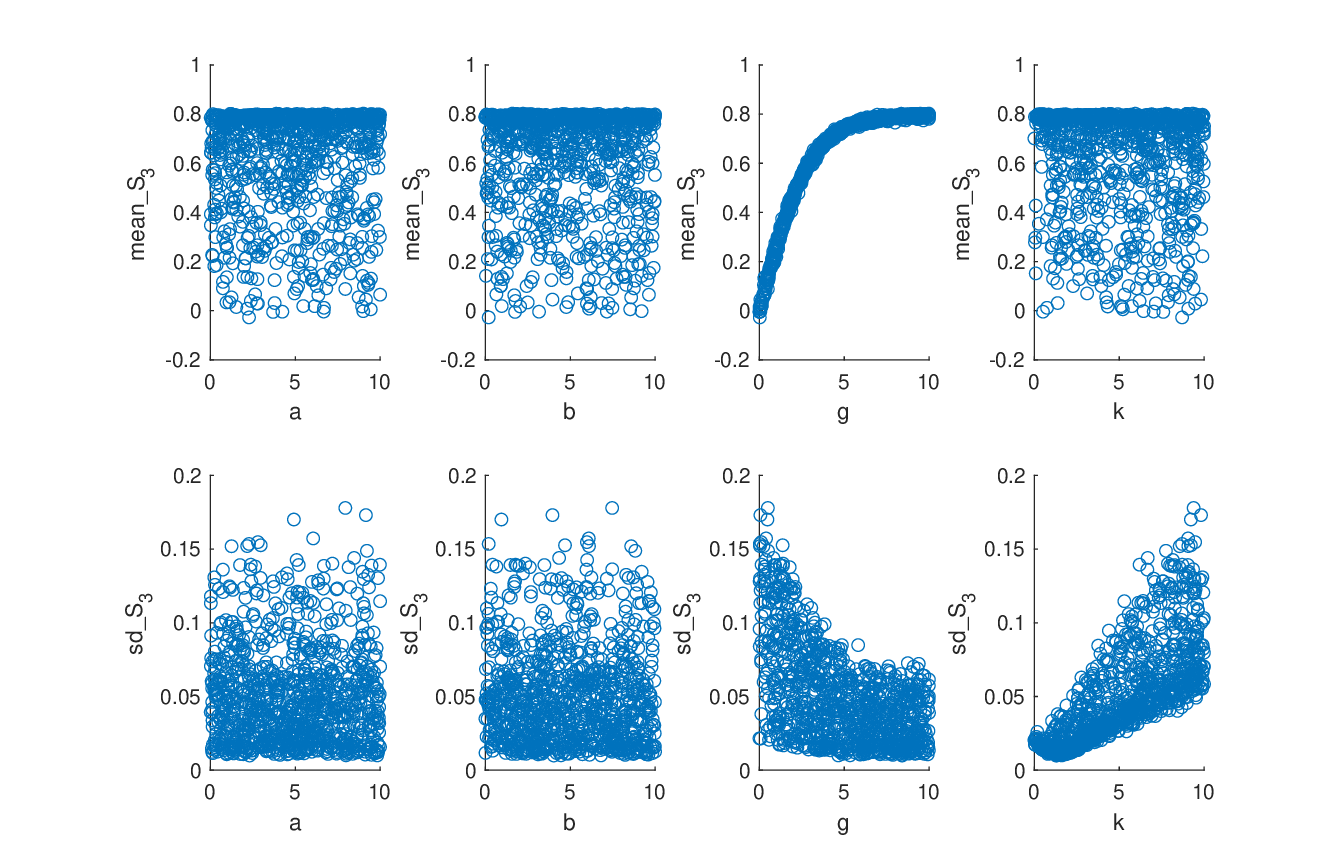} 
	\caption{Scatter plots of the g-and-k parameters against the estimated mean (top row) and standard deviation (bottom row) of $S_{3}$} 
	\label{fig_S3_gnk} 
\end{figure}

\begin{figure}[h!]
	\centering 
	\includegraphics[width=0.55\linewidth]{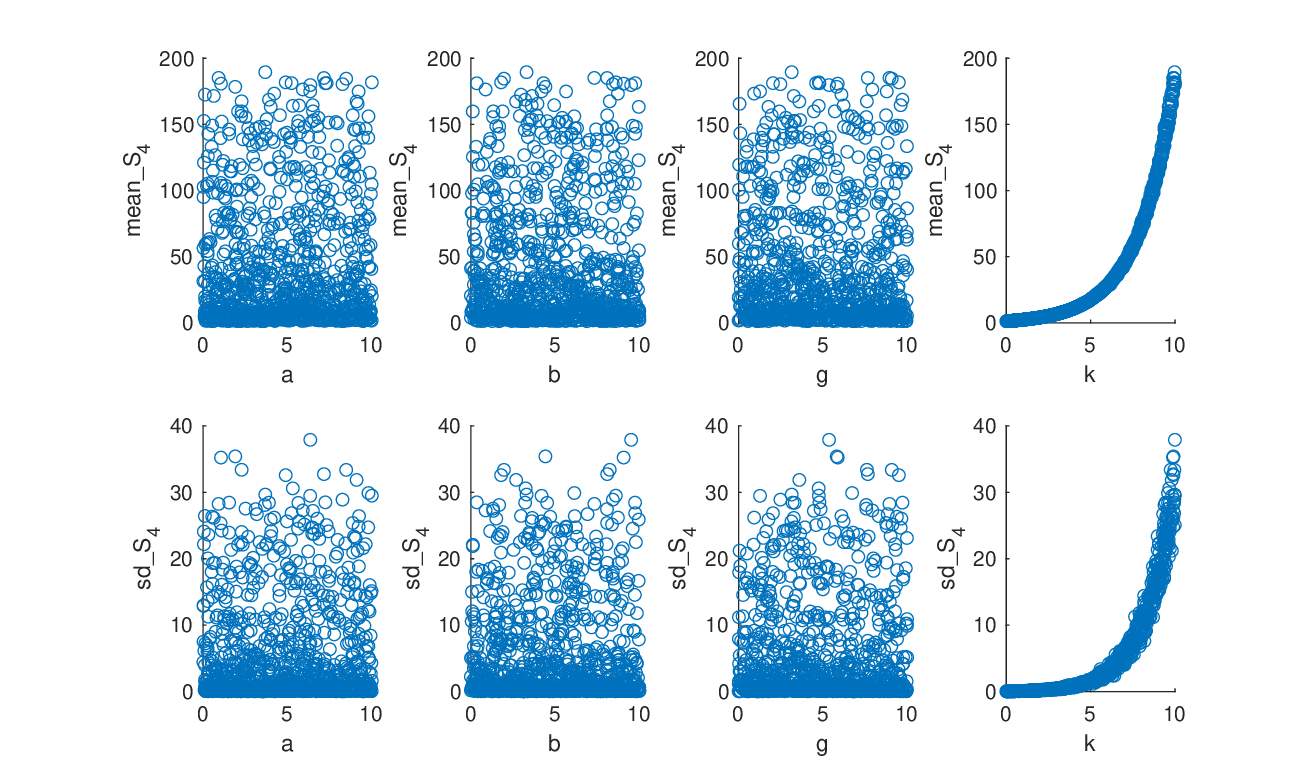} 
	\caption{Scatter plots of the g-and-k parameters against the estimated mean (top row) and standard deviation (bottom row) of $S_{4}$} 
	\label{fig_S4_gnk} 
\end{figure}

To understand the structure of how summaries can be partitioned in this g-and-k example for our proposed R-ABC approach, we perform a straightforward exercise. We generate 50 independent replicates of datasets based on 1,000 samples from the prior predictive distribution. For each dataset, we calculate the summaries $S_{1}, S_{2}, S_{3}$ and $S_{4}$. For each of the 1,000 prior predictive samples, we separately calculate the sample mean and standard deviation over the 50 values for each summary. Then we investigate how the mean and standard deviation of each of the summary is influenced by the g-and-k parameters. Figures \ref{fig_S1_gnk}-\ref{fig_S4_gnk} display scatter plots showing the estimated mean and standard deviation of $S_{1}, S_{2}, S_{3}$ and $S_{4}$ in relation to the g-and-k parameters.

As noted in \cite{drovandi2024improving}, Figure \ref{fig_S1_gnk} shows that the parameter $a$ has a linear impact on the mean of $S_{1}$, however, it is also evident that the parameter $b$ significantly affects the standard deviation of  $S_{1}$. Additionally, Figure \ref{fig_S2_gnk} indicates that both parameters $b$ and $k$ influence $S_{2}$. Figure \ref{fig_S4_gnk} reveals that only the parameter $k$ affects $S_{4}$, but as shown in Figure \ref{fig_S2_gnk}, $k$ also depends on the summary $S_{2}$.  According to Figure \ref{fig_S3_gnk}, both the mean and standard deviation of $S_{3}$ depends on $g$, while the standard deviation of $S_{3}$ also depends on the parameter $k$. Given these dependencies, it is noteworthy that $g$ is the only parameter that relies solely on its robust summary, $S_{3}$. Therefore, we choose $S_{3}$ as the step one summary in our proposed R-ABC approach.

\subsection*{C.2 Calculation of the pseudo-true value}
To obtain the pseudo-true value of $\theta$ for the given simulation experiment, we need to calculate $b_{0}$ and $b(\theta)$, the probability limits of observed summaries ($\eta(\y)$) and simulated summaries ($\eta(\z)$) respectively. To determine $b_{0}$, we must calculate the population quartiles and octiles of the Gaussian mixture distribution. Given that the quantile function of the Gaussian mixture has no closed form, the value of $b_{0}$ can be obtained by numerically inverting the corresponding CDF of the Gaussian mixture. Given the value of $b_{0}$ and the fact that the quantiles of the g-and-k distribution have an analytical form in terms of the standard normal quantile function, we can numerically solve for the value of $\theta=(a, b, g, k)^{\prime}$ that minimizes $\|b(\theta)-b_0\|$. Thus, the pseudo-true value under this Monte Carlo design when using $\eta(\y)$ is given by 
$$
\theta^{*} = (a^{*}, b^{*}, g^{*}, k^{*})^{T} =(2.3663, 4.1757, 1.7850, 0.1001)^{\prime}.
$$

\subsection*{C.3 Adjustment components}
\begin{figure}[h!]
	\centering 
	\includegraphics[width=0.65\linewidth]{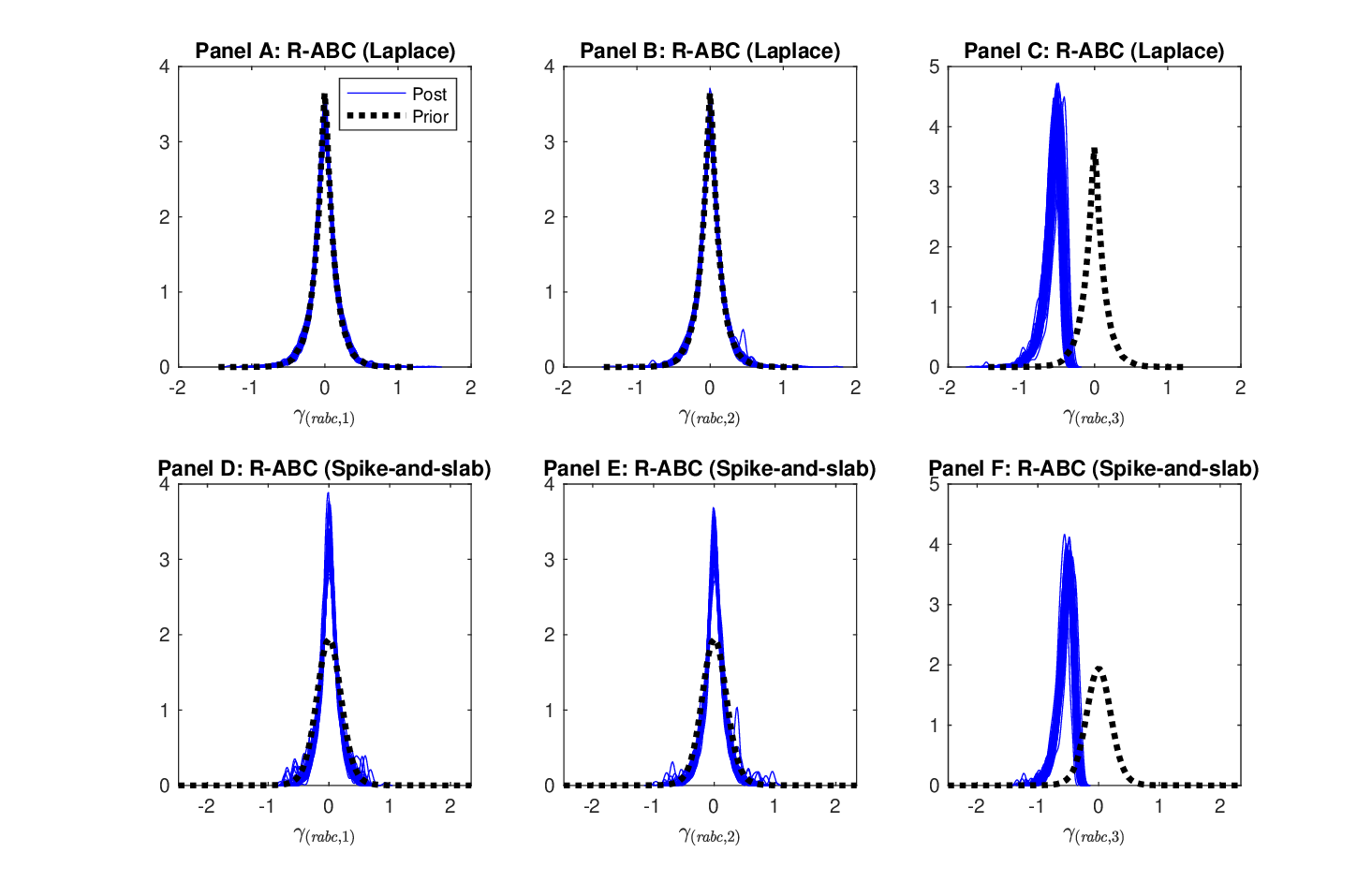} 
	\caption{Panels A, B and C respectively provide the posteriors for adjustment components of the R-ABC approach when using Laplace prior while Panels D, E and F provide the same information about the R-ABC approach when using spike-and-slab prior. As indicated in the key, the marginal posterior of $\Gamma_{\text{R-ABC}}$ component (Post) is represented in a blue solid line. The black dotted line in the top row represents the La(0,0.125) prior while the one in the bottom row corresponds to the spike-and-slab prior.} 
	\label{fig_gnk_gamRABC_random} 
\end{figure}

\begin{figure}[h!]
	\centering 
	\includegraphics[width=0.65\linewidth]{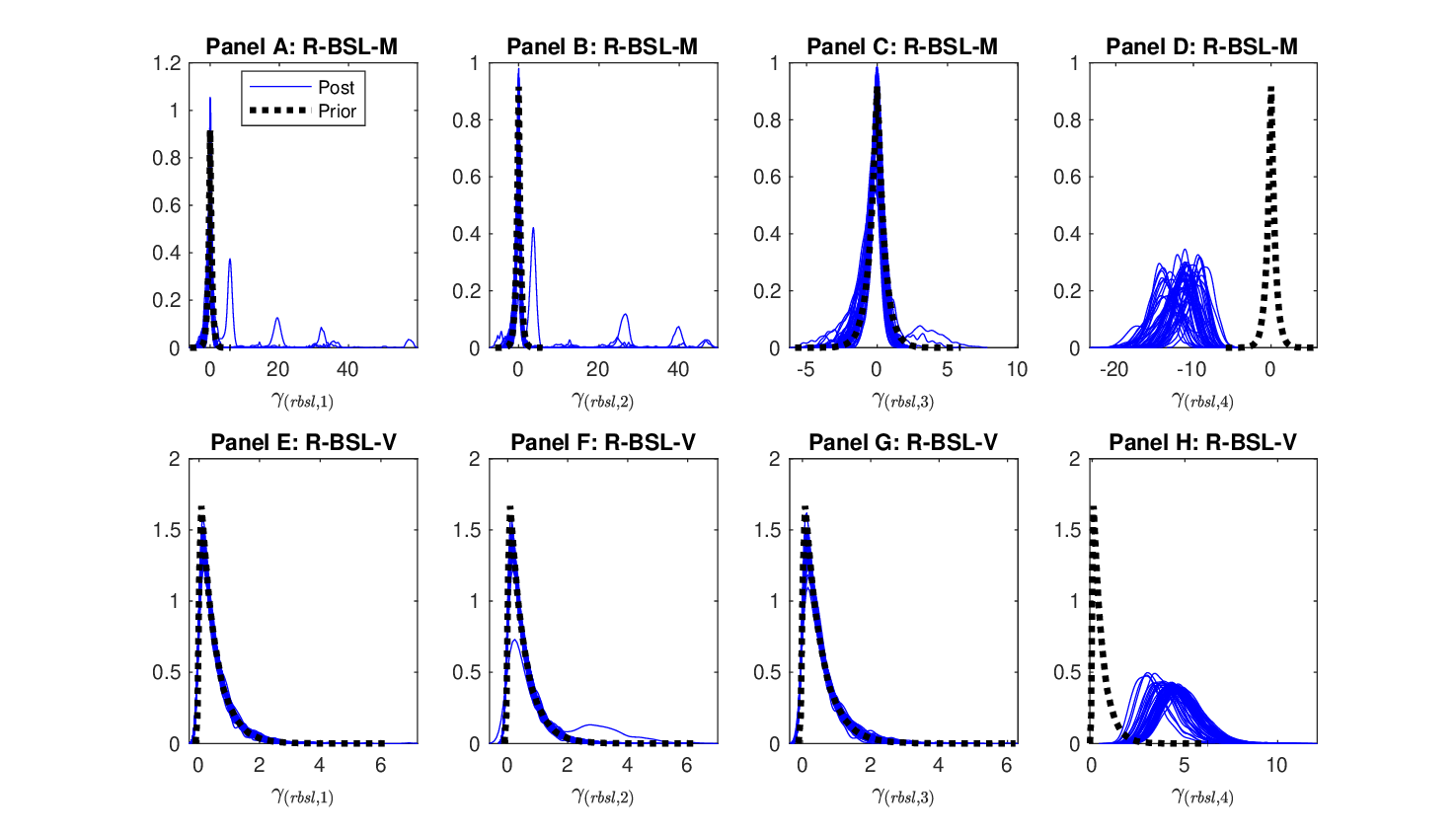} 
	\caption{Panels A, B, C and D respectively provide the posteriors for adjustment components of the R-BSL-M approach while Panels E, F, G and H provide the same information about R-BSL-V approach. As indicated in the key, the marginal posterior of $\Gamma_{\text{R-BSL}}$ component (Post) is represented in a blue solid line. The black dotted line in the first row represents the La(0,0.5) prior while the one in the bottom row corresponds to the exponential prior with $\lambda=0.5.$} 
	\label{fig_gnk_gamRBSL_random} 
\end{figure}

Figure \ref{fig_gnk_gamRABC_random} presents the posteriors for the adjustment components of the R-ABC approach across 50 replications. Panels A-C in the top row display the adjustment components when the Laplace prior is used for $\Gamma$, while Panels D-F in the bottom row correspond to the adjustment components under the spike-and-slab prior for $\Gamma$. In Panels A, B, D, and E, across all replications, we observe that the posterior densities of $\gamma_{(rabc,1)}$ and $\gamma_{(rabc,2)}$ components closely align with their respective priors, showing no distinguishable deviation. In contrast, Panels C and F show that the posterior densities of $\gamma_{(rabc,3)}$ exhibit a significant deviation from their corresponding prior across all replications, indicating that the model cannot reliably match $S_{4}$, robust summary related to kurtosis.\footnote{This is also confirmed by the results of the two-sample randomization test of location over the replicated datasets. Under R-ABC (Laplace), the proportion of times the test rejects $H_{0}$ for $\gamma_{(rabc,1)}, \gamma_{(rabc,2)}$ and $\gamma_{(rabc,3)}$ respectively are 0.16, 0.14 and 1. The same information under spike-and-slab prior are 0.3, 0.26 and 1.}

We now examine the posterior distributions for the adjustment components across R-BSL-M and R-BSL-V, shown in Figure \ref{fig_gnk_gamRBSL_random}, across the 50 replications. The top row, Panels A-D, represent the posterior densities of $\Gamma_{\text{R-BSL}}$ for R-BSL-M, while the bottom row, Panels E-H, correspond to the same components for R-BSL-V. In Panels A and B, $\gamma$ components of R-BSL-M show some deviations from their respective priors, though in most replications, $\gamma_{(rbsl,1)}$, $\gamma_{(rbsl,2)}$ and $\gamma_{(rbsl,3)}$ do not significantly depart from the prior. In contrast, for R-BSL-V, $\gamma_{(rbsl,1)}$, $\gamma_{(rbsl,2)}$ and $\gamma_{(rbsl,3)}$ are indistinguishable from the prior across replications as observed in Panels E, F, and G. However, in both R-BSL versions, Panels D and H reveal that $\gamma_{(rbsl,4)}$, the adjustment component for $S_{4}$, deviates significantly from the prior. This finding, consistent with the results from R-ABC, suggests that the model struggles to accurately match $S_{4}$, the robust summary related to kurtosis.

\subsection*{C.4 Additional results for the g-and-k example}

\begin{figure}[h!]
	\centering 
	\includegraphics[width=0.55\linewidth]{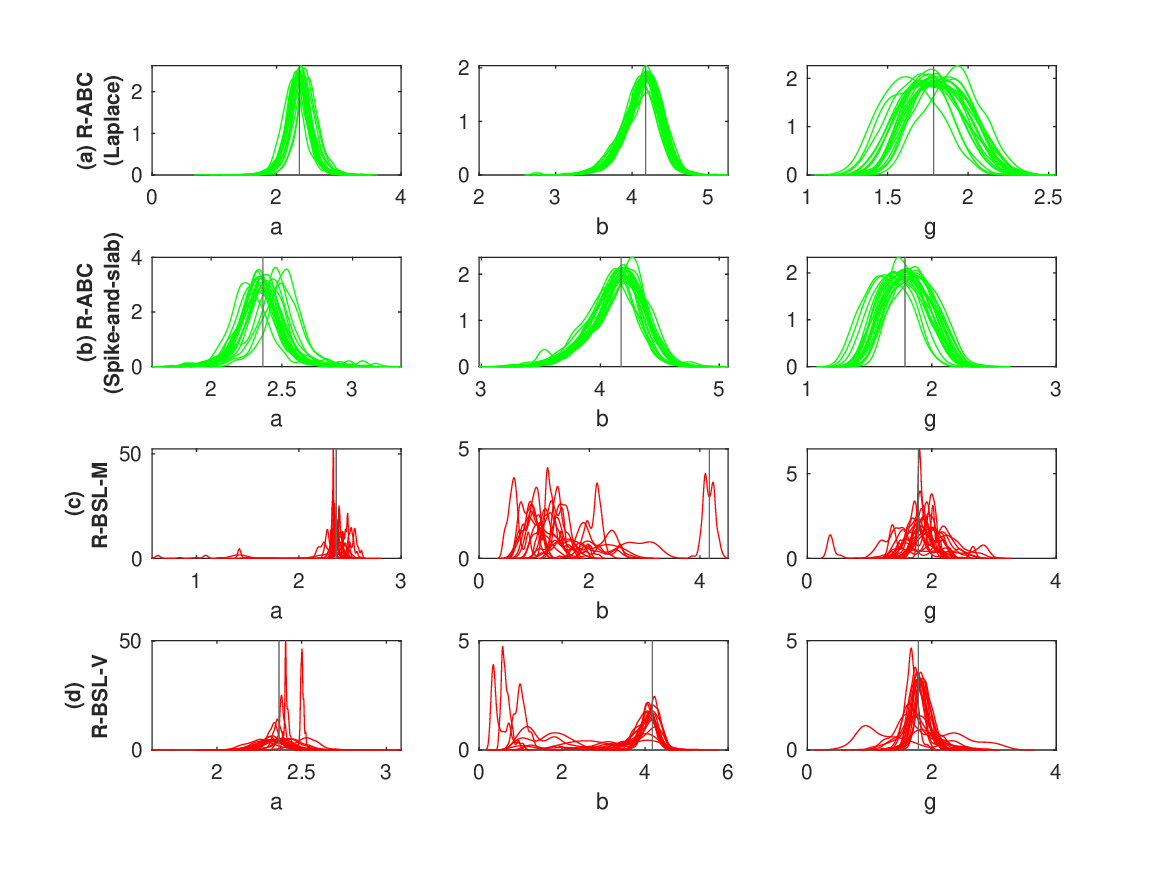} 
	\caption{Rows (a) and (b) respectively provide the posteriors for $\theta=(a, b, g)^{\prime}$ using the R-ABC approach when using Laplace and spike-and-slab priors for the adjustment components. Rows (c) and (d) respectively provide the same information for R-BSL-M and R-BSL-V. The vertical line represents the pseudo-true values. The hyperparameter for the priors associated with the adjustment components of R-ABC is $\lambda=0.125.$ Here, the MCMC sampler is started with random values.} 
	\label{gnk_theta_rabc_rbsl_abg} 
\end{figure}

\begin{figure}[h!]
	\centering 
	\includegraphics[width=0.85\linewidth]{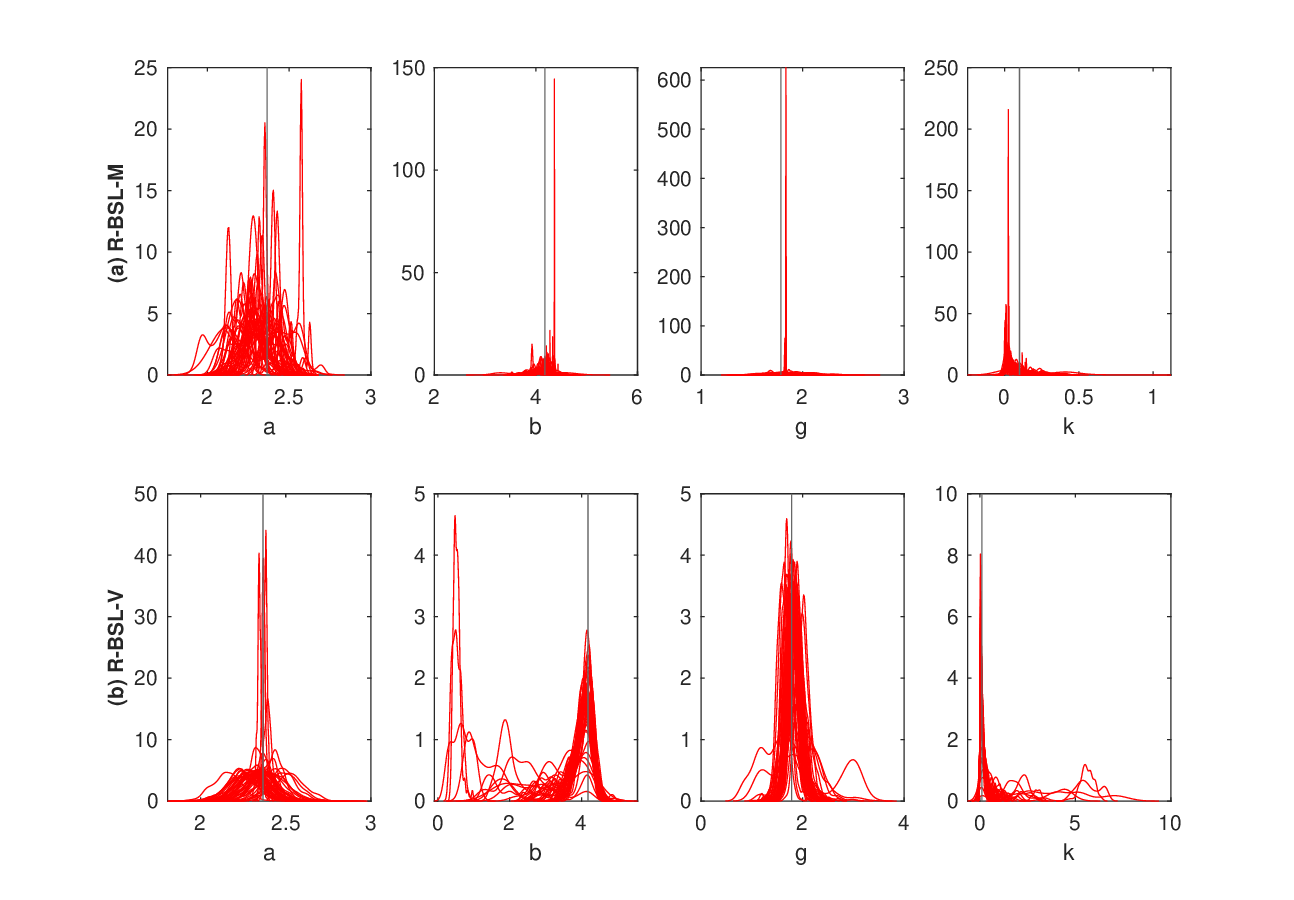} 
	\caption{Rows (a) and (b) respectively provide the posteriors for $\theta=(a, b, g, k)^{\prime}$ using R-BSL-M and R-BSL-V. The vertical lines represent the pseudo-true values. Here, the MCMC sampler is started with pseudo-true values.} 
	\label{gnk_theta_rbsl_pstv} 
\end{figure}

\begin{figure}[h!]
	\centering 
	\includegraphics[width=0.85\linewidth]{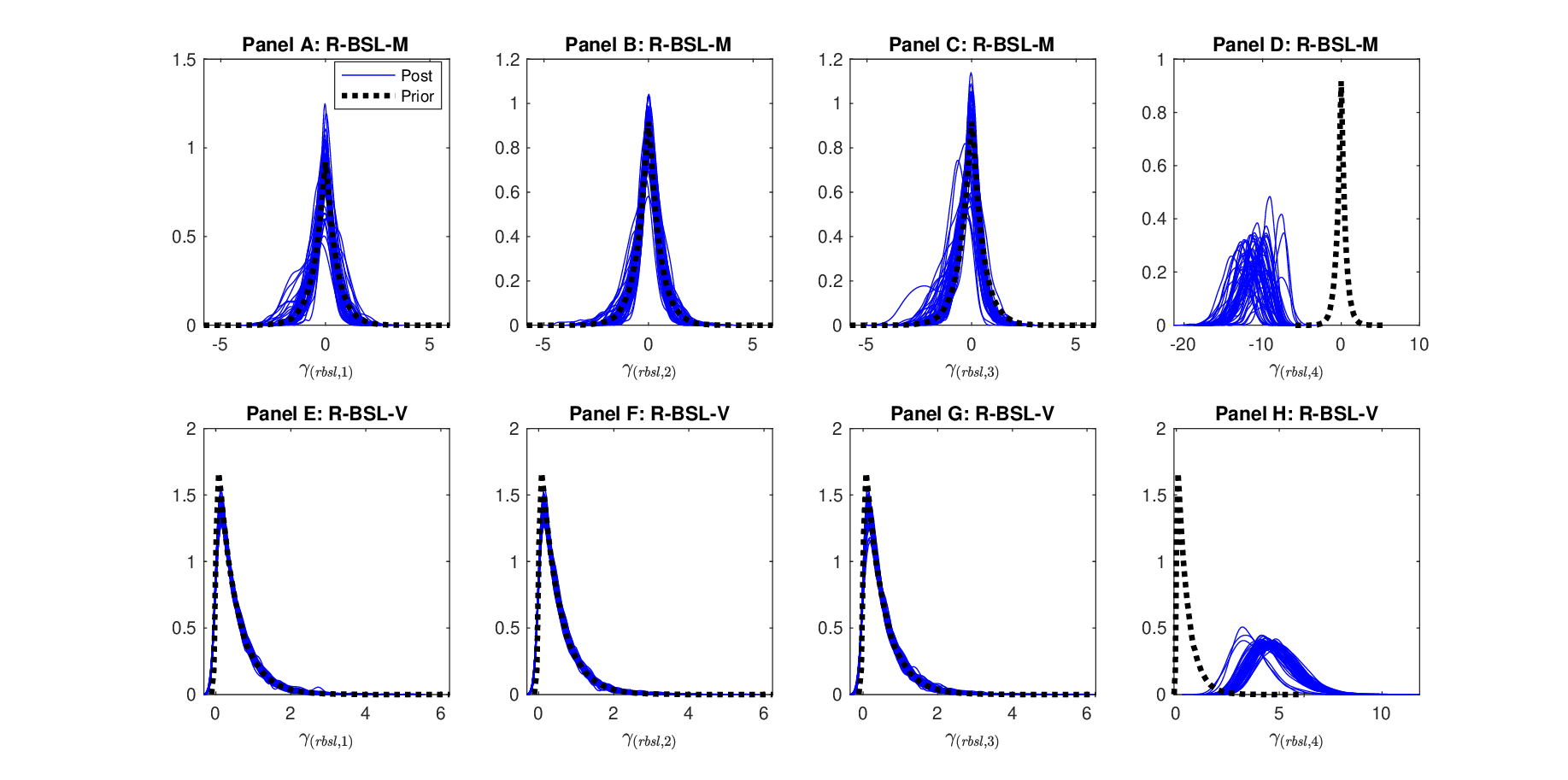} 
	\caption{Panels A, B, C and D respectively provide the posteriors for adjustment components of the R-BSL-M approach while Panels E, F, G and H provide the same information about R-BSL-V approach. As indicated in the key, the marginal posterior of $\Gamma_{\text{R-BSL}}$ component (Post) is represented in a blue solid line. The black dotted line in the first row represents the La(0,0.5) prior while the one in the bottom row corresponds to the exponential prior with $\lambda=0.5.$ Here, MCMC sampler is started with pseudo-true values.} 
	\label{gnk_gam_rbsl_pstv} 
\end{figure}

\begin{figure}[h!]
	\centering 
	\includegraphics[width=0.5\linewidth]{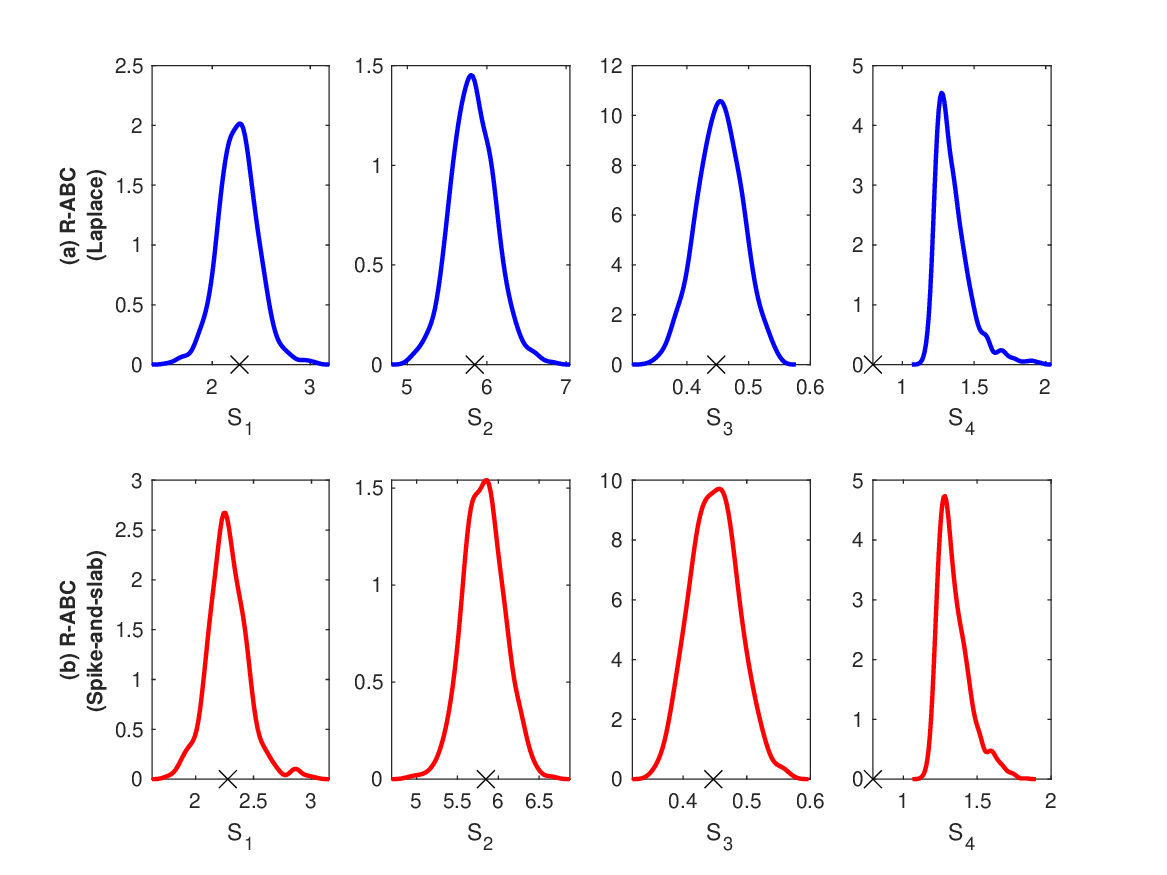} 
	\caption{Rows (a) and (b) respectively presents the posterior predictive densities obtained under R-ABC (Laplace) and R-ABC (Spike-and-slab) for the summary statistics in the g-and-k example. The value of the observed summary is marked with an $\times$ in the figure. The hyperparameter for the priors associated with the adjustment components is $\lambda=0.125.$} 
	\label{gnk_post_pred_sum} 
\end{figure}

\newpage
\subsection*{C.5 Animated figures}

\begin{figure}[h!]
	\centering
	\includemedia[
	width=0.7\linewidth, % Adjust size
	height=0.35\linewidth, % Adjust size
	activate=onclick,
	addresource=Anim_gnk_theta_tau0_125_randomSt_val.mp4,
	flashvars={
		source=Anim_gnk_theta_tau0_125_randomSt_val.mp4   % Video source
		&autoPlay=true              % Automatically play when clicked
		&loop=true                  % Loop the video
	}
	]{}{VPlayer.swf} % Use the default video player
	\caption{Animation plot corresponding to Figures \ref{gnk_theta_rabc_rbsl} and \ref{gnk_theta_rabc_rbsl_abg} across the replications. Rows (a) and (b) respectively provide the posteriors for $\theta=(a, b, g, k)^{\prime}$ using the R-ABC approach when using Laplace and spike-and-slab priors for the adjustment components. Rows (c) and (d) respectively provide the same information for R-BSL-M and R-BSL-V. The vertical lines represent the pseudo-true values. The hyperparameter for the priors associated with the adjustment components is $\lambda=0.125.$}%
	\label{anim_gnk_1} 
\end{figure}

\begin{figure}[h!]
	\centering
	\includemedia[
	width=0.7\linewidth, % Adjust size
	height=0.35\linewidth, % Adjust size
	activate=onclick,
	addresource=Anim_gnk_theta_rbsl_pseudoTval.mp4,
	flashvars={
		source=Anim_gnk_theta_rbsl_pseudoTval.mp4   % Video source
		&autoPlay=true              % Automatically play when clicked
		&loop=true                  % Loop the video
	}
	]{}{VPlayer.swf} % Use the default video player
	\caption{Animation plot corresponding to Figure \ref{gnk_theta_rbsl_pstv} across the replications. Rows (a) and (b) respectively provide the provide the posteriors for $\theta=(a, b, g, k)^{\prime}$ using R-BSL-M and R-BSL-V. The vertical lines represent the pseudo-true values. Here, MCMC sampler is started with pseudo-true values.}%
	\label{anim_gnk_2} 
\end{figure}

\newpage
\subsection*{C.6 Posterior bias, average posterior standard deviation, and Monte Carlo coverage for each estimation method across replications for the parameters $a,b$ and $g$.}

In Table \ref{tab_gnk_abg}, we record the posterior bias, average posterior standard deviation, and Monte Carlo coverage for each estimation method across replications for the parameters $a,b$ and $g$ of the g-and-k model.

{\renewcommand{\arraystretch}{0.8}
	\begin{table}[h!]
		\caption{\scriptsize Monte Carlo coverage (Cov), bias of the posterior mean (Bias), and average posterior standard deviation (Std) for $\theta=(a, b, g)'$ in the g-and-k example. Cov is the percentage of times that the marginal 95\% credible set contains $\theta^*$. Std is the average posterior standard deviation across the Monte Carlo trials. The rows refer to either the ABC or BSL approach. The last four rows refer to the R-BSL approach, where R-BSL-M, R-BSL-V refer to the case where random starting values are used for the MCMC sampler, while R-BSL-M (PSTV), R-BSL-V (PSTV) refer to the case where pseudo-true values are used as starting values of MCMC sampler.}
		\centering%
		\small  
		\begin{tabular}{lrrrrrrrrrr}
			\hline\hline
			&   \multicolumn{3}{c}{$a$}     & &  &   \multicolumn{1}{c}{$b$}  &  &\\
			\cline{2-4} \cline{6-8}
			& \multicolumn{1}{c}{Cov} & \multicolumn{1}{c}{Bias} & \multicolumn{1}{c}{Std} &       & \multicolumn{1}{c}{Cov} & \multicolumn{1}{c}{Bias} & \multicolumn{1}{c}{Std} \\
			{\textbf{Method}} \\
			ABC-SMC	&	98\%	&	-0.0798	&	0.1100	&&	100\%	&	-0.0137	&	0.1160	\\
			ABC-SMC-Reg	&	0\%	&	-0.4820	&	0.0712	&&	0\%	&	-0.5858	&	0.0939	\\
			R-ABC (Laplace)	&	100\%	&	-0.0165	&	0.1948	&&	100\%	&	-0.0562	&	0.2384	\\
			R-ABC (Spike-and-slab) 	&	100\%	&	-0.0165	&	0.1498	&&	100\%	&	-0.0540	&	0.2197	\\
			R-BSL-M	&	76\%	&	-0.0390	&	0.0753	&&	18\%	&	-2.1710	&	0.2579	\\
			R-BSL-V	&	80\%	&	-0.0195	&	0.0617	&&	62\%	&	-1.4754	&	0.4138	\\
			R-BSL-M (PSTV)	&	86\%	&	-0.0581	&	0.0781	&&	94\%	&	-0.0530	&	0.1421	\\
			R-BSL-V (PSTV)	&	98\%	&	-0.0287	&	0.0808	&&	92\%	&	-0.5683	&	0.3741	\\
			&  &  &  &  &  &  & \\
			&   \multicolumn{3}{c}{$g$}\\
			\cline{2-4} 
			& \multicolumn{1}{c}{Cov} & \multicolumn{1}{c}{Bias} & \multicolumn{1}{c}{Std} \\
			ABC-SMC	&	100\%	&	0.3834	&	0.6655	\\
			ABC-SMC-Reg	&	0\%	&	1.5031	&	0.2242\\
			R-ABC (Laplace)	&	100\%	&	0.0238	&	0.1764\\
			R-ABC (Spike-and-slab) 	&	100\%	&	0.0209	&	0.1752	\\
			R-BSL-M	&	98\%	&	0.0886	&	0.2108	\\
			R-BSL-V	&	100\%	&	0.0575	&	0.2238\\
			R-BSL-M (PSTV)	&	88\%	&	0.0714	&	0.1067	\\
			R-BSL-V (PSTV)	&	100\%	&	0.0401	&	0.1556	\\
			\hline\hline
		\end{tabular}%
		\label{tab_gnk_abg}%
	\end{table}	
}

In terms of the bias of the posterior mean, ABC-SMC-Reg exhibits a high bias for all parameters, particularly for parameters $a, b$ and $g$, indicating a significant under or over-estimation of the pseudo-true true values. ABC-SMC shows relatively lower bias for most parameters, except for $g$. When employing random starting values for MCMC in R-BSL, both versions have a large bias for parameters $b$ and $g$. However, when pseudo-true values are used as starting points in the MCMC sampler, both R-BSL approaches show considerably smaller bias for all parameters. In comparison, the R-ABC approach — especially with the spike-and-slab prior for the adjustment components — achieves the lowest bias for all parameters except the parameter $b$.

When comparing posterior variability across the different methods, we find that ABC-SMC exhibits the highest posterior uncertainty for parameter $g$, while ABC-SMC-Reg display the low posterior variability for most of the parameters. R-ABC yields relatively small posterior uncertainties across all parameters, although these are relatively larger than those of ABC-SMC and ABC-SMC-Reg, except for parameter $g$. Additionally, with the exception of parameter $a$, R-ABC also shows reduced posterior uncertainties compared to R-BSL-M and R-BSL-V.

R-BSL-M and R-BSL-V, which initialized the MCMC sampler with random starting values, exhibit very poor coverage, particularly for the parameter $b$. Notably, the credible sets produced by ABC-SMC-Reg do not contain the pseudo-true values for any parameter in any of the Monte Carlo replications. While ABC-SMC-Reg achieves the lowest posterior variability for most of the parameters, it leads to a misleading sense of accuracy, resulting in a coverage of 0\%. In contrast, we observe that for all parameters, R-ABC always contains the pseudo-true value (and thus has 100\% Monte Carlo coverage).

\clearpage
\section{Empirical illustration}
\label{appendix_D}

\begin{figure}[h!]
	\centering 
	\includegraphics[width=0.65\linewidth]{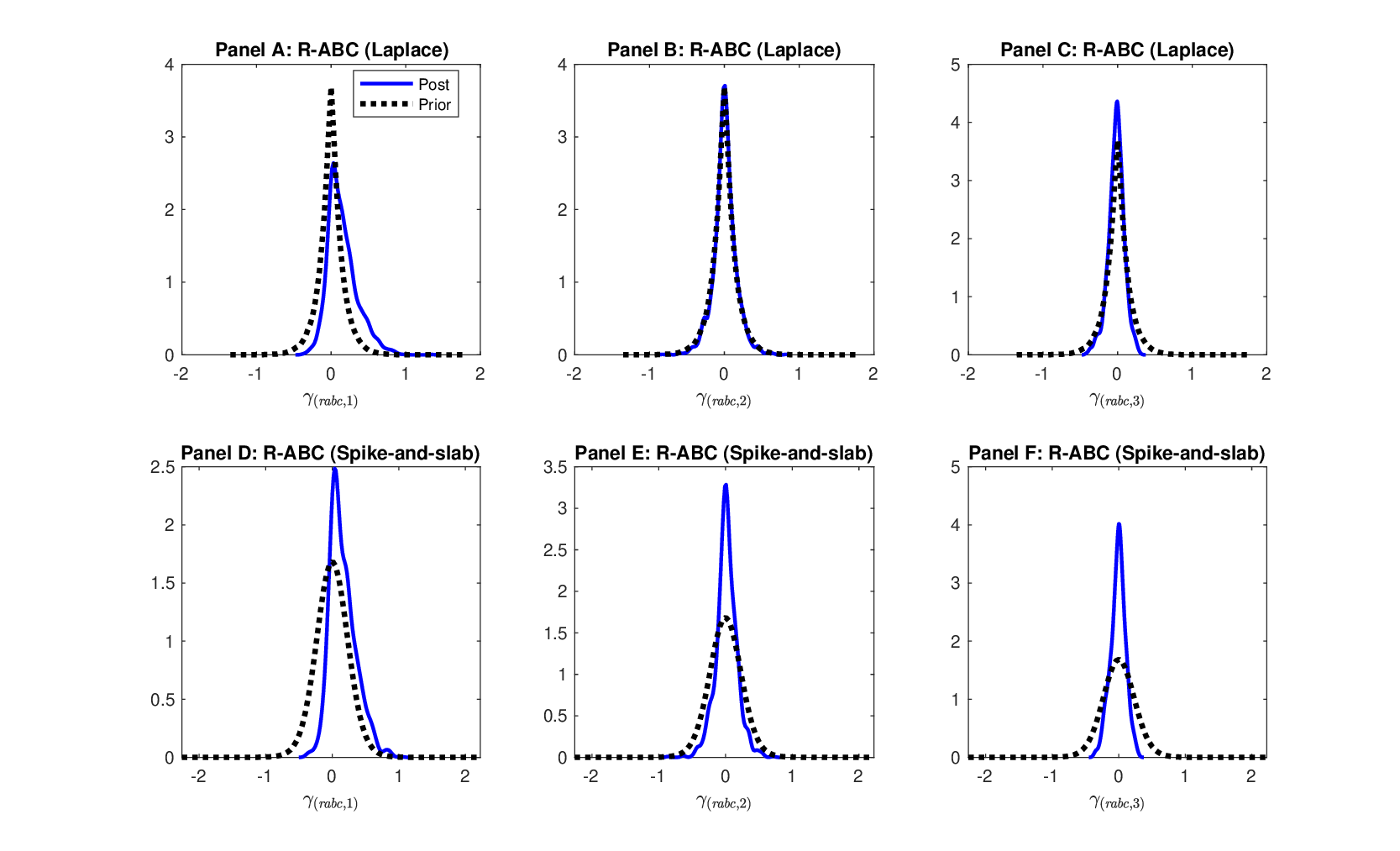} 
	\caption{Panels A, B and C respectively provide the posteriors for adjustment components of the R-ABC approach when using Laplace prior while the Panels D, E and F provide the same information about R-ABC approach when using spike-and-slab prior. As indicated in the key, the marginal posterior of $\Gamma_{\text{R-ABC}}$ component (Post) is represented in a blue solid line. The black dotted line in the top row represents La(0,0.125) prior while the one in the bottom row corresponds to the spike-and-slab prior.}
	\label{emp_gam_rabc} 
\end{figure}

\begin{figure}[h!]
	\centering 
	\includegraphics[width=0.65\linewidth]{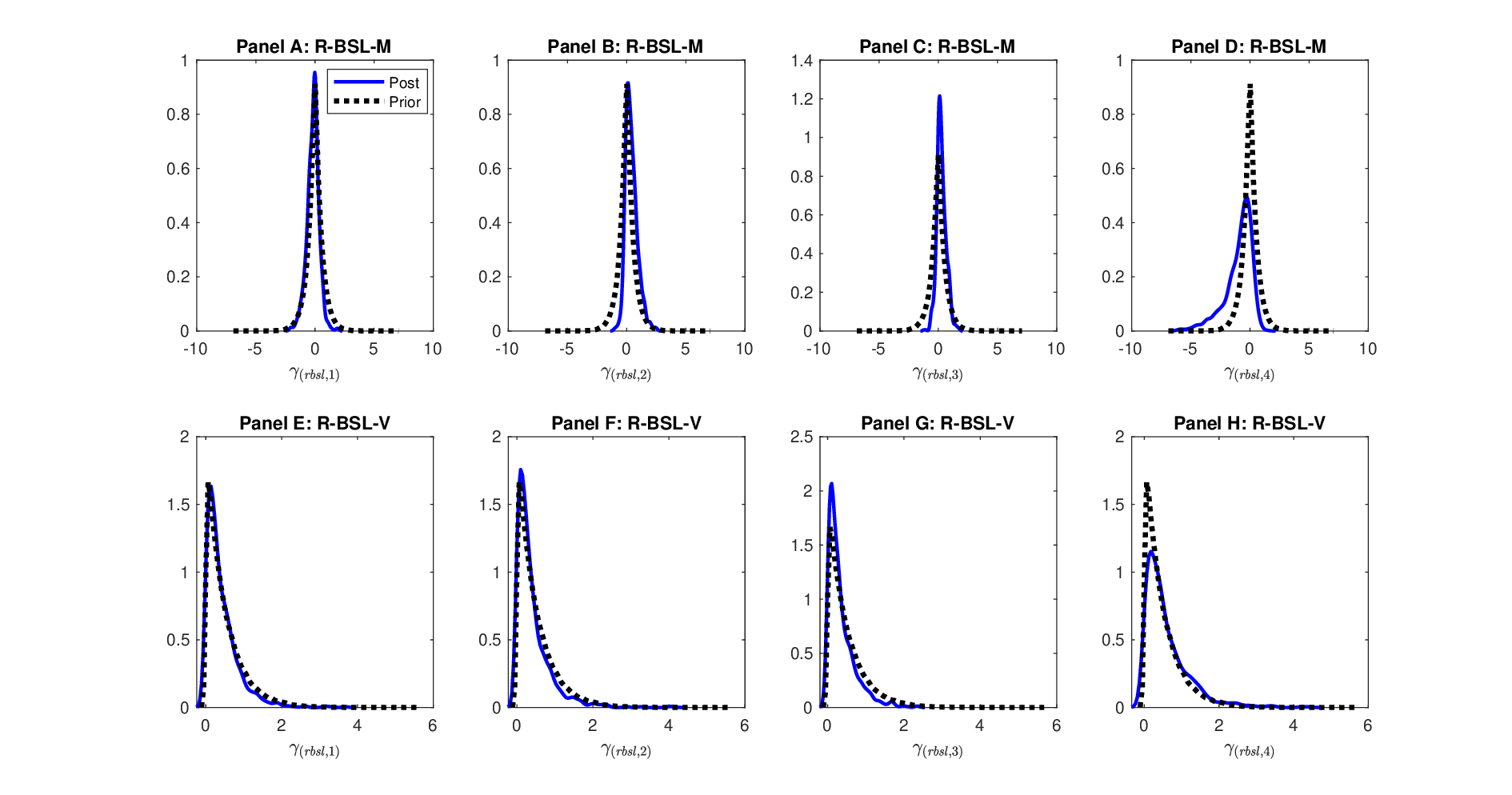} 
	\caption{Panels A, B, C and D respectively provide the posteriors for adjustment components of R-BSL-M approach while the Panels D, E and F provide the same information of R-BSL-V approach. As indicated in the key, the marginal posterior of $\Gamma_{\text{R-BSL}}$ component (Post) is represented in blue solid line. The black dotted line in the first row represents the La(0,0.5) prior while the black dotted line in the second row represents the exponential prior with $\lambda=0.5.$} 
	\label{emp_gam_rbsl} 
\end{figure}

\begin{figure}[h!]
	\centering 
	\includegraphics[width=0.6\linewidth]{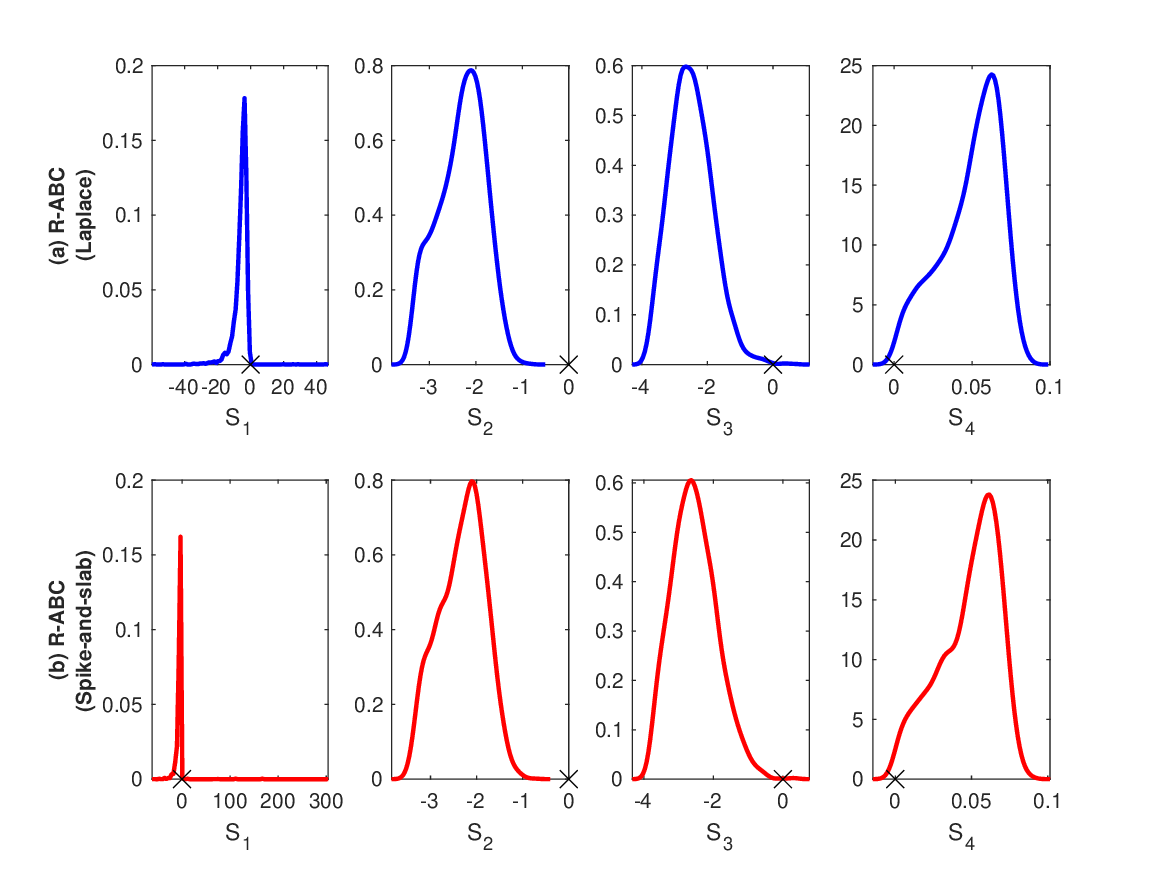} 
	\caption{Rows (a) and (b) respectively presents the posterior predictive densities obtained under R-ABC (Laplace) and R-ABC (Spike-and-slab) for the summary statistics in the empirical example. The value of the observed summary is marked with an $\times$ in the figure.} 
	\label{emp_post_pred_sum} 
\end{figure}

\begin{figure}[h!]
	\centering 
	\includegraphics[width=0.8\linewidth]{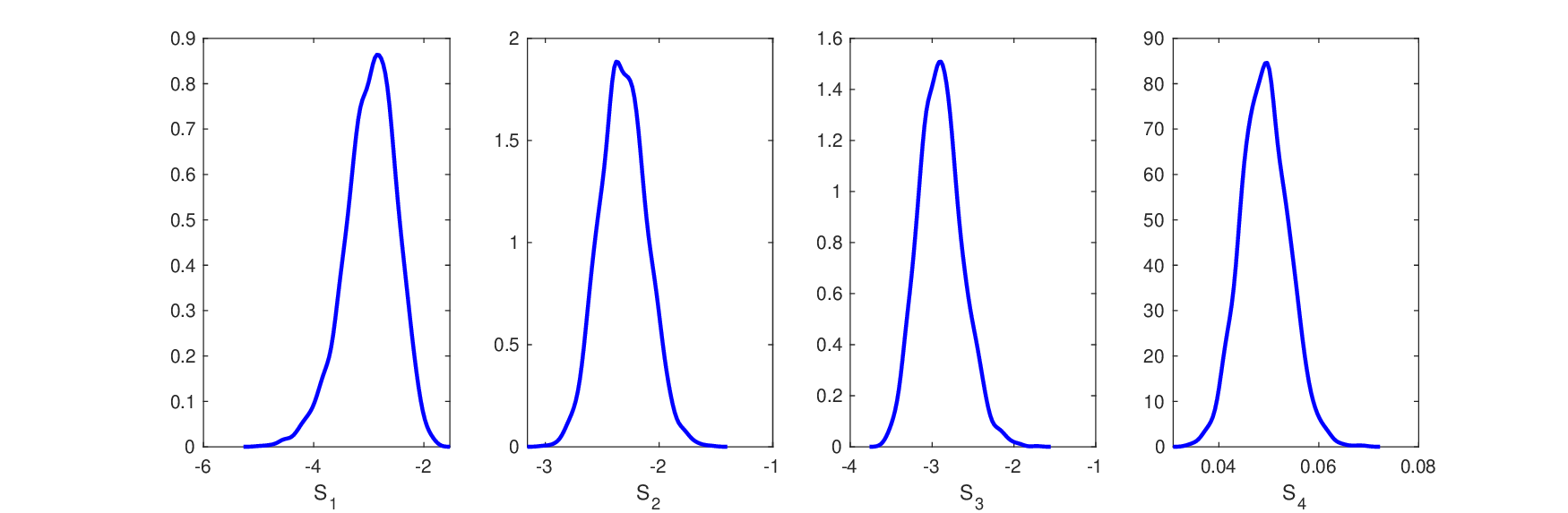} 
	\caption{Estimated density plots of the marginal summaries in empirical example simulated from the $\alpha-$stable SV model based on parameter values: $\theta_{2}=0.95; \theta_{3}=0.5; \theta_{4}=1.2.$ } 
	\label{emp_est_sum_goodparam} 
\end{figure}

Figure \ref{emp_est_sum_goodparam} represents estimated density plots of the marginal summaries in empirical example, which were simulated from the $\alpha-$stable SV model based on parameter values $\theta_{2}=0.95, \theta_{3}=0.5$ and $\theta_{4}=1.2.$ To assess the normality of these summaries, we use Shapiro–Wilk test (\citealp{shapiro1965analysis}), a hypothesis test that evaluates whether the data follows a normal distribution. At 5\% level of significance, the test rejects $H_{0}$ for $S_{1}, S_{3}$ and $S_{4}$ (p-values: 0, 0, 0.0035) but does not reject $H_{0}$ for $S_{2}$ (p-value=0.1592).\footnote{This result is also further confirmed by the Anderson-Darling normality test (\citealp{anderson1954test}), which also tests the null hypothesis that the data follows a normal distribution. At 5\% level of significance, the test rejects $H_{0}$ for $S_{1}, S_{3}$ and $S_{4}$ (p-values: 0.00005, 0.00005, 0.0324) but does not reject $H_{0}$ for $S_{2}$ (p-value=0.1430).}

\newpage
\section{Additional simulation example}
\label{appendix_E}

Consider the following example where the researcher is interested in conducting inference on $\theta$ in the model
\begin{equation*}
	y_{i} = \theta + \nu_{i};\quad  \nu_{i}\stackrel{iid}{\sim} N(0, \sigma^{2}),
\end{equation*}
where it is assumed that $\sigma=1$, and pseudo-data in ABC is generated according to 
$$z_i = \theta+\nu_{i}; \quad  \nu_{i}\stackrel{iid}{\sim}\mathcal{N}(0,1).$$
The summary statistics chosen for inference in this example are the sample mean and variance. That is, $\eta(\y)=(\eta_{1}(\y), \eta_{2}(\y))^{'}$ where $\eta_{1}(\y)=\frac{1}{n}\sum_{i=1}^{n}{y}_{i}$ and $\eta_{2}(\y)=\frac{1}{n-1}\sum_{i=1}^{n}({y}_{i}-\eta_{1}(\mathbf{y}))^{2}.$

\begin{figure}[h!]
	\centering 
	\includegraphics[width=0.8\linewidth]{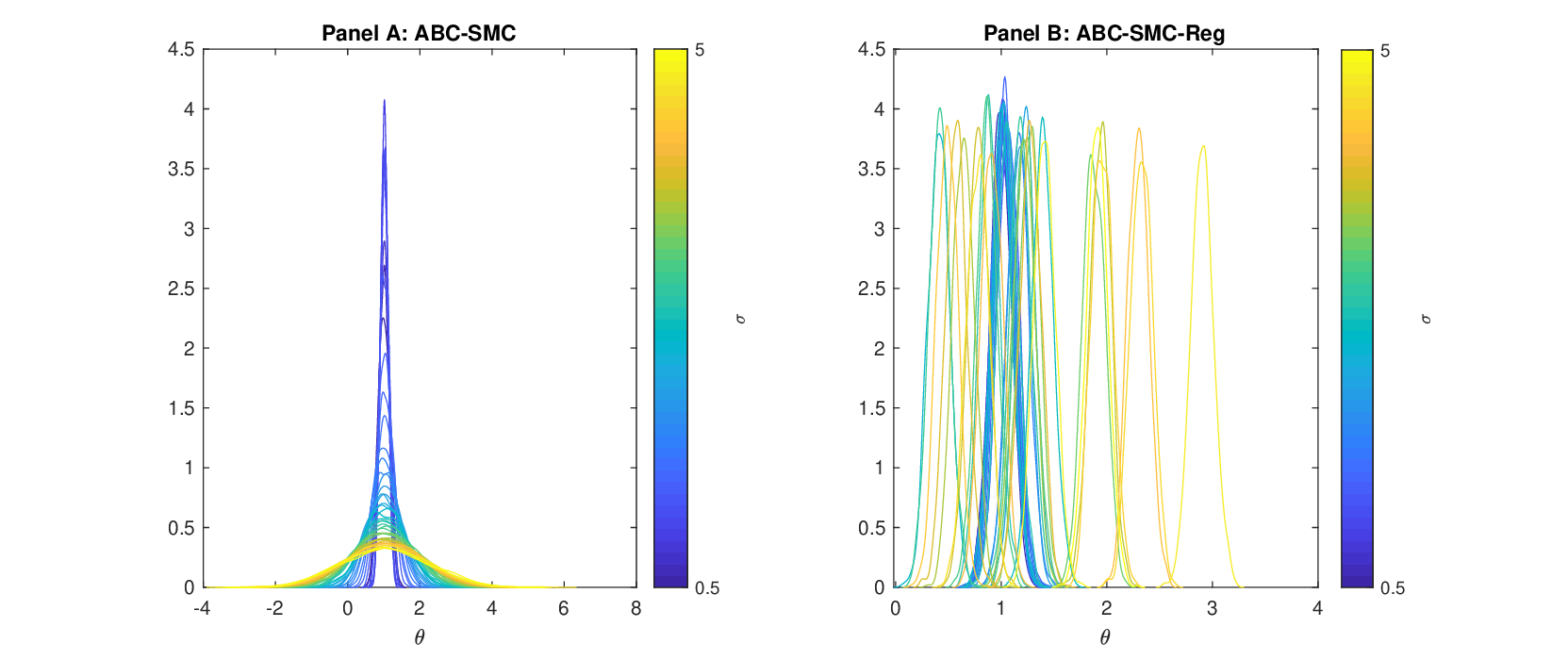} 
	\caption{Comparison of ABC-SMC and ABC-SMC-Reg posteriors across varying levels of model misspecification under the toy example. The color scheme indicates the level of model misspecification, with darker colors encoding less misspecification.} 
	\label{norm_theta_stdabc} 
\end{figure}

In order to demonstrate the behaviour of ABC-SMC and ABC-SMC-Reg approaches under varying levels of model misspecification, we fix $\theta=1$ in the true model and simulate observed data $\mathbf{y}$ according to different values of $\sigma$. The sample size across the experiments is taken to be $n=100$. We consider a sequence of simulated data sets for $\y$ such that each corresponds to a different value of $\sigma$, with $\sigma$ taking values from ${\sigma}^{}=0.5$ to ${\sigma}^{}=5$ with evenly spaced increments of $0.1$. Across all the data sets, we fix the random numbers used to generate the observed data and only change the value of $\sigma$ to isolate the impact of model misspecification. The prior on $\theta$ is given by $\mathcal{N}(0,25)$. Here, for each ABC approach, we stop the SMC-ABC algorithm when the acceptance rate in the MCMC step drops below 1\%, as flagged in the main text. The proposal distribution in the MCMC ABC kernel is a Gaussian random walk, with a covariance tuned based on the population of SMC particles.

Figure \ref{norm_theta_stdabc} displays the resulting ABC-SMC and ABC-SMC-Reg posteriors for various values of ${\sigma}$. The color of the density represents the level of model misspecification, with $\sigma=1$ encoded as blue, and the color becoming lighter as misspecification increases, i.e., as $\sigma$ increases. The results show that model misspecification causes significant differences between the two ABC approaches.  Specifically, while the posterior mean of ABC-SMC stays centered around the pseudo-true value $\theta=1$, the posterior mean of ABC-SMC-Reg becomes unstable by shifting towards larger or smaller values of $\theta$ as the degree of misspecification increases.

Even though ABC-SMC maintains robust performance under varying levels of model misspecification in this example, it is evident that the regression adjustment technique can lead to unreliable inference under model misspecification.

Next, we analyze the performance of R-ABC in this example. We use the same simulation design discussed above and precisely the same observed data and apply R-ABC to this data. Since the sample mean should be able to recover the population mean $\theta$, it is reasonable to assume
$$\psi(\y)= \eta_{1}(\y)=\frac{1}{n}\sum_{i=1}^{n}{y}_{i} \quad \text{and} \quad    \varphi(\y) = \eta_{2}(\y)=\frac{1}{n-1}\sum_{i=1}^{n}({y}_{i}-\eta_{1}(\mathbf{y}))^{2}$$
as the partition of the summary statistics in this example in conducting the R-ABC approach.

To implement R-ABC, we first conduct inference on $\theta$ using only the sample mean as the summary statistic in the ABC accept/reject algorithm (see Algorithm \ref{R-ABC-SMC-step1} in Appendix \ref{appendix_A}). We use $N_{1}=100000$ draws from prior and retain the draws that lead to the smallest 5\% of the overall simulated distances. The threshold $\epsilon_{1}$ is then set to the maximum of these retained distances. We then implement the second ABC step using the summary $\varphi(\z)$, where a single adjustment component only is necessary for the sample variance summary statistic.

As discussed in Section \ref{R-ABC_alg}, when implementing the second step of R-ABC, we use the modified ABC-SMC replenishment algorithm as detailed in Algorithms \ref{R-ABC-SMC-lap} and \ref{R-ABC-SMC-spk} (see Appendix \ref{appendix_A}) respectively with the Laplace and spike-and-slab priors for the adjustment components, with the prior hyperparameters set to default values. Both algorithms are initialized with the ABC draws retained from step one, and are run until the MCMC acceptance rate drops below 1\%. Additionally, we ensure that each proposal satisfies the joint selection condition in (\ref{joint_cond}) with a tolerance of $\epsilon_{2}.$

\begin{figure}[h!]
	\centering 
	\includegraphics[width=0.7\linewidth]{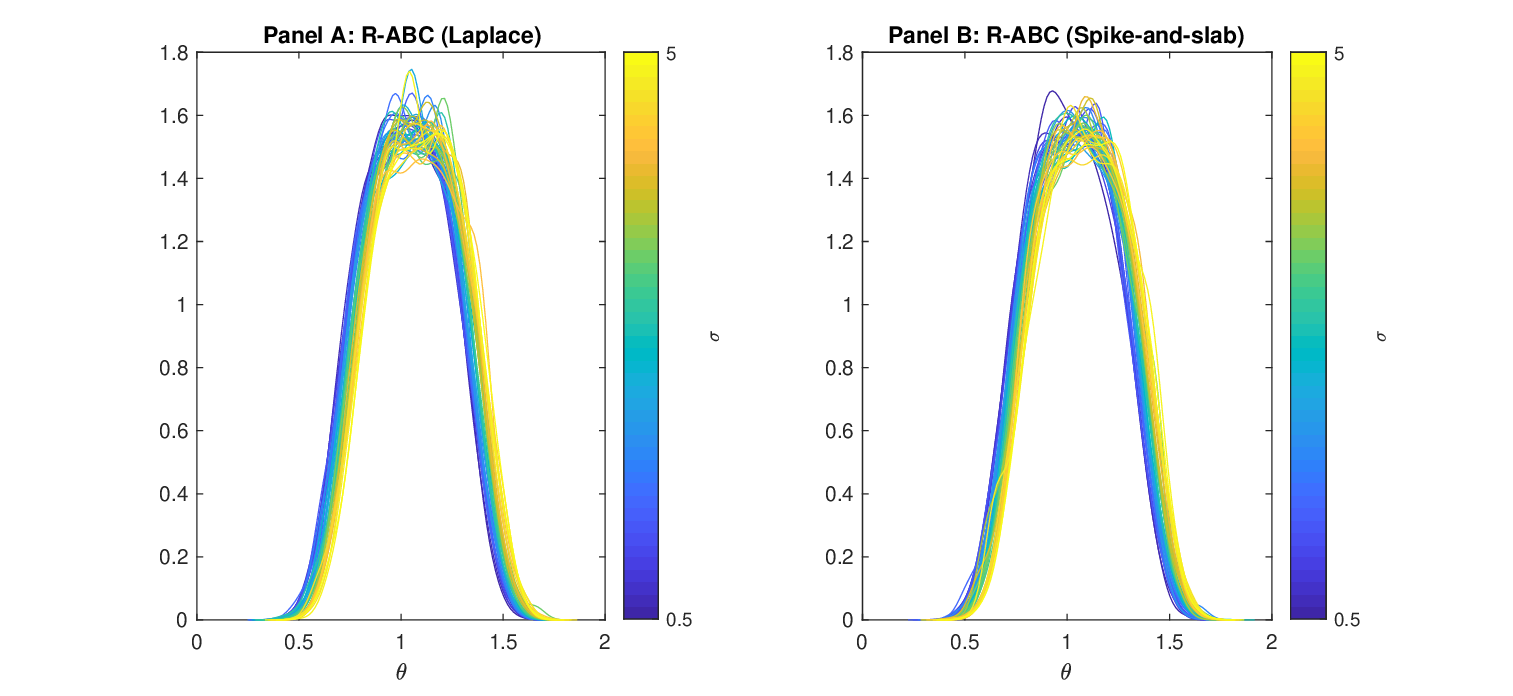} 
	\caption{Panels A and B respectively provide the posteriors of $\theta$ from step two of the R-ABC approach across varying levels of model misspecification when using Laplace and spike-and-slab priors for adjustment components. The color scheme indicates the level of model misspecification, with darker colors encoding less misspecification.} 
	\label{norm_theta_rabc} 
\end{figure}

The results in Panels A and B of Figure \ref{norm_theta_rabc} respectively display the posteriors of $\theta$ obtained in step two when using Laplace and spike-and-slab priors for the adjustment components. In contrast to the ABC-SMC-Reg posterior in Figure \ref{norm_theta_stdabc}, the R-ABC posteriors in Figure \ref{norm_theta_rabc} are now not heavily influenced by model misspecification, and are centered over the true value $\theta=1$ regardless of the level of model misspecification. Specifically, the posterior `drift' that was previously in evidence for ABC-SMC-Reg is no longer in evidence for R-ABC.

More specifically, the ABC-SMC posteriors in Figure \ref{norm_theta_stdabc} are highly concentrated around $\theta=1$ when $\sigma$ is low, but as $\sigma$ increases, the posteriors become wider and less peaked, indicating that a high level of model misspecification leads to increased uncertainty in parameter estimation. In contrast, the R-ABC posterior distributions are not sharply concentrated around $\theta=1$ when $\sigma$ is low. Moreover, as $\sigma$ increases, R-ABC posteriors maintain a similar level of concentration around $\theta=1$, showing a stable distribution even under higher levels of model misspecification.

\begin{figure}[h!]
	\centering 
	\includegraphics[width=0.7\linewidth]{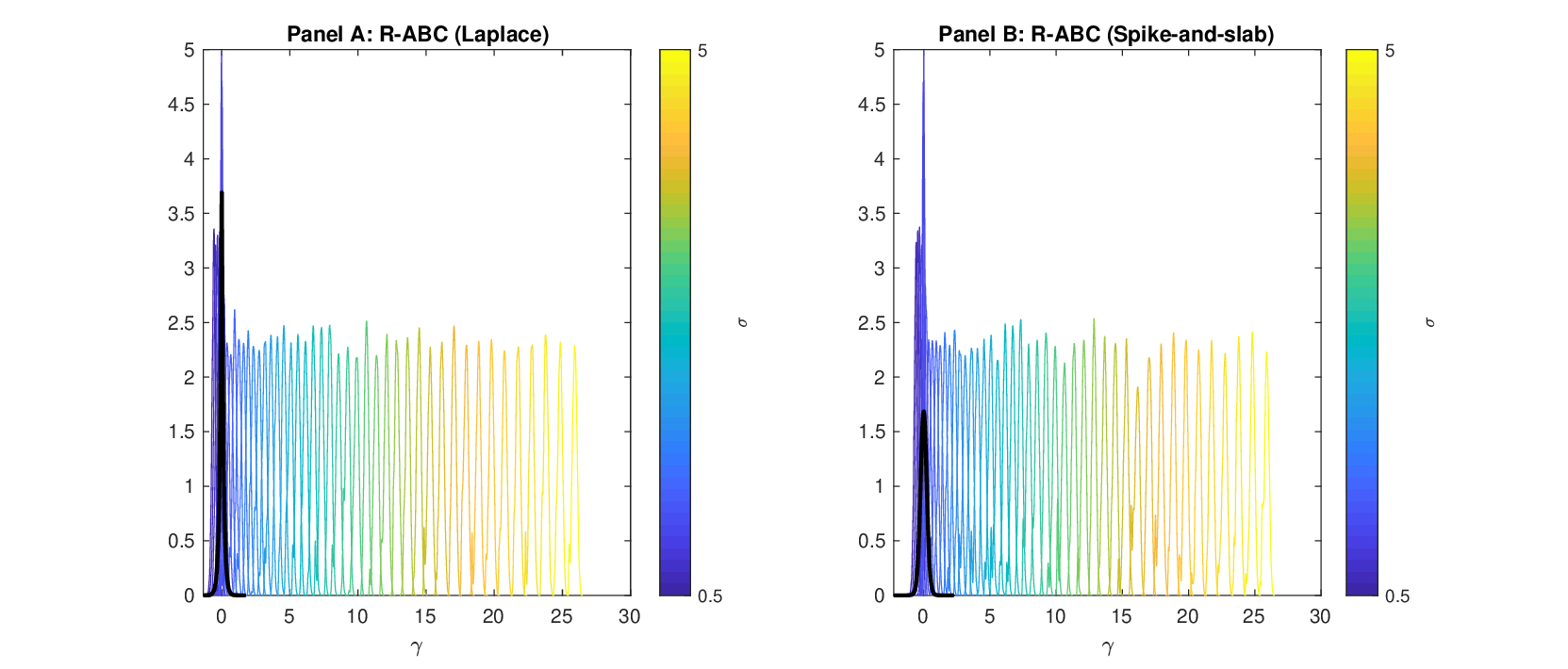} 
	\caption{Posteriors for the adjustment component $\gamma$ across varying levels of model misspecification. Panel A shows results using Laplace prior for $\gamma$, while Panel B presents those with the spike-and-slab prior. The color scheme indicates the level of model misspecification, with darker colors encoding less misspecification. The black solid lines in Panels A and B respectively denote the Laplace and spike-and-slab priors, both with the default hyperparameter $\lambda = 0.125$.} 
	\label{norm_gam_rabc} 
\end{figure}

Figure \ref{norm_gam_rabc} plots the posteriors for the adjustment components obtained from R-ABC when using Laplace and spike-and-slab priors. The results in both panels demonstrate that the posterior mean for the $\gamma$ component shifts further away from the prior mean of zero, as the level of misspecification increases. This behavior allows the R-ABC posteriors of $\theta$ to remain centered around the true value ($\theta = 1$) even as model misspecification increases. This result highlights that, as suggested, the model cannot reliably match the second summary statistic, the sample variance. Therefore, it is evident that the behavior of the posterior components for $\gamma$ can be used to detect which summary statistics the assumed model cannot match.

The above results are not a function of any given sample but are persistent across different samples. To demonstrate this feature we consider a repeated sampling version of the above Monte Carlo experiment, for $\sigma=2.$ For $\sigma=2$, we simulate 50 replications of the observed data and run the above-mentioned different ABC procedures across each of the different data sets. For each replication, we record the bias of the posterior mean, the posterior standard deviation, and across the replications we calculate the Monte Carlo coverage of each procedure. The results for the different procedures are presented in Table \ref{tab_norm}.

{\renewcommand{\arraystretch}{0.8}
\begin{table}[h!]
	\caption{\scriptsize Monte Carlo coverage (Cov), bias of the posterior mean (Bias), and average posterior standard deviation (Std) for $\theta$ in the simple normal example under model misspecification when $\sigma=2$. Cov is the percentage of times that the marginal 95\% credible set contains $\theta^*=1$. Std is the average posterior standard deviation across the Monte Carlo trials. The rows refer to the ABC approach where the last two rows represent R-ABC with different priors for the adjustment components, Laplace and spike-and-slab, respectively.}
	\centering%
	\small
	\begin{tabular}{lrrrrrrr}
		\hline\hline
		&   \multicolumn{3}{c}{$\theta$}     & & \\
		\cline{2-4} 
		& \multicolumn{1}{c}{Cov} & \multicolumn{1}{c}{Bias} & \multicolumn{1}{c}{Std} \\
		{\textbf{ABC method}} \\
		ABC-SMC	&92\%	&	-0.0597	&	0.3725	\\
		ABC-SMC-Reg	&52\%	&	-0.0728	&	0.0919	\\
		R-ABC (Laplace)	&96\%	&	0.0170	&	0.2098	\\
		R-ABC (Spike-and-slab) 	&96\%	&	0.0148	&	0.2100	\\
		\hline\hline
	\end{tabular}%
	\label{tab_norm}%
\end{table}
}

The results in Table \ref{tab_norm} demonstrate that ABC-SMC-Reg has very poor coverage at $\sigma=2$. However, we see that the R-ABC posteriors (either using Laplace or spike-and-slab priors for the adjustment components) display much more reasonable coverage compared to ABC-SMC-Reg. ABC-SMC also achieves a reasonable coverage, that is slightly lower than nominal level. The bias of the posterior mean is roughly similar across R-ABC approaches and is the smallest bias compared to ABC-SMC and ABC-SMC-Reg. Comparing posterior variability across the different methods, we see that ABC-SMC displays the largest posterior uncertainty, while ABC-SMC-Reg displays the smallest posterior uncertainty. Notably, R-ABC yields posterior uncertainties that are smaller than ABC-SMC, but larger than ABC-SMC-Reg.

%Overall, given the reasonable coverage, relatively small posterior uncertainty, and the well-centered nature of the posterior, we argue that the R-ABC approach performs best compared to other ABC approaches under model misspecification.

\bibliographystyle{apalike} 
\bibliography{ref}

\end{document}